%% file: Revised_kondo_impurity.tex
\RequirePackage{ifpdf}
\documentclass[letterpaper]{article}
\pdfoutput=1

\usepackage[usenames,dvipsnames]{pstricks}
\usepackage{jheppub}
\usepackage[utf8]{inputenc}
\usepackage[english]{babel}
\usepackage[babel]{csquotes}
\usepackage{amsmath}
\usepackage{amsfonts}
\usepackage{amssymb}
\usepackage{mathtools,colonequals}
\usepackage{mathrsfs}
\usepackage{bm}
\usepackage{bbold}
\usepackage{braket}
\usepackage{cases}
\usepackage{graphicx}
\usepackage{graphics,multicol}
\usepackage{tikz}
\usepackage{physics}
\usepackage{soul}
\usetikzlibrary{decorations.pathmorphing}
\usetikzlibrary{decorations.markings}
\usepgflibrary{shapes.geometric}
\usepgflibrary{shapes.symbols}
\usetikzlibrary{arrows.meta}
\usetikzlibrary{shapes.misc}
\usetikzlibrary{fadings}
\newcommand{\sign}{\text{sign}}
\tikzset{cross/.style={cross out, draw=black, minimum size=2*(#1-\pgflinewidth), inner sep=0pt, outer sep=0pt},
cross/.default={1pt}}

\usetikzlibrary{arrows}

\usepackage{verbatim}
\usepackage{booktabs}
\usepackage{array}
\usepackage{color}
\usepackage{subcaption}
\usepackage[percent]{overpic}
\usepackage{multirow}

\input{defs}

\title{The Kondo impurity in the large spin limit}

\author{Abijith Krishnan and}
\author{Max A. Metlitski}
\affiliation{Department of Physics, Massachusetts Institute of Technology, Cambridge, MA  02139, USA}
\emailAdd{mmetlits@mit.edu}

\abstract{
The Kondo problem, which describes the interaction of a spin $s$ magnetic impurity with a free Fermi gas,
is a classic example of strongly coupled physics. Historically, the problem {has been} solved by Wilson’s numerical renormalization group and later by  Bethe ansatz.  In this paper, we present an alternate analytic solution of the Kondo problem that combines an expansion in $1/s$ with { {the}} renormalization group.  We study both the case of { an impurity interacting with a single channel $K=1$ of fermions} 
in the $s \to \infty$ limit and the {{case}} 
with $K$ channels in the double-scaling limit $K\to\infty$, $s \to \infty$, $K/s$ fixed. Our approach allows us to {describe analytically} 
intermediate scales of the Kondo problem at large $s$ and compute thermodynamic observables such as the impurity entropy  and susceptibility. 
{We find these observables to agree with the} Bethe ansatz results. We also compute the impurity spectral function, finite temperature resistivity{ and the Kondo screening cloud profile}, properties that are not easily accessible from Bethe ansatz. Notably, in the regime $K > 2s$ {we access} 
the ``non-Fermi-liquid" overscreened fixed point of the multichannel Kondo problem. 

}

\begin{document}

\maketitle

\section{Introduction} \label{sec:intro}
 The single impurity Kondo problem, a classic example of strongly coupled physics, has been studied since the 1930s, when experiments revealed that gold wires with dilute magnetic impurities exhibited a resistance minimum as a function of temperature \cite{gold_experiment}. This system can be modeled with the Kondo Hamiltonian, which treats each impurity as a spin coupled through an antiferromagnetic exchange interaction to free conduction electrons. The name originates from the work of Jun Kondo, who found a logarithmic upturn with temperature in a perturbative calculation of resistance in this model\cite{Jun_Kondo}.
 
 Over the past sixty years, the Kondo problem has served as a testing ground for new ideas and methods in theoretical physics. In particular, it has played an important role in the development of renormalization group (RG). First, Anderson used perturbative RG to show that the weak coupling (free impurity spin) fixed point in the Kondo model is unstable, and that an infrared (IR) energy scale, the Kondo energy $T_K$, is dynamically generated \cite{Anderson_1970}. Wilson later used  the numerical renormalization group (NRG) to understand the full crossover from the weak coupling fixed point at high energy to the strong coupling fixed point at low energy \cite{Wilson}. At the strong coupling fixed point, a spin-$1/2$ impurity is exactly screened by one conduction electron and gives rise to a paramagnetic state. An impurity of spin $s > 1/2$ is underscreened in the IR by one conduction electron, and the remaining spin $s-1/2$ moment has a weak residual ferromagnetic coupling to the conduction electrons. These findings of NRG were later confirmed by an exact Bethe ansatz solution \cite{Andrei1, Wiegmann_1981,Kondoarbitrarys,Kondothermo,KondoReview,Kondoarbitraryscorrection1,numerics}, which exists for an arbitrary impurity spin $s$ and most directly accesses thermodynamic observables, such as impurity magnetization and entropy. Experiments have provided evidence for these theoretical findings through scanning tunneling microscopy (STM) measurements of a Kondo resonance near the Fermi surface \cite{Kondo_resonance_1,Kondo_resonance_2}, measurements of the Kondo screening cloud \cite{cloud_2011,cloud_2020}, and direct measurements of the impurity magnetization \cite{screening_direct}. 

In the 1990s, the framework of boundary conformal field theory (BCFT) and modern RG language provided further understanding of the ultraviolet (UV) and IR fixed points of the Kondo problem \cite{AFFLECK1991641,affleck1993,Affleck:1995ge,Affleck2000}. 
However, to date, only NRG and the Bethe ansatz have accessed the crossover between these fixed points.\footnote{See, however, our discussion of Ref.~\cite{hu2022kondo} below.} In this paper, {we fully describe} 
this crossover using analytical RG calculations when the impurity spin $s$ is large. 
In particular, we compute thermodynamic quantities such as the finite temperature impurity entropy, finite temperature impurity magnetic susceptibility, and the zero temperature impurity magnetization in an external magnetic field. These results agree analytically with the Bethe ansatz solution at large $s$. We also compute the impurity spectral function, finite temperature resistivity, and the profile of the Kondo screening cloud --- properties that are not easily accessible from Bethe ansatz.

We also extend our treatment to the $K$-channel Kondo model. It is known that depending on the ratio of $K/s$ the model admits three regimes: the underscreened regime, $K < 2s$, the fully screened regime, $K = 2s$, and finally the ``non-Fermi-liquid" overscreened regime, $K > 2 s$, which is qualitatively different from the single channel case. We consider the double-scaling limit when $K \to \infty$, $s \to \infty$ and $K/s$ is finite, and show that the crossover from the free spin UV fixed point to the IR fixed point can be fully described both in the underscreened and overscreened regimes. Again, our findings for the impurity entropy, susceptibility and magnetization agree with 
the exact Bethe ansatz solution of the multichannel Kondo problem \cite{PhysRevLett.52.364, WiegmannMultiShort, WiegmannMultiLong}  in the double-scaling limit. As in the single channel case, in addition to the thermodynamic observables, we compute the impurity spectral function, finite temperature resistivity, and Kondo cloud profile and find agreement with the CFT predictions at the overscreened fixed point \cite{affleck1993, cloud_1998}. While the existence of the non-trivial double-scaling limit $K \to \infty$, $s \to \infty$, $K/s$ --- fixed was already observed in the Bethe ansatz solution \cite{WiegmannMultiShort, WiegmannMultiLong}, our work provides a RG understanding of this limit. We note that this double-scaling limit is much richer than the limit $K \to \infty$, $s$ --- fixed, where the IR fixed point is accessible from the UV fixed point by a ``short" RG flow.\cite{NozierBlandin, GanAndrei}  

Our treatment of the multichannel Kondo model is related to that of Ref.~\cite{hu2022kondo}, which considered a generalization of the Kondo model with bosonic and fermionic bulk degrees of freedom in the same double-scaling limit. We briefly discuss the relation between our work and Ref.~\cite{hu2022kondo} when the bosonic degrees of freedom are absent. Our non-perturbative $\beta$-function exactly agrees with that of Ref.~\cite{hu2022kondo}. However, Ref.~\cite{hu2022kondo} focused on the fully screened regime $K = 2s$ and did not address that this approach can access the overscreened non-Fermi-liquid fixed point. In fact, we find that the fully screened regime is the only one where the RG treatment breaks down at and below the Kondo temperature $T_K$.\footnote{Since physical observables in the double scaling limit are generally a function of $ (1/s) \log T/T_K$, the crossover behavior in the regime $T\gg T_K$ is still  non-trivial and described correctly by the RG treatment. Furthermore, as $T$ approaches $T_K$ from above, the RG results suggest an approach to a fully screened fixed point, as pointed out in Ref.~\cite{hu2022kondo}.}   In addition, Ref.~\cite{hu2022kondo} did not compute physical observables, such as susceptibility, specific heat, and resistivity along the crossover. Finally, our treatment's technical details slightly differ from those of Ref.~\cite{hu2022kondo}: we believe that our adaptation of the methods of Refs.~\cite{Metlitski:2020cqy, Cuomo2022} described below elucidates the physics of the crossover.

The main insight of our approach is the following: when the magnitude of the impurity spin $s$ is infinite, the impurity spin does not fluctuate, acts on the electrons as a static  exchange field, and produces a simple boundary condition characterized by opposite scattering phase shifts for up and down electrons. Thus, at $s = \infty$ we have a line of impurity fixed points characterized by the phase shift $\rho$. When the bare dimensionless Kondo coupling $J \ll 1$, the UV value of $\rho \ll 1$. Once $s$ is taken to be large but finite, small impurity spin fluctuations lead to an analytically tractable RG flow along this line of fixed points. In the single channel case, the phase shift flows to $\rho = \pi$ in the IR; this is exactly the strong coupling fixed point with impurity spin $s -1/2$. In the multichannel case, for $K < 2s$, the phase shift again flows to $\rho = \pi$, corresponding to the underscreened impurity with remnant spin $s - K/2$. However, for $K > 2s$, there is an intermediate coupling stable fixed point at $\rho \approx 2 \pi s/K$ --- this is the non-Fermi-liquid overscreened fixed point.

This paper is inspired by recent advances in boundary and defect conformal field theory, in particular, on   
a number of problems where a critical bulk state described by conformal field theory (CFT) hosts a boundary or defect that is almost ordered magnetically. One example is the previously unknown extraordinary-log boundary universality class of the 3D classical O(N) model \cite{Metlitski:2020cqy, Toldin2020,  DengEOL,  Padayasi:2021sik, ToldinMM, Krishnan:2023cff, Cuomo:2023qvp}. Another example is a magnetic impurity with large spin in a 2+1d critical O(3) model \cite{Cuomo2022}. In both examples, one treats the defect degrees of freedom semi-classically with a non-linear $\sigma$-model and considers small fluctuations of the defect spin(s) about an ordered state that  breaks the spin-rotation symmetry of the system. The crux of the construction is that, through carefully coupling the defect spin fluctuations to the bulk critical fields, one restores the  spin-rotation symmetry.
In the present paper, we adapt this construction, developed in \cite{Metlitski:2020cqy, Padayasi:2021sik,  Cuomo2022, Krishnan:2023cff, Cuomo:2023qvp}, to the Kondo problem.

This paper is organized as follows. Section \ref{sec:action} constructs the Kondo impurity effective action using the above BCFT techniques. Section \ref{sec:RG} computes the RG flow for the single impurity Kondo model in the large $s$ limit. Section \ref{sec:thermo} uses the results of section  \ref{sec:RG} to compute thermodynamic observables for the impurity. Then, section \ref{sec:check} compares these results to the Bethe ansatz analytic and numeric results. In section \ref{sec:multi}, we extend our results to the multichannel Kondo problem. Then, section \ref{sec:spectral_function} computes the impurity spectral function and the finite temperature resistivity, and section \ref{sec:cloud} computes the profile of the Kondo screening cloud. Finally, we conclude with an outlook in section \ref{sec:future}.

\section{The Kondo impurity action in the large spin limit}\label{sec:action}
We begin with the single channel Kondo model. As in the standard treatment of the Kondo problem \cite{Affleck:1995ge}, we use the free nature of the conduction electrons and the assumption of $s$-wave scattering of electrons by the magnetic impurity to reduce the 3d Kondo model to an effective 1d theory where the spin $s$ impurity sits at the boundary of a semi-infinite 1d free electron CFT. The resulting action is
\begin{equation}\label{eq:folded1}
    S = S_B + \int \dd \tau \int_0^\infty \dd x [\psi_{L\alpha}^\dagger(\partial_\tau + iv_F \partial_x) \psi_{L\alpha} +\psi_{R\alpha}^\dagger(\partial_\tau - iv_F \partial_x) \psi_{R\alpha}] +  J v_F\int_{x=0} \dd \tau  \psi_{R}^\dagger\frac{\sigma^a}{2} \psi_{R}(x) S^a.
\end{equation}
Here, the term $S_B$ is the Berry phase action for the impurity spin, the operator $S^a$ (where $a \in \{1, 2, 3\}$) is the impurity spin operator, the fields $\psi_{R/L, \alpha}$ are the 1+1d free electron right/left moving fields, $J  >0$ is an effective dimensionless antiferromagnetic Kondo coupling, and $\alpha \in \{\uparrow, \downarrow\}$ runs over electron spin indices. The electrons live on a semi-infinite line $x > 0$, and the impurity spin is located at $x = 0$. The electron fields are subject to a boundary condition $\psi_{R \alpha}(x=0) = \psi_{L \alpha}(x=0)$. For the rest of this paper, we work in units where the Fermi velocity is $v_F=1$.

In this section, we first consider the Kondo impurity action in the $s=\infty$ limit, where the direction of the impurity spin is frozen. In this limit, we have a simple boundary condition on the bulk electrons, characterized by opposite phase shifts for up and down electrons. We then ``unfreeze" the impurity spin and use the techniques of Refs.\ \cite{Metlitski:2020cqy,Padayasi:2021sik, Krishnan:2023cff, Cuomo2022, Cuomo:2023qvp} to write an effective action for impurity spin fluctuations, which is fixed order by order in $1/s$ by the $SU(2)$ spin-rotation symmetry.

\subsection{The infinite \texorpdfstring{$s$}{s} limit}
We first begin with the $s=\infty$ limit, where the impurity spin is frozen. The Kondo interaction term is thus effectively a static impurity potential. Without loss of generality, we take the impurity spin to point along the positive $z$ axis. Then, the impurity acts as an opposite sign delta function potential for the spin-up electrons and spin-down electrons. In other words, the effective action at $s=\infty$ is 
\bea
        \int \dd \tau \int_{0}^\infty \dd x [\psi_{L\alpha}^\dagger(\partial_\tau + i \partial_x) \psi_{L\alpha} + \psi_{R\alpha}^\dagger(\partial_\tau - i \partial_x) \psi_{R\alpha}]+ sJ  \int \dd \tau \psi_{R\alpha}^\dagger(0,\tau)\frac{\sigma^3_{\alpha \beta}}{2} \psi_{R\beta}(0,\tau).
\eea
Here and below, we take $s \to \infty$ keeping $s J$ fixed. The frozen spin induces a phase shift $\rho$ between the spin-up left and right movers, and $-\rho$ between the spin-down left and right movers:\footnote{Note that our normalization is not standard: typically, the scattering phase shift $\delta$ is defined as half of our $\rho$.}
\begin{equation}
    \psi_{L\uparrow}(0^+,\tau) = e^{i\rho}\psi_{R\uparrow}(0^+,\tau), \quad \psi_{L\downarrow}(0^+,\tau) = e^{-i\rho}\psi_{R\downarrow}(0^+,\tau). \label{def_phase_shift}
\end{equation}
We use $x=0^+$ to denote an infinitesimal distance from the origin. We call $\rho$ the Kondo phase shift. For $s J \ll 1$, $\rho \approx s J/2$. For general $J$, the relation between $\rho$ and $J$ depends on the regularization of the $\delta$-function potential. For a particular $\delta$-function regularization choice used in Refs.\ \cite{KondoReview, hu2022kondo}, $\rho = 2\arctan(sJ/4)$.\footnote{Ref.\ \cite{hu2022kondo}'s $J$ is equivalent to our $sJ/4$.} We derive this relation in App.\ \ref{app:unfolded} in order to make contact with these references. In what follows, all our results {are} 
expressed in terms of the phase shift $\rho$, rather than $J$.

From \eqref{def_phase_shift}, we find the following correlation functions (for spin-down electrons, we swap $\rho$ with $-\rho$):
\begin{align}
    \ev{\psi_{R\uparrow}(x,\tau) \psi^\dagger_{L\uparrow}(x',0)} &=  \frac{e^{-i\rho}}{2\pi (\tau - i(x+x'))},\quad x,x' > 0, \nn\\
    \ev{\psi_{L\uparrow}(x,\tau) \psi^\dagger_{R\uparrow}(x',0)} &=  \frac{e^{i\rho}}{2\pi (\tau + i(x+x'))},\quad x,x' > 0, \nn\\
    \ev{\psi_{R\uparrow}(x,\tau) \psi^\dagger_{R\uparrow}(x',0)} &=  \frac{1}{2\pi (\tau - i(x-x'))},\quad x,x' > 0, \nn \\
    \ev{\psi_{L\uparrow}(x,\tau) \psi^\dagger_{L\uparrow}(x',0)} &=  \frac{1}{2\pi (\tau + i(x-x'))},\quad x,x' > 0, \label{psipsi1}
\end{align}
Thus, the action at $s=\infty$, $S_\text{frozen}$, is effectively parameterized by $\rho$:
\begin{equation}
    S_\text{frozen}(\rho) = \int \dd \tau \int_0^\infty \dd x [\psi_{L\alpha}^\dagger(\partial_\tau + i \partial_x) \psi_{L\alpha} + \psi_{R\alpha}^\dagger(\partial_\tau - i \partial_x) \psi_{R\alpha}], \quad \psi_{R\alpha}(0^+,\tau) = e^{-i\alpha \rho}\psi_{L\alpha}(0^+,\tau), \label{Sfrozen1}
\end{equation}
where $\alpha \in \{1,-1\}$ for the spin index. The boundary condition preserves the full $SU(2)$ spin-rotation symmetry when $\rho$ is a multiple of $\pi$. Otherwise, only a $U(1)$ subgroup of $SU(2)$ is preserved.

The action $S_\text{frozen}(\rho)$ has four  marginal bosonic boundary operators: 
\begin{itemize}
\item $\psi^{\dagger}_{R\alpha} \sigma^3_{\alpha \beta} \psi_{R\beta}(0^+,\tau)$. This operator tunes the system along the line of fixed points parameterized by $\rho$. Indeed, turning on a perturbation 
\beq S = S_\text{frozen}(\rho) + c \int d \tau \, \psi^{\dagger}_{R\alpha} \sigma^3_{\alpha \beta} \psi_{R\beta}(0^+,\tau), \label{drho}\eeq 
is equivalent to changing $\rho \to \rho + c + O(c^2)$, as can be checked by computing the corrections to fermion propagators to first order in $c$.
\item $\psi^{\dagger}_{R\alpha} \sigma^i_{\alpha \beta} \psi_{R\beta}(0^+,\tau)$. Here and below $i, j, k$ run over $1, 2$. These operators form a vector under the $U(1)$ rotation symmetry that remains after fixing the direction of the impurity spin. Physically, adding them to the action corresponds to ``tilting" the direction of the frozen impurity spin. We discuss in the next section the linear coupling of these operators to spin fluctuations.
\item $\psi^{\dagger}_{R \alpha} \psi_{R\alpha}(0^+,\tau)$. We note in passing the (unitary) particle-hole symmetry of the Kondo problem:
\beq 
\label{eq:particle_hole}
\psi_{R \alpha} \to \epsilon_{\alpha \beta} \psi^{\dagger}_{R \beta}, \quad \psi_{L \alpha} \to \epsilon_{\alpha \beta} \psi^{\dagger}_{L \beta}, \quad \vec{S} \to \vec{S}, \quad i \to i.\eeq
\end{itemize}
For simplicity, we assume that the UV regulator of the Kondo model (\ref{eq:folded1}) does not break this symmetry. Because, $\psi^{\dagger}_{R\alpha} \psi_{R\alpha}(0^+,\tau)$ is odd under particle-hole symmetry, it cannot appear in the effective impurity action. The other three fermion bilinears $\psi^{\dagger}_{R\alpha} \vec{\sigma}_{\alpha \beta} \psi_{R\beta}(0)$  above are particle-hole even.\footnote{When particle-hole symmetry is broken and a potential scattering term $\psi^{\dagger} \psi(0)$ is added to the action, it simply introduces an overall phase-shift in the charge channel without affecting the spin channel \cite{affleck1993}.}

\subsection{Restoring \texorpdfstring{$SU(2)$}{SU(2)} symmetry for large, finite \texorpdfstring{$s$}{s}}
\label{SU2restore}
At large finite $s$, spin fluctuations are restored and, as we show in Sec.\ \ref{sec:RG}, induce an RG flow in the parameter $\rho$. 
Parameterizing the spin $\vec{S} = s \vec{n}$, $\vec{n}^2 = 1$ we consider small fluctuations of $\vec{n}$ around the North pole: $\vec{n} = (\pi_1, \pi_2, \sqrt{1-\vec{\pi}^2})$.\footnote{More generally, we may divide the two sphere on which $\vec{n}$ lives into a set of small patches and for each patch consider the fluctuations around the patch center. We do not need to do this explicitly in this paper.} We expect the effective action to be 
\beq S_{\rm eff} = S_{\rm frozen}(\rho) + S_B[\vec{n}] + S_{\rm int} + S_{\rm cont}, \label{Sefffull}  \eeq
where $S_{\rm frozen}$ is given by \eqref{Sfrozen1}, $S_B[\vec{n}]$ is the Berry phase action of the spin:
\begin{equation}
    S_B[\vec{n}] = is\int \dd \tau \int_0^1 \dd u [\vec{n} \cdot \left(\partial_u \vec{n} \times \partial_\tau \vec{n}\right)] = \frac{i s}{2} \int d\tau (\pi_1 \d_\tau \pi_2 - \pi_2 \d_\tau \pi_1) + O(\pi^4),
\end{equation}
and $S_{\rm int}$ couples $\pi^i$ to marginal boundary operators of $S_\text{frozen}$: $\psi^{\dagger}_{R} \sigma^i \psi_{R}(0^+,\tau)$ and $\psi^{\dagger}_{R} \sigma^3 \psi_{R}(0^+,\tau)$. This coupling is exactly fixed by $SU(2)$ spin-rotation symmetry and can be determined order by order in $\vec{\pi}$. Below, we fix the coupling to order $\vec{\pi}^2$. The term 
$S_{\rm cont}$ includes boundary contact terms involving the spin fluctuations $\pi^i$ and is also fixed by $SU(2)$ symmetry.

\subsubsection*{The interaction terms}
We begin by applying the methods of \cite{Padayasi:2021sik, Cuomo2022} to fix $S_{\rm int}$ to linear order in $\pi$. Freezing the impurity spin along the North pole breaks $SU(2)$ symmetry on the boundary. Therefore, $S_{\rm frozen}(\rho)$ transforms non-trivially under infinitesimal global $SU(2)$ rotations around the $x$ and $y$ axes:
\bea \delta \psi_{R/L}(x) &=& -i \theta^a \frac{\sigma^a}{2} \psi_{R/L}(x), \quad\quad \delta n^a = \epsilon^{a b c} \theta^b n^c, \nn\\
\delta S_{\rm frozen}(\rho) &=& - \theta^i \int d \tau \, j^i_{x}(0^+, \tau). \label{dSfroz}\eea
Here $\theta^a$ are infinitesimal rotation angles, $j^a_{x}( x, \tau)$ is the bulk $SU(2)$ spin current:
\begin{equation}\label{eq:spin_current}
    j^{a}_x(x, \tau) = \frac{1}{2}  \left[\psi_{R}^\dagger \sigma^a \psi_{R} - \psi_{L}^\dagger \sigma^a \psi_{L}\right],
\end{equation}
and we should think of (\ref{dSfroz}) as implementing an $SU(2)$ rotation by integrating the spin current over a contour enclosing the entire bulk. For a BCFT where the boundary explicitly breaks a symmetry,  the bulk to boundary OPE of the currents corresponding to broken symmetry generators has a nonsingular leading term that 
yields a marginal boundary operator, known as the ``tilt" \cite{Herzog_2017,Metlitski:2020cqy, Padayasi:2021sik}. In fact, using the boundary condition of $S_{\rm frozen}$ in Eq.~(\ref{def_phase_shift}), the $j^i_x$ OPE is
\beq j^i_x( 0^+, \tau) = \sin\rho \left[\sin \rho \, \psi^{\dagger}_R \sigma^i \psi_R (0^+, \tau)- \cos \rho\, \epsilon^{ij} \psi^{\dagger}_R \sigma^j \psi_R(0^+, \tau)\right], \quad i =1,2. \label{jpsiR}\eeq 
The current corresponding to the unbroken rotation generator satisfies $j^3_x(x, \tau) \to 0$ as $x \to 0$. The variation (\ref{dSfroz}) can be cancelled by the coupling
\beq S_{\rm int} = \int d \tau \, \epsilon^{ij} \pi^i(\tau) j^j_x(0^+, \tau) + O(\pi^2). \label{Sint1} \eeq

Another way to reproduce (\ref{Sint1}) is to initially treat $\vec{\pi}$ as a constant in time, corresponding to a small rigid rotation of the direction of the impurity spin. The fermion two point functions (\ref{psipsi1}) then undergo a corresponding rigid rotation, which must be reproduced by treating $S_{\rm int}$ (\ref{Sint1}) in perturbation theory on top of $S_{\rm frozen}$. This method also fixes higher order terms in $\pi$ in $S_{\rm int}$. In App.\ \ref{app:kappa_term}, we show that the $O(\pi^2)$ contribution to $S_{\rm int}$ is
\beq \label{eq:kappa_term} \delta S_{\rm int} = r \int d \tau\, \vec{\pi}^2(\tau)  \psi^{\dagger}_R\sigma^3 \psi_R( 0^+, \tau), \quad r = -\frac{\sin 2 \rho}{4}.\eeq 
The precise value of $r$ does not affect our calculation below in the single channel problem but is important for the analysis of the multichannel problem in section \ref{sec:multi}.

Finally, note that $j^i_x( 0^+, \tau)$ and $r$ vanish if $\rho$ is a multiple of $\pi$. This is not a coincidence, as at these points $S_{\rm frozen}$ preserves the full $SU(2)$ symmetry. In fact, we expect  $S_{\rm int}$ to vanish at all orders in $\pi^i$ when $\rho$ is a multiple of $\pi$ --- the spin fluctuations fully decouple from the bulk fermion fields.

\subsubsection*{The contact terms}\label{sec:contact_terms}
We conclude this section by fixing the contact terms $S_{\rm cont}$ in the effective action. To quadratic order in $\pi^i$, the allowed contact terms are:
\beq  S_{\rm cont} =  \delta m_{\pi} \int d \tau \,  \vec{\pi}^2 + \frac{i \delta s_B}{2} \int d \tau \, \left(\pi_1 \d_\tau \pi_2 - \pi_2 \d_\tau \pi_1\right). \label{sBact} \eeq
The coefficient $\delta m_{\pi}$ is proportional to the UV cut-off, and to respect rotational invariance should be chosen so that the effective potential is independent of $\vec{\pi}$. We do this explicitly in App.\ \ref{app:berry_phase_counter}. The coefficient $\delta s_B$ of the Berry phase countact term can be fixed as follows. Consider a system of large, finite length $L$ with the impurity at $x = 0$ and a boundary condition $\psi_{R \alpha} = -\psi_{L \alpha}$ at $x  = L$. If we fix the direction of the impurity spin $\vec{n}$, the fermion ground state is separated from the excited states by a finite gap of order $1/L$: indeed, our choice of boundary condition at $x = L$ ensures that there are no level crossings for  $|\rho| < \pi$. Therefore, we can integrate the fermions out and obtain an effective action for $\vec{n}$ valid at energy scales $\omega \ll 1/L$. The total Berry phase term in this action should be quantized. Furthermore, because it is quantized as the microscopic impurity spin $s$ at $\rho=0$, it should be quantized as $s$ throughout by continuity. The effective action (\ref{Sefffull}) must reproduce this finite-size quantization of the spin Berry phase term. As shown in App.\ \ref{app:berry_phase_counter}, this requirement fixes 
\beq \delta s_B = -\frac{1}{2\pi}\left(\rho - \frac{1}{2}\sin 2 \rho\right). \label{deltasB}\eeq
At $\rho = \pi$, the strong coupling (under)screened fixed point, $\delta s_B = -1/2$. The bulk fermions decouple from boundary spin fluctuations, so we are left with a free boundary spin of magnitude $s-1/2$. This reduction of the effective Berry phase term at the strong coupling fixed point was also pointed out in Ref.~\cite{hu2022kondo}.

\section{RG} \label{sec:RG}

\begin{figure}[t]
\centering
  \includegraphics[width = 0.6\textwidth]{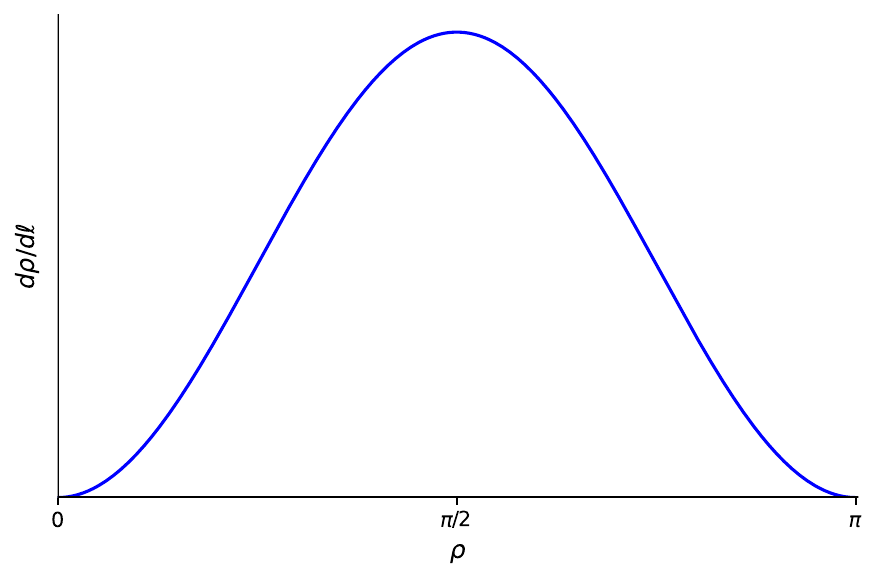}
 \caption{The $\beta$-function for the phase shift $\rho$ in the single channel Kondo model to leading order in $1/s$.} \label{fig:beta}
\end{figure}

We now use the action (\ref{Sefffull}) to derive the RG flow of the Kondo phase shift $\rho$ to order $1/s$. 
Integrating out high-energy modes to leading order in $1/s$ effectively generates a perturbation to the action:
\beq \delta S = -\frac{1}{2}\int_{e^{-d\ell} a < |\tau_1 - \tau_2| < a}  d\tau_1 d \tau_2 \,\, \epsilon^{ij} \epsilon^{kl} \pi^i(\tau_1) \pi^k(\tau_2) j^j_x(0^+, \tau_1) j^l_x(0^+, \tau_2) \label{dSRG}\eeq
where $a$ is a short-distance cut-off and $\ell$ is the RG flow parameter. Using $S_B$, the $\pi$-propagator  is:\footnote{We treat the zero mode of $\pi$ more carefully when we discuss the finite temperature properties in Sec.\ \ref{sec:thermo}.}
\beq \langle \pi^i(\tau) \pi^j(0)\rangle = \frac{i}{2s} \epsilon^{ij} {\rm sgn}(\tau). \label{piprop}\eeq
Since $S_{\rm frozen}$ is quadratic, we can use \eqref{psipsi1} to compute the two and three point functions of boundary fermion bilinears and find the boundary OPE:
\beq \psi^{\dagger}_R \sigma^a \psi_R (0^+,\tau) \psi^{\dagger}_{R} \sigma^b \psi_R(0^+, 0) \sim \frac{\delta^{ab}}{2 \pi^2 \tau^2} + \frac{i}{\pi \tau} \epsilon^{a b c} \psi^{\dagger}_R \sigma^c \psi_R (0^+,0) + \ldots. \label{psibiOPE}\eeq
so that
\beq \label{eq:ope_j} j^i_x(0^+, \tau) j^j_x(0^+, 0) \sim \sin^2 \rho \left( \frac{\delta^{ij}}{2 \pi^2 \tau^2} + \frac{i}{\pi \tau} \epsilon^{ij} \psi^{\dagger}_R \sigma^3 \psi_R(0^+, 0) + \ldots \right). \eeq
Inserting this into \eqref{dSRG} and using the $\pi$-propagator (\ref{piprop}) yields
\beq \delta S = \frac{\sin^2 \rho}{\pi s} d \ell \int d \tau \, \psi_R^{\dagger} \sigma^3 \psi_R(0^+, \tau). \eeq
As noted below \eqref{drho}, turning on $\psi_R^{\dagger} \sigma^3 \psi_R$ at the boundary precisely shifts the phase shift $\rho$, so
\beq \frac{d \rho}{d \ell} = - \beta (\rho) = \frac{1}{\pi s}  \sin^2 \rho + O(1/s^2). \label{beta} \eeq
The $\beta$-function is sketched in Fig.~\ref{fig:beta}. If we start with a small positive phase shift $0<\rho_0 \ll 1$ (i.e., weak antiferromagnetic coupling $0 < s J \ll 1$) in the UV, it flows to $\rho = \pi$ in the IR -- the strongly coupled (under)screened fixed point -- as 
\beq \rho(\ell) = \frac{\pi}{2} - \tan^{-1}\left(\cot \rho_0 - \frac{\ell}{\pi s}\right), \quad\quad \rho(\ell = 0) = \rho_0. \label{rhoell}\eeq
As $\ell \to \infty$, $\rho(\ell) \approx \pi (1 - s/\ell)$ -- the $\rho = \pi$ fixed point is approached logarithmically. This logarithmic flow corresponds to an  effective ferromagnetic coupling between the remnant impurity spin $s - 1/2$ and the bulk electrons.
We also observe dimensional transmutation: the dependence on the UV cut-off $\Lambda$ and the UV phase-shift $\rho_0$ can be traded for an energy scale $T_0 = \Lambda \exp{-\pi s \cot \rho_0}$
chosen so that $\rho(\omega = T_0) = \pi/2$.  Then
\beq \rho(\omega) = \frac{\pi}{2} - \tan^{-1}\left(\frac{\log{(\omega/T_0)}}{\pi s}\right)\label{eq:dim_transmutation}.\eeq
The energy scale $T_0$ can be thought of as the Kondo temperature. We note that for a spin-$s$ Kondo model several definitions of the Kondo temperature exist in the literature: the resulting temperatures differ by an $\mathcal{O}(1)$ numerical factor \cite{KondoReview}. As we discuss below, to the order in $1/s$ that we pursue here, we are not able to fix this pre-factor.

Note that the $\beta$-function (\ref{beta}) is fully non-perturbative in the coupling $J$ (which is traded for the phase shift $\rho$). Instead, the expansion parameter is $1/s$, which controls the impurity spin fluctuations. Recalling that for $s J \ll 1$, $\rho \approx  s J/2$, we can take the weak coupling limit of (\ref{beta}),
\beq \frac{d J}{d \ell} \approx \frac{J^2}{2 \pi}, \eeq
and $T_0 \approx \Lambda \exp{- 2\pi/J}$, which agrees with the leading perturbative result \cite{Affleck:1995ge}.

\section{Thermodynamics of the large-\texorpdfstring{$s$}{s} Kondo impurity}\label{sec:thermo}
\subsection{Entropy and specific heat}
To compute the entropy and specific heat, we consider a finite system of length $L$ at a temperature $T\gg 1/L$. We compute the free energy $F = - T \log Z$ and subtract the bulk contribution $ F_{\rm bulk} = -(\pi c/6) L T^2 $, where $c = 2$ is the bulk central charge, to obtain the impurity free energy $F_{\rm imp} = F - F_{\rm bulk}$. From the impurity free energy, we compute the impurity entropy and specific heat.

\subsubsection*{Finite temperature formalism }

We first treat the zero-mode of the impurity spin fluctuations.  
Following Ref.~\cite{Hasenfratz}, we write
\beq \vec{n}(\tau) = \vec{n}_0 \sqrt{1- \pi_\alpha(\tau) \pi_\alpha(\tau)} + \vec{e}_\alpha \pi_\alpha(\tau), \quad\quad \int_0^\beta d \tau \,\pi_\alpha(\tau) = 0 \label{spinfluctuations}.\eeq
where $\alpha$ runs over $1, 2$ and  $\vec{n}_0$, $\vec{e}_1$, $\vec{e}_2$ is a set of three time-independent orthonormal vectors. The path integral is then divided into an integral over $\vec{n}_0$ (the  average direction of $\vec{n}(\tau)$) and an integral over the fluctuations $\pi_\alpha(\tau)$. By rotational invariance, the integral over $\vec{n}_0$ contributes the surface area of the two-sphere and enters the partition function of a free spin. To perform the integral over the fluctuations $\pi_\alpha$, we freeze $\vec{n}_0$ along the North pole. Then, we recover the action (\ref{Sefffull}) with the understanding that the (constant in time) zero-mode part of $\pi_i(\tau)$ is removed from the path integral. The action also must be supplemented by a Jacobian term resulting from the transformation in the path integral from $\vec{n}(\tau)$ to $\vec{n}_0$ and $\pi_\alpha(\tau)$ \cite{Hasenfratz},
\beq  S_J = -2  \log\left[\int_0^{\beta} \frac{d \tau}{\beta} \sqrt{1 - \vec{\pi}^2(\tau)}\right]. \label{Jacobian}\eeq
To the order in $1/s$ that we consider, $S_J$  only affects the partition function of the free spin.\footnote{This term should not be confused with the term arising from the measure on the sphere, $\delta S = \frac{1}{2} \int \frac{d\tau}{a} \log \left(1 - \vec{\pi}^2(\tau)\right)$, where $a$ is the UV cut-off. As is clear, this last term is highly sensitive to the regularization procedure. Again to the order in $1/s$ we consider, it only affects the free spin partition function.}
With these remarks in mind, we obtain the $\pi$ propagator:
\beq \langle \pi^i(\tau) \pi^j(0)\rangle = - \frac{1}{\beta s} \epsilon^{ij} \sum_{\omega_n\ne 0} \frac1{\omega_n } e^{-i \omega_n \tau} = \frac{i}{s}\epsilon^{ij} \left(\frac{1}{2} {\rm sgn}(\tau) - \frac{\tau}{\beta}\right), \quad |\tau| < \beta. \label{pipropT} \eeq

\subsubsection*{Impurity entropy}
We now compute the partition function of the system from the action in \eqref{Sefffull}. To first subleading order in $1/s$, the partition function is
\begin{align}
    \mathcal{Z} = \mathcal{Z}_\text{frozen}(\rho) \mathcal{Z}_\text{spin}\bigg[1 &+ \frac{1}{2}\int_0^{\beta} \int_0^{\beta} \dd \tau \dd \tau' \epsilon^{ij}\epsilon^{kl} \ev{j^{j}_x(0^+, \tau) j^{l}_x(0^+, \tau')}_0 \langle \pi^i(\tau) \pi^k(\tau')\rangle_0 \nonumber\\  &-i\frac{\delta s_B}{2} \int_0^{\beta} \dd \tau \langle \pi_1 \partial_\tau \pi_2 - \pi_2 \partial_\tau \pi_1\rangle_0\bigg], \label{Zfull}
\end{align}
where all expectation values are evaluated in the decoupled theory.\footnote{We do not include the term in \eqref{eq:kappa_term} because the $\pi$ propagator (\ref{pipropT}) is off-diagonal. Furthermore, at finite temperature, $\ev{\psi^\dagger_{R} \sigma^3 \psi_{R}(0^+,\tau)} = 0$, up to exponentially small corrections in $LT$.} Here $\mathcal{Z}_\text{frozen}(\rho)$ is the partition function of the free fermion theory (\ref{Sfrozen1}) and $Z_{\rm spin} = 2s +1$ is the partition function of free spin $s$. In fact, $Z_{\rm frozen}(\rho)/e^{- \beta F_{\rm bulk}} = g_{\rm frozen} = 1$. Here $g_{\rm frozen} = 1$ is the boundary entropy of the free fermion theory. Because $\rho$ is an exactly marginal perturbation, this entropy does not depend on $\rho$ \cite{boundaryentropy}.

We now consider the $1/s$ corrections originating from the last two terms in brackets in \eqref{Zfull}. As shown in App.\ \ref{app:current_current}, $\langle j^i_x(0^+, \tau) j^j_x(0^+, \tau')\rangle_0 \propto \delta_{ij}$, so contracting this correlator with the $\pi$ propagator in \eqref{pipropT} yields zero. 
The last term in brackets can be evaluated from  the $\pi$ propagator, $\epsilon_{ij} \langle \pi^i \d_{\tau} \pi^j\rangle = -(2i/s) (\delta(0) - 1/\beta)$, where $\delta(0)$ only contributes to the non-universal impurity energy. Thus, 
\begin{equation}
\mathcal{Z}_{\rm imp} =  \mathcal{Z}/e^{-F_\text{bulk}/T} \approx 2s + 1 + 2 \delta s_B = 2s + 1 - \frac{\rho}{\pi }+ \frac{\sin(2\rho)}{2\pi }.
\end{equation}
Since $\mathcal{Z}_{\rm imp}$ is an RG invariant quantity, we may RG improve the above expression by replacing $\rho \to \rho(T)$ as defined in \eqref{eq:dim_transmutation}. In this way,
\beq \mathcal{Z}_{\rm imp} \approx 2s + 1 - \frac{\rho(T)}{\pi }+ \frac{\sin(2\rho(T))}{2\pi } + {\cal O}(1/s). \label{Zimp1}\eeq
The impurity free energy is thus $F_{\rm imp} = - T \log {\cal Z}_{\rm imp}$, and the 
the impurity entropy is 
\begin{equation}
\mathcal{S}_\text{imp} = -\frac{dF_\text{imp}}{dT} = \log\left[2s + 1 - \frac{\rho(T)}{\pi }+ \frac{\sin(2\rho(T))}{2\pi }\right] + \mathcal{O}(1/s^2).\footnote{Even though the $\log$ expression contributes to $\mathcal{S}_\text{imp}$ to $\mathcal{O}(1/s^2)$, we leave $\mathcal{S}_\text{imp}$ in this form to reproduce the correct behavior at the weak and strong coupling fixed points.} \label{eq:imp_entropy}
\end{equation}
Here, we use $T(\pdv*{\rho}{T}) =  \mathcal{O}(1/s)$. 
The impurity entropy at the strong coupling fixed point is $\mathcal{S}_\text{imp}(\rho = \pi) \approx \log(2s)$. We expect this result for the strong coupling fixed point to hold at all orders in $1/s$:  the impurity ``absorbs" an opposite spin-$1/2$ electron at the strong coupling fixed point.

In App.\ \ref{app:g_theorem} we present another derivation of the impurity entropy that applies the $g$-theorem of Ref.\ \cite{boundaryentropy}.

\subsubsection*{Specific Heat}
Finally, we derive the specific heat of the impurity noting that $C = T (\dv*{\mathcal{S}_\text{imp}}{T})$.
\bea 
C(T) &=& T \dv{\mathcal{S}_\text{imp}}{T}  = \beta(\rho) \dv{\mathcal{S}_\text{imp}}{\rho} =\frac{4\sin^4 \rho(T)}{\pi s(2\pi + 4\pi s - 2\rho(T) + \sin (2\rho(T)))} + {\cal O}(1/s^3) \label{eq:sp_heat_long} \\
&\approx&  \frac{\sin^4 \rho(T)}{\pi^2 s^2} + {\cal O}(1/s^3) \label{eq:sp_heat_rho}\eea
This function peaks at $\rho(T) = \pi/2$ as $s\to \infty$. From \eqref{eq:dim_transmutation}, we see that $C(T) \sim s^2 \pi^2 \log(T/T_0)^{-4}$ at low and high temperatures. This behavior agrees with the Bethe ansatz \cite{Kondothermo}.

\subsection{Magnetization and magnetic susceptibility} 
 \label{sec:magnet}
We now discuss the system's response to a uniform external magnetic field $\vec{h}$. We compute the zero field susceptibility at finite temperature, and we compute the magnetization as a function of $h = |\vec{h}|$ at zero temperature. 

\subsubsection*{Impurity spin absorption}

The field $\vec{h}$ enters the original Kondo action (\ref{eq:folded1}) as 
\beq \delta S_h = - \int d\tau h^a S^a(\tau) - \int d\tau dx\, h^a j^a_0(x, \tau), \label{hcoupl0}\eeq
where 
\beq j^a_0(x, \tau) = \frac{1}2\left(\psi^{\dagger}_R \sigma^a \psi_R + \psi^{\dagger}_L \sigma^a \psi_L\right) \label{hcouple0} \eeq
is the fermion spin density. 

We now discuss how to incorporate the magnetic field $\vec{h}$ into the large $s$ effective action (\ref{Sefffull}). 
The first term in (\ref{hcoupl0})  appears in the effective action as the leading coupling of $\vec{h}$ in the large $s$ limit. The second term in (\ref{hcoupl0})  also appears in the effective action  since it is fixed by the bulk spin-rotation symmetry, however, an issue of UV regulating this term as $x \to 0$ arises. We show that the following regularization  is needed to respect the $SU(2)$ rotation symmetry:
\beq S_{\rm eff}(h) = S_{\rm eff}(\rho, h=0)  - (s + \delta s_M) \int d \tau h^a n^a(\tau) - \int d\tau \int_{0^+}^\infty dx \, h^a j_0^a(x, \tau), \quad\quad \delta s_M = -\frac{\rho}{2 \pi}.  \label{Seffh} \eeq  
Here $j_0^a(x, \tau)$ in the last term is understood as an operator in the effective theory $S_{\rm frozen}(\rho)$ and the domain of the $x$ integral is understood as entirely described by $S_{\rm frozen}(\rho)$. The term proportional to $\delta s_M$ is a contact term describing the spin ``sucked in" by the impurity at finite $\rho$. Its magnitude can be fixed similarly to how $\delta s_B$ was determined in section \ref{SU2restore}. 

Indeed, consider the case when the spin is frozen along the $z$ direction and imagine changing the phase shift $\rho$ in $S_{\rm frozen}$. In general, some spin is expected to accumulate over a microscopic length scale near $x =0$. To find the magnitude of this sucked in spin we again consider a finite system of size $L$ at zero temperature and impose the boundary condition $\psi_{R} = -\psi_L$ at $x = L$. In App.\ \ref{app:finite_density}, we compute the expectation value of the bulk spin density with the effective action $S_{\rm frozen}(\rho)$ and find
\beq \langle j^3_0(x)\rangle = \frac{1}{L} \frac{\rho}{2\pi}, \label{j30}\eeq
i.e. there is a spin $\rho/2\pi$ uniformly smeared over the system away from the impurity. Since the ground state electron spin is zero at $\rho = 0$, no level crossings occur as $\rho$ is dialed from $0$ to $\pi$, and the electron spin in the $z$-direction is quantized, there must be an opposite spin
\beq \Delta S^z = \delta s_M = - \frac{\rho}{2\pi}, \label{sM}\eeq
accumulated at the impurity. This result is a special case of the Friedel sum rule for impurities \cite{Friedel,doniach_and_sondheimer,KondoReview, andrei_magnetoresistance}. In App.\ \ref{app:spin_current}, we also derive \eqref{sM} in a different way, by adiabatically changing the phase shift $\rho$ and calculating the spin current flowing into the impurity.

Note that at $\rho = \pi$, $\delta s_M  =-1/2$, i.e. the impurity spin is reduced by $1/2$ as is expected at the strong coupling fixed point. Curiously, for general $\rho$, $\delta s_M \neq \delta s_B$ in (\ref{deltasB}). While, for an isolated spin, rotational invariance indeed implies that the coefficient of the Berry phase term and the spin coupling to the magnetic field are equal, the lack of manifest $SU(2)$ rotational symmetry in our effective description generally allows the contact terms $\delta s_M$ and $\delta s_B$ to be different.

\subsubsection*{Zero temperature magnetization}
We now compute the magnetization of the impurity at $T = 0$ in finite magnetic field $h \hat{z}$. To leading order in $1/s$ and in the limit $L \to \infty$, the bulk magnetization $\langle j^3_0(x) \rangle = h/\pi$ is just given by the free bulk expression. Since the impurity spin is, to leading order, decoupled from the fermions we have $\langle n^3 \rangle_h \approx \langle 1 - \vec{\pi}^2/2 \rangle_h  = 1 + O(1/s^2)$. Therefore, to order $s^0${, the impurity magnetization is}
\begin{equation}
    M = -\dv{F_\text{imp}}{h} = (s + \delta s_M) \ev{n^3}_h + {\cal O}(1/s)= s - \frac{\rho}{2\pi} + {\cal O}(1/s),
\end{equation}
Because $h$ is a dimension-$1$ relevant coupling and it acts as a mass term for the spin fluctuations $\pi^i$, it becomes an IR regulator. Thus, we can RG improve $\rho \to \rho(\ell)$ with $\ell = \log(\Lambda/h)$:
\begin{equation}\label{eq:magnetization}
    M(h) =  s - \frac{1}{4} + \frac{1}{2\pi} \tan^{-1}\left(\frac{\log(h/\Lambda)}{\pi s} + \cot (\rho_0)\right) = s - \frac{1}{4} + \frac{1}{2\pi} \tan^{-1}\left(\frac{\log(h/T_0)}{\pi s} \right) + {\cal O}(1/s).
\end{equation}
We use that the total magnetization is an RG invariant. Because $M(h)$ decreases from $M = s$ at large $h$ to $M = s-1/2$ at $h \to 0$, it exhibits the (under)screened nature of the strong-coupling fixed point.

\subsubsection*{Magnetic susceptibility}
We now compute the $h=0$ magnetic susceptibility, $\chi_\text{imp} = -\frac{1}{3} \sum_a (\pdv*[2]{F_\text{imp}}{h_a})|_{h=0}$ for $T>0$. The full susceptibility of the entire system is 
\bea
    \chi  &=&  \frac{(s+\delta s_m)^2}{3} \int d \tau \langle n^a(\tau) n^a(0)\rangle
    + \frac{2(s + \delta s_m) }{3}   \int_0^\beta \dd \tau \int_{0^+}^\infty \dd x \langle n^a(\tau)   j^a_0(x,0)\rangle \nn\\
    &+& \frac{1}{3} \int_0^{\beta} d\tau \int_{0^+}^{\infty}  dx dx' \, \langle j_0^a(x, \tau) j_0^a(x',0)\rangle. \label{chi}
\eea
The last term in (\ref{chi}) is ${\cal O}(s^0)$ and contributes to the bulk susceptibility. To ${\cal O}(s^0)$, the bulk spin density $\langle j^a_0(x)\rangle$ is exponentially small in $LT$ at finite temperature. We are left with the first term:
\beq \chi_{\rm imp} = \chi_{\rm free}(s) + \frac{2 s \delta s_m + \delta s_m^2}{3} \int d \tau \langle n^a(\tau) n^a(0)\rangle + {\cal O}(s^0) = \frac{(s + \delta s_m)(s+1+\delta s_m)}{3 T}+{\cal O}(s^0). \eeq 
Because the total magnetization is an RG invariant, we may RG improve our expression:
\begin{equation} \label{eq:susc_rho}
    \chi_\text{imp}(T) = \frac{[s  - \rho(T)/(2\pi)][s  - \rho(T)/(2\pi) + 1]}{3 T} + \mathcal{O}(s^0).\footnote{Even though one of the cross-terms in the numerator on the right-hand-side contributes to $\chi_\text{imp}$ at $\mathcal{O}(s^0)$, we leave $\chi_\text{imp}$ in this form to reproduce the correct behavior at the weak and strong coupling fixed points. We expect the $\mathcal{O}(s^0)$ contribution to vanish at both fixed points.}
\end{equation}
At the strong coupling fixed point, we find that $\chi_\text{imp}$ approaches $(s^2 - 1/4)/3T$, which is exactly the susceptibility of a free spin of magnitude $s-1/2$. We again expect this result for the strong coupling fixed point to hold at all orders in $1/s$ to reflect the impurity absorbing an opposite spin-$1/2$ electron. 

From \eqref{eq:dim_transmutation}, we find that the behavior of $\chi(T)$ at low and high temperatures is 
\begin{align}
    \chi(T) = \begin{cases}
        \frac{1}{3T} \left\{ (s^2-1/4) + s^2 \log\left(T_0/T\right)^{-1}\right\},  &T \ll T_0, \quad \log(T_0/T) \gg s  \\
        \frac{1}{3T} \left\{ s(s+1)- s^2 \log\left(T/T_0\right)^{-1}\right\},  &T \gg T_0, \quad \log(T/T_0) \gg s.
    \end{cases}
\end{align}
These low and high temperature results agree with the Bethe ansatz results \cite{Wiegmann_1981} at large $s$.

\section{Comparison to Bethe ansatz results}\label{sec:check}
We now compare our results to those from the Bethe ansatz at arbitrary $s$. The observables computed in the Bethe ansatz are functions of $T/T_0^B$ and $h/T_0^B$, where $T_0^B$ is a reference temperature that depends on the microscopics of the Bethe ansatz Hamiltonian.  We find analytic agreement between the Bethe ansatz results and our results at large $s$ and also good numeric agreement for $s\gtrsim 5$.

\subsection{Analytic comparisons}

\subsubsection*{Zero temperature magnetization}
We now compare \eqref{eq:magnetization} to the Bethe ansatz result for the $T=0$ impurity magnetization \cite{Kondoarbitrarys,Kondoarbitraryscorrection1, PhysRevB.70.174432}: 
\begin{equation}
M(h) = s-\frac14+ \frac{1}{2 \pi^{3/2}} \int_0^{\infty} \frac{d y}{y} {\rm Im}\left\{ \Gamma\left(\frac12-i y\right) e^{i y \left(\log y - 1 + 2 \log \frac{h}{2 T_1}\right)} \right\} e^{-2 \pi (s-1/4) y}  \label{MhBethe0}
\end{equation}
The energy scale $T_1$ is $\sqrt{2 \pi/e} \, T_0^B$.\footnote{We refer to the temperature $T_0$ in \cite{KondoReview} as $T_0^B$. Our $T_1$ matches $T_1$ in \cite{KondoReview}. Our normalization of magnetic field $h$ differs from that of magnetic field $H$ in \cite{KondoReview}: $h_{\rm our} = 2 H$.} In the large $s$ limit, this expression evaluates to (see App. \ref{app:larges})
\begin{equation}
\label{eq:mag_approx}
    M(h) \approx s - \frac{1}{4} + \frac{1}{2\pi} \tan^{-1}\left(\frac{\log(h/(2T_1))}{\pi s}\right). 
\end{equation}
This expression is equivalent to \eqref{eq:magnetization} for $T_0=2T_1$.
However, to the order in $1/s$ that we are working, we cannot reliably fix the ${\cal O}(1)$ proportionality constant between our $T_0$ and $T_1$ of Bethe ansatz. Indeed, rescaling $T_0$ in \eqref{eq:magnetization} by an ${\cal O}(1)$ constant leads to an $O(1/s)$ correction to $M(h)$ -- beyond the precision of our calculation.

\begin{figure}
\centering
\includegraphics[width = 0.48\textwidth]{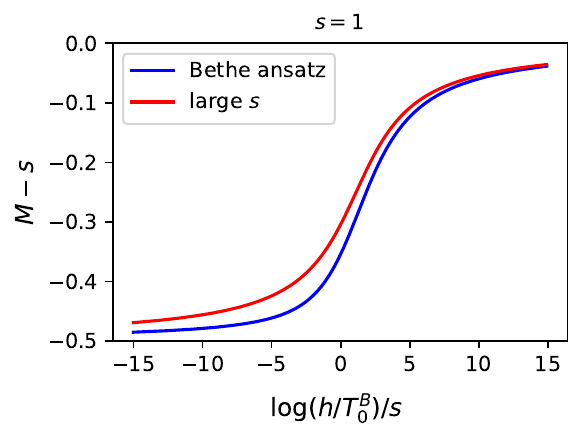}
\hfill
\includegraphics[width = 0.48\textwidth]{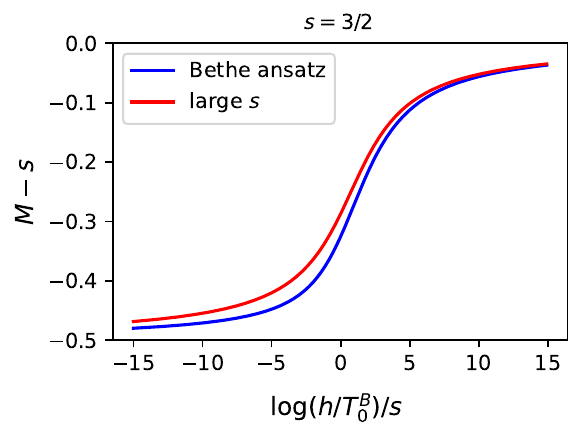}\\
\includegraphics[width = 0.48\textwidth]{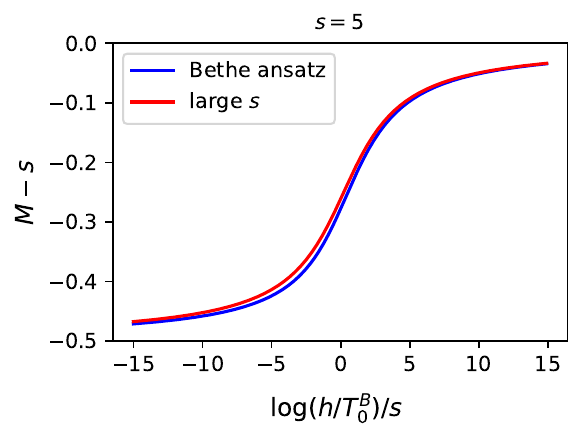}
\hfill
\includegraphics[width = 0.48\textwidth]{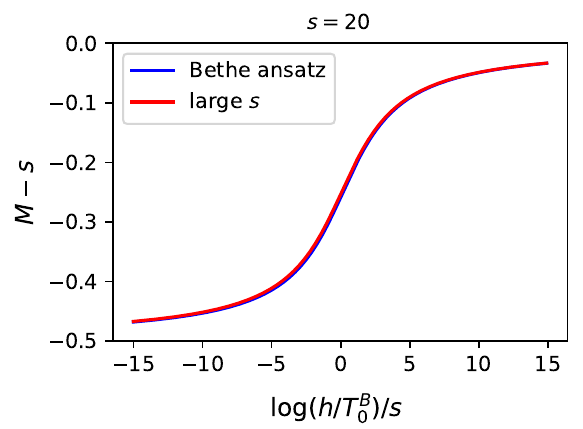}
\caption{
Single channel Kondo model: the change in impurity magnetization $M$ from its free value $s$ as a function of applied field for $s=1$, $3/2$, $5$, and $20$. We rescale the $x$-axis by $1/s$ for easy comparison between plots.}\label{fig:mag}
\end{figure}

We plot the exact Bethe ansatz result \eqref{MhBethe0} and our large $s$ result (\ref{eq:magnetization}) in Fig.\ \ref{fig:mag}. For definiteness, we take the relationship between $T_0$ in (\ref{eq:magnetization}) and $T_1$ of Bethe ansatz (\ref{MhBethe0}) to be $T_0 = 2T_1$. 

\subsubsection*{Finite temperature free energy}

While there exists no closed form solution for the free energy in the Bethe ansatz, there are recursive convolution equations for the Bethe ansatz free energy. The Bethe ansatz free energy satisfies
\begin{align}
    F_B = -T \int_{-\infty}^\infty \dd \ell \, G\left[\ell - \log(T_0^B/T)\right] \log\left[1 + \eta_{2s}(\ell)\right], \label{F_B}
\end{align}
where the functions $\eta_n(\ell)$ satisfy the following recursion relation:
\begin{align}\label{eq:recurse}
    \log \eta_n = G \star \log[(1 + \eta_{n+1})(1 + \eta_{n-1})] - 2\delta_{n,1} e^{\ell}, \\
    \eta_0 = 0, \quad \lim_{n\to \infty}\left(\frac{1}{n}\log \eta_n(\ell)\right) = 2x_0 
\end{align}
the function $G(\ell) = (1/2\pi) \sech(\ell)$, $x_0 = h/(2T)$, and $\star$ represents convolution \cite{numerics, KondoReview, Kondoarbitrarys}.
We solve this recursion equation for $s \gg 1$ and $x_0 \ll 1/s$ in App.\ \ref{app:larges}. We find that 
\bea \label{eq:largesf} F_B(x_0 = 0) = -T \left[\log 2 s + \frac{1}{4s} - \frac{1}{2\pi s} \left(\tan^{-1}\left(\frac{\ell}{\pi s}\right) + \frac{\ell/\pi s}{1 + (\ell/\pi s)^2}\right)  + \mathcal{O}(s^{-1})\right] , \eea
and that 
\beq
\label{eq:largesfx0}
F_B(x_0) - F_B(0) = -\frac{2x_0^2 T}{3}\left(s^2 + \frac{s}{2}\left\{1 - \frac{2}{\pi}\tan^{-1}\left(\frac{\ell}{\pi s}\right)\right\}  + \mathcal{O}(s^0)\right) + \mathcal{O}(x_0^3),
\eeq
where $\ell = \log (T^B_0/T)$. At $x_0 = 0$, \eqref{eq:largesf} exactly matches \eqref{Zimp1} with the identification $T_0 \sim T^B_0$ up to an overall $\mathcal{O}(1)$ factor. As in our discussion of the magnetization below (\ref{eq:mag_approx}), we cannot fix this numerical factor at this order in $1/s$. Computing the susceptibility from \eqref{eq:largesfx0} yields 
\beq
  \chi_B =  \frac{1}{3T}\left(s^2 + \frac{s}{2}\left\{1 - \frac{2}{\pi}\tan^{-1}\left(\frac{\ell}{\pi s}\right)\right\} +  \mathcal{O}(s^0)\right).
  \eeq 
This result exactly matches \eqref{eq:susc_rho}.

\subsection{Numeric comparisons}
In general, the recursion relations Eqs.\ \eqref{F_B} and \eqref{eq:recurse} compute a free energy of the form 
$F_B(T, x_0) = -T \tilde{f}(\log(T/T_0^B), x_0)$, where $x_0 = h/2T$. The $h=0$ specific heat is
\begin{equation}
    C_B(T,0) = \tilde{f}^{1,0}(\log(T/T_0^B),0) +\tilde{f}^{2,0}(\log(T/T_0^B),0),
\end{equation}
where the superscript refers to the order of derivative applied to each entry of $\tilde{f}$. 
To compute the susceptibility, note that 
\begin{equation}
\chi_B(T,0) = -\left.\pdv[2]{F_B}{h}\right|_{h=0} = -\frac{1}{4T^2}\left.\pdv[2]{F_B}{x_0}\right|_{x_0 = 0} = \frac{1}{4T} \tilde{f}^{0,2}(\log(T/T_0^B),0).
\label{susccalc}
\end{equation}

\begin{figure}
\centering
\includegraphics[width = 0.48\textwidth]{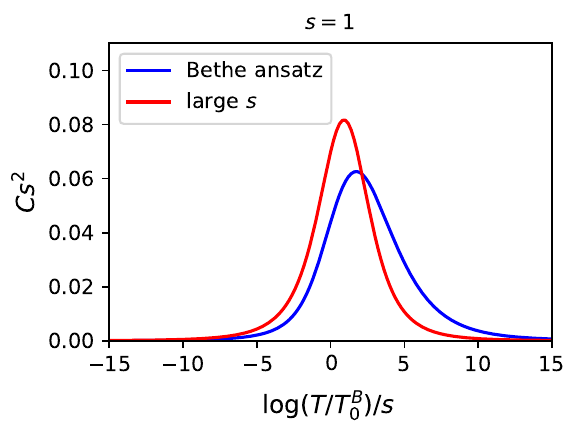}
\hfill
\includegraphics[width = 0.48\textwidth]{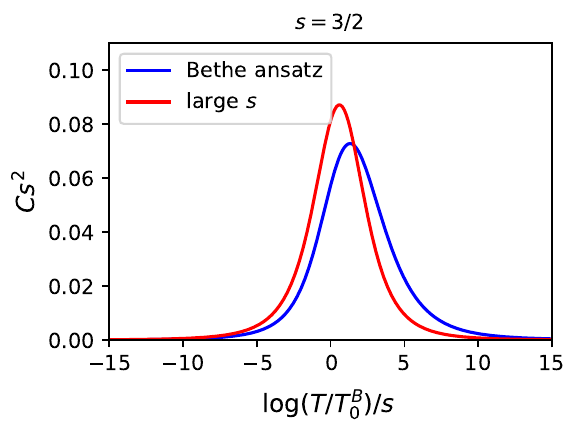}\\
\includegraphics[width = 0.48\textwidth]{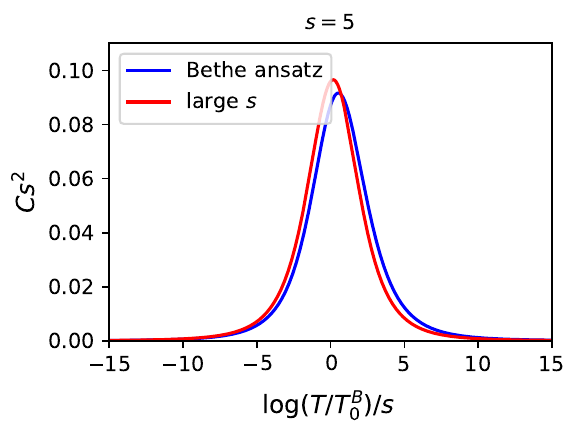}
\hfill
\includegraphics[width = 0.48\textwidth]{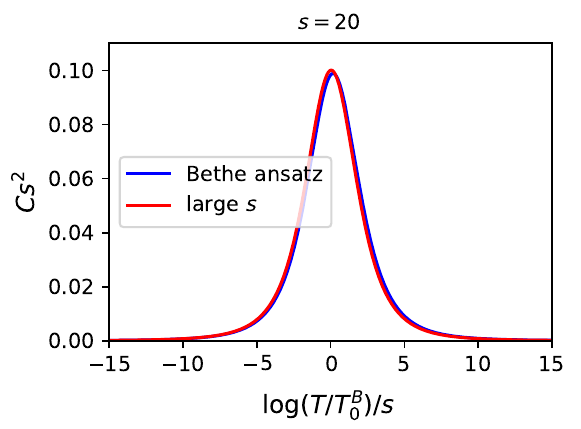}
\caption{
Single channel Kondo model: the specific heat as a function of temperature for $s=1$, $3/2$, $5$, and $20$. We rescale the $x$-axis by $1/s$ and the $y$-axis by $s^2$ for easy comparison.}\label{fig:sp_heat}
\end{figure}

\begin{figure}
\centering
\includegraphics[width = 0.48\textwidth]{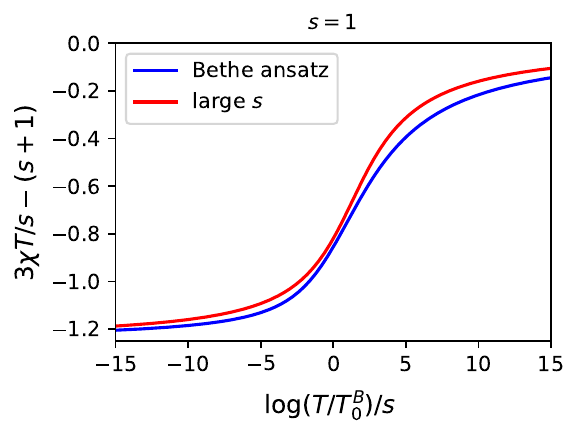}
\hfill
\includegraphics[width = 0.48\textwidth]{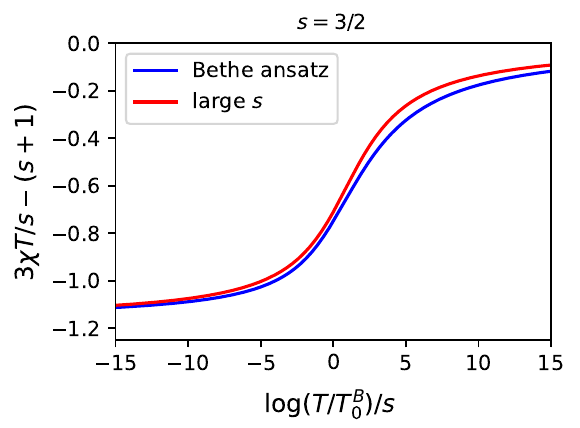}\\
\includegraphics[width = 0.48\textwidth]{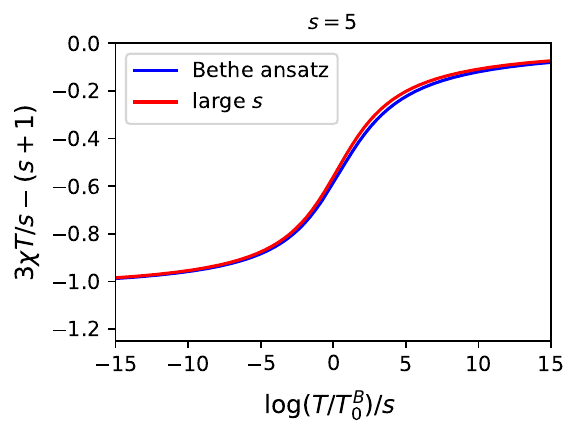}
\hfill
\includegraphics[width = 0.48\textwidth]{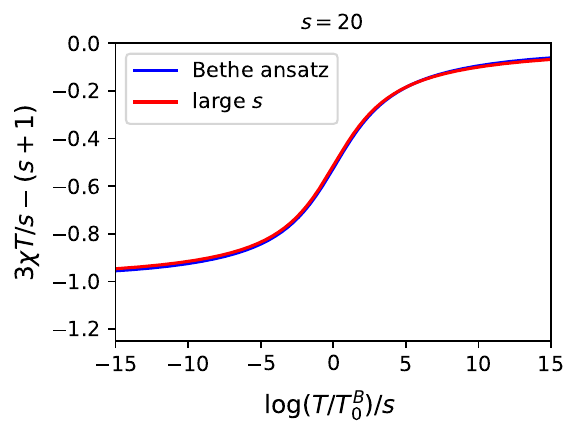}
\caption{
Single channel Kondo model: the magnetic susceptibility as a function of temperature for $s=1$, $3/2$, $5$, and $20$. We rescale the $x$-axis by $1/s$ and we shift and rescale the $y$-axis for easy comparison.}\label{fig:susc}
\end{figure}

We use the recursion relations to compute the values of $\eta$, as described in App.\ \ref{app:numerics}. Then, we use numeric differentiation to compute $C_B(T,0)$ and $\chi_B(T,0)$ and compare them with the large $s$ results, Eqs.\ \eqref{eq:sp_heat_long} and \eqref{eq:susc_rho}, in Figs.\ \ref{fig:sp_heat} and \ref{fig:susc}. 
To remain consistent with our zero temperature magnetization plots, we assume that our $T_0$ as defined in \eqref{eq:dim_transmutation} is equal to $2T_1 = \sqrt{8\pi/e}\,T_0^B$ (see the discussion below \eqref{eq:mag_approx}). Any error in this assumption once again appears as a $\mathcal{O}(1/s)$ constant horizontal  shift in the plots in Figs.\ \ref{fig:mag}, \ref{fig:sp_heat} and \ref{fig:susc}.

We see good alignment between the large $s$ results and the exact Bethe ansatz results for the specific heat at $s=5$ and $s=20$. The discrepancy in specific heat arises in part 
due to the asymmetry in the exact Bethe ansatz specific heat curves at small $s$.

For the magnetic susceptibility, we see good alignment between the large $s$ results and the exact Bethe ansatz results for all plotted values of $s$. At large values of $|\log(T/T_0^B)|$, there is a small discrepancy between the two curves at $s=20$. We suspect this discrepancy arises from how we truncate the numeric calculation of the recursion relations (see App.\ \ref{app:numerics}).\footnote{In order to iterate over the recursion relations numerically, we assume that the functions $\eta$ are constant for $|\log(T_0^B/T)| > 500$. This assumption is less accurate for larger values of $s$, and there are small numeric discrepancies in the behavior of $\eta$ for larger values of $s$ and large values of $|\log(T/T_0^B)|$.}

\section{The multichannel Kondo problem}
\label{sec:multi}
We now consider the case when the impurity of spin $s$ is coupled to $K$ channels of conduction electrons. We focus on the {double-scaling} limit $K \to \infty$, $s \to \infty$, $\kappa = K/s$ - finite. Our treatment is similar to that of Ref.~\cite{hu2022kondo}; however, while Ref.~\cite{hu2022kondo} focuses on the fully screened case $\kappa = 2$, we instead consider $\kappa \neq 2$. In particular, we access the full crossover between the free fixed point and the non-trivial overscreened impurity fixed point for $\kappa > 2$. For the fully screened case $\kappa = 2$, we point out that the $1/s$ expansion breaks down sufficiently close to the screened fixed point.

We begin with the UV action:
\begin{equation}\label{eq:foldedK}
    S = S_B + \int \dd \tau \int_0^\infty \dd x [\psi_{L\alpha w}^\dagger(\partial_\tau + i \partial_x) \psi_{L\alpha w} +\psi_{R\alpha w}^\dagger(\partial_\tau - iv_F \partial_x) \psi_{R\alpha w}] +  J \int_{x=0} \dd \tau  \psi_{R w}^\dagger\frac{\sigma^a}{2} \psi_{R w}(x) S^a,
\end{equation}
where $w$ is the channel index that runs over $w = 1 \ldots K$. The UV action (\ref{eq:foldedK}) is supplemented by a boundary condition $\psi_{R \alpha w}(0) = \psi_{L \alpha w}(0)$. We again first consider the frozen impurity spin limit, which gives a line of fixed points characterized by the fermion phase shift $\rho$:
\bea
    &&S_\text{frozen}(\rho) = \int \dd \tau \int_0^\infty \dd x [\psi_{L\alpha w}^\dagger(\partial_\tau + i \partial_x) \psi_{L\alpha w} + \psi_{R\alpha w}^\dagger(\partial_\tau - i \partial_x) \psi_{R\alpha w}],\nn\\
    && \psi_{R\alpha w}(0^+,\tau) = e^{-i\alpha \rho}\psi_{L\alpha w}(0^+,\tau), \label{SfrozenK}
\eea
The fermion two-point functions in the frozen theory are the same as in Eq.~(\ref{psipsi1}) and are diagonal in channel index. 

We derive an effective theory for small fluctuations of the impurity spin in the same way as we did for the single channel case. As before, the action takes the same form
\beq S_{\rm eff} = S_{\rm frozen}(\rho) + S_B[\vec{n}] + S_{\rm int} + S_{\rm cont}, \label{SefffullK}  \eeq
 where 
\beq S_{\rm int} = \int d \tau \,\left[ \epsilon^{ij} \pi^i(\tau) j^j_x(0^+, \tau) + r \, \vec{\pi}^2(\tau)  \psi^{\dagger}_{Rw} \sigma^3 \psi_{Rw}( 0^+, \tau) + O(\pi^3) \right], \quad r = -\frac{\sin 2 \rho}{4}, \eeq

\bea j^i_x( 0^+, \tau) &=& \frac{1}{2} \left[ \psi^{\dagger}_{R w} \sigma^i \psi_{R w} (0^+, \tau) - \psi^{\dagger}_{L w} \sigma^i \psi_{L w} (0^+, \tau)\right] \nn\\
&=& \sin\rho \left[\sin \rho \, \psi^{\dagger}_{Rw} \sigma^i \psi_{Rw} (0^+, \tau)- \cos \rho\, \epsilon^{ij} \psi^{\dagger}_{R w} \sigma^j \psi_{Rw}(0^+, \tau)\right], \quad i =1,2, \label{jpsiRK}\eea

\beq  S_{\rm cont} =  \frac{i K \delta s_B}{2} \int d \tau \, \left(\pi_1 \d_\tau \pi_2 - \pi_2 \d_\tau \pi_1\right) + O(\pi^4), \quad\quad \delta s_B = -\frac{1}{2\pi}(\rho - \frac{1}{2}\sin 2 \rho). \label{sBactK} \eeq
Note the extra factor of $K$ in front of $\delta s_B$ compared to the single channel case.
Similarly, $K$ now multiplies  the magnetization contact term:
\beq S_{\rm eff}(h) = S_{\rm eff}(\rho, h=0)  - (s + K \delta s_M) \int d \tau h^a n^a(\tau) - \int d\tau \int_{0^+}^\infty dx \, h^a j_0^a(x, \tau), \quad\quad \delta s_M = -\frac{\rho}{2 \pi},  \label{SeffhK} \eeq 
where $j^a_0(x, \tau) =\frac{1}{2} (\psi^{\dagger}_{Rw} \sigma^a \psi_{Rw} + \psi^{\dagger}_{Lw} \sigma^a \psi_{Lw})$. 

If we integrate the fermions out from the action Eq.~(\ref{SefffullK}), we obtain a non-local effective action for the spin fluctuations $\vec{\pi}$ with a factor of $K$ in front. Because the bare spin action $S_B[\vec{n}]$ contains a factor of $s$ and we are taking the limit $K \to \infty$, $s \to \infty$, $K/s$ fixed, we can expand in $1/s$. As a starting point, we compute the quadratic piece in the non-local effective action for $\vec{\pi}$:
\beq S_2[\pi] =  \frac{i (s+ K\delta s_B)}{2} \int d \tau \, \left(\pi_1 \d_\tau \pi_2 - \pi_2 \d_\tau \pi_1\right) 
-\frac{1}{2}\int d\tau_1 d \tau_2 \,\,  \pi^i(\tau_1) \Pi(\tau_1 - \tau_2) \pi^i(\tau_2)\label{eq:nonlocalK} \eeq
where
\beq \Pi(\tau) \delta_{ij} = -\langle j^i_x(0^+, \tau) j^j_x(0^+, 0)\rangle_0. \eeq
We work at finite temperature $T$. Repeating the calculation in appendix \ref{app:current_current}, we obtain
\beq \Pi(i \omega_n) = \int_0^{1/T} d \tau\,  \Pi(\tau) e^{i \omega_n \tau} = \frac{ K \sin^2 \rho }{2\pi} |\omega_n|. \eeq
Here we have dropped a UV divergent term that is constant in $\omega_n$ and absorbed by the $m_\pi \vec{\pi}^2$ counterterm. Thus, we obtain the $\pi$ propagator:
\beq \langle \pi_i(\tau) \pi_j(0) \rangle =  D_{\pi,ij}(\tau) = T \sum_{n \neq 0} D_{\pi,{ij}}(i \omega_n) e^{-i \omega_n \tau},  \eeq
\bea D_{\pi}(i \omega_n) &=& \frac{1}{s |\omega_n|} d(\rho)\left(\begin{array}{cc} \kappa \sin^2 \rho/2\pi &\,\, - {\rm sgn} (\omega_n) (1 + \kappa \delta s_B) \\ {\rm sgn}(\omega_n) (1 + \kappa \delta s_B) & \,\, \kappa \sin^2 \rho /2\pi\end{array}\right), \nn\\
  d(\rho) &=& \frac{1}{(1 + \kappa \delta s_B)^2 + (\kappa/2\pi)^2 \sin^4 \rho}. \label{eq:DPiK}\eea
  Fourier transforming,
  \beq  D_{\pi, ij}(\tau) = \frac{1}{s} d(\rho) \left(- \frac{\kappa \sin^2 \rho}{2\pi^2} \log(2 \sin \pi T |\tau|) \delta_{ij} + \frac{i}{2} \epsilon_{ij} (1+\kappa \delta s_B)({\rm sgn}(\tau) - 2 \tau/\beta)\right). \label{eq:PiPropK} \eeq 
For $\kappa = 2$, $s d(\rho)^{-1}$ approaches zero as $\rho$ approaches $\pi$, so the $1/s$ expansion breaks down near the screened fixed point as stated at the beginning of the section. For other values of $\kappa$, $s d(\rho)^{-1} \gg 1$.

Next, we compute the RG flow of the phase shift $\rho$ to order $1/s$. We can follow the same strategy as Ref.~\cite{hu2022kondo} and extract the renormalization of $\rho$ from the $1/s$ correction to the fermion two-point function $\langle \psi_R(x, \tau) \psi^{\dagger}_L(x', \tau)\rangle$. Such correction is represented by the Feynman diagrams in Fig.~(\ref{fig:SelfEnergy}). In practice, this approach is equivalent to expanding the partition function to second order in $\pi$ and considering only the high-energy fluctuations:
\beq \delta S = r \int d \tau  \vec{\pi}^2  \psi^{\dagger} \sigma^3 \psi -\frac{1}{2}\int_{e^{-d\ell} a < |\tau_1 - \tau_2| < a}  d\tau_1 d \tau_2 \,\, \epsilon^{ij} \epsilon^{kl} \pi^i(\tau_1) \pi^k(\tau_2) j^j_x(0^+, \tau_1) j^l_x(0^+, \tau_2) \label{dSRGK}\eeq
We use the analogue of the OPE (\ref{psibiOPE}), (\ref{eq:ope_j}), 
\beq \label{eq:ope_jmultichannel} j^i_x(0^+, \tau) j^j_x(0^+, 0) \sim \sin^2 \rho \left( \frac{K \delta^{ij}}{2 \pi^2 \tau^2} + \frac{i}{\pi \tau} \epsilon^{ij} \psi^{\dagger}_{Rw} \sigma^3 \psi_{Rw}(0^+, 0) + \ldots \right). \eeq
We replace the two insertions of $\pi$ in (\ref{dSRGK}) with the $\pi$ propagator (\ref{eq:PiPropK}). Because 
\beq
\langle \vec{\pi}^2\rangle = \frac{-\kappa d(\rho) \sin^2 \rho }{s \pi^2} \log (T a)
\eeq 
is UV divergent, the high-energy mode contribution in an RG step $d\ell$ is $\langle \vec{\pi}^2\rangle_> =  d \ell(\kappa d(\rho) \sin^2 \rho)/(\pi^2 s) $. After combining this term with the contribution from the second term in Eq.~(\ref{dSRGK}),
\beq \frac{d\rho}{d \ell} = -\beta(\rho) =  \frac{d(\rho) \sin^2 \rho}{\pi s}\left(1 - \frac{\kappa \rho}{2\pi}\right). \label{betaK} \eeq
This result exactly agrees with the $\beta$-function obtained by Ref.~\cite{hu2022kondo} after identifying the coupling constant $J$ of \cite{hu2022kondo} with $\tan \rho/2$.

\subsection*{Underscreened fixed point}
Let us discuss the consequences of the RG flow (\ref{betaK}). When $K < 2 s$, i.e. $\kappa < 2$, an initially small $\rho$ flows to a fixed point with $\rho_* = \pi$. This is the underscreened fixed point. Indeed, if we compute the zero temperature impurity magnetization in a uniform magnetic field $h$, from Eq.~(\ref{SeffhK}) to leading order in $1/s$ we simply get
\beq M(h) \approx (s - K \rho/2\pi), \label{MK} \eeq
i.e. $M = s - K/2$ at $\rho_* = \pi$. The approach to the underscreened fixed point is logarithmic and corresponds to a weak ferromagnetic coupling of the remnant spin to the electrons. The physics is qualitatively similar to the underscreened single channel case, so we do not discuss it further in the main text. We discuss in more detail the solution to the RG equation (\ref{betaK}) in the underscreened and overscreened cases in App. \ref{app:multichannelW}.

\subsection*{Overscreened fixed point}
In the overscreened case $K > 2s$, i.e. $\kappa > 2$, an initially small $\rho$ flows to a fixed point with $\rho_* = 2 \pi/\kappa$. This is the overscreened fixed point. Note that the correction to scaling exponent $\omega = \beta'(\rho_*) = \frac{2}{K}$ at this fixed point agrees with the exact result $\omega = \frac{2}{K+2}$ for $K \to \infty$.\cite{AFFLECK1991641}\footnote{Recall, the correction to scaling exponent $\omega$ is related to the scaling dimension of the leading irrelevant operator via $\Delta = 1 + \omega$.}

The crossover from the free fixed point to the  overscreened fixed point in our large $s$ solution can be accessed by  integrating Eq.\ \eqref{betaK}:
\beq
H(\rho_0)-H(\rho(\ell)) = \frac{2}{K} \ell, \quad H(\rho) = \left(\frac{2\pi}{\kappa} - \rho \right) \cot \rho + 
 \log \left(\frac{\frac{2 \pi}{\kappa} -  \rho}{\sin \rho}\right), \quad \rho(\ell = 0) = \rho_0.
   \label{implicitK}
\eeq 
Let us define an infrared energy scale $T_M$ via\footnote{As we discuss below, this definition matches the scale $T_H$ in the Bethe ansatz solution\cite{WiegmannMultiShort, WiegmannMultiLong} in the large $s$ limit.}
\beq \frac{2}{K} \log \frac{\Lambda}{T_M} = H(\rho_0) - 1. \label{tmdef}\eeq
Then the dependence on the UV cut-off and the bare value $\rho_0$ disappears from Eq.~(\ref{implicitK}):
\beq H(\rho(\omega)) = -\frac{2}{K} \log \frac{T_M}{\omega} - 1, \label{Hrho}\eeq
where $\omega$ denotes an energy scale. This equation has an explicit solution (see App. \ref{app:multichannelW}):
\beq
\rho(\omega) = \frac{2\pi}{\kappa} - \Im\left\{W\left[\left(\omega/T_M\right)^{2/K}e^{\frac{2\pi i}{\kappa} - 1}\right]\right\}, \label{eq:rhoW}
\eeq 
where $W$ is the Lambert W function (also known as the product-log function). For $\kappa < 2$ we must choose the main (zeroth) branch of the $W$-function as defined in Ref.\ \cite{NIST:DLMF}, section 4.13. Likewise, the zero temperature magnetization to leading order in $s$ is  
\beq
M(h) = \frac{K}{2\pi}\Im\left\{W\left[\left(h/T_M\right)^{2/K}e^{\frac{2\pi i}{\kappa} - 1}\right]\right\}
\label{MagK}
\eeq 

We now compare some properties of the overscreened fixed point following from our large-$s$ analysis to the results in the literature. Near the strong coupling fixed point,
\beq
M(h) \approx \frac{K e^{-1}}{2\pi}\left(h/T_M\right)^{2/K} \sin(2\pi/\kappa), \quad h \ll T_M. \label{MstrongK}
\eeq
Eq. \eqref{MstrongK} implies that the impurity carries no remnant magnetic moment at the strong coupling fixed point. The power law $M(h) \sim h^{2/K}$ agrees with the exact exponent $2/K$\cite{WiegmannMultiShort, WiegmannMultiLong, AFFLECK1991641}. 
As we show in App.\ \ref{app:Product_Log}, the full crossover behavior of the impurity magnetization in  Eq.\ \eqref{MagK} agrees with the Bethe ansatz result in the overscreened case at large $s$ \cite{WiegmannMultiShort, WiegmannMultiLong}:
\beq
M_\text{Bethe}(h) = -\frac{iK}{4\pi^{3/2}}\int_{-\infty}^\infty \dd y \frac{e^{2i y \log(h/T_H)/K}}{y -i \epsilon}\frac{\Gamma(1 + i y/K)\Gamma(1/2 - i y/K)}{\Gamma(1 + iy)} \left(\frac{iy + \epsilon}{e}\right)^{i y}\frac{\sinh\left(\frac{2\pi}{\kappa}y\right)}{\sinh(\pi y)}.\label{BetheMagK}
\eeq 
Importantly, our choice of $T_M$ ensures that $|\log(T_H/T_M)|$ is $\mathcal{O}(K^0)$.

We also compute finite temperature properties of the overscreened large $s$ Kondo model. In appendix \ref{app:entropyK}, we compute the impurity entropy and find
\beq \mathcal{S}_\text{imp} = \log 2 s - \frac{1}{2} \log d(\rho) + O(1/s),
\label{multichannelentropy}\eeq
which can be RG improved by the substitution  $\rho \to \rho(T)$. As $T \to 0$ we find
\beq  \mathcal{S}_{\text{imp}}(T =0) = \log \left[\frac{K}{\pi} \sin(\frac{2 \pi}{\kappa})\right] + O(1/s),  \eeq
which agrees with the exact result\cite{PhysRevLett.52.364, AffleckLudwigg},
\beq
\mathcal{S}_{\text{imp}}(T =0) =  \log \left[\frac{\sin\left((2 s+1) \pi/(K+2)\right)}{\sin\left(\pi/(K+2)\right)}\right],
\eeq
in the large $s$ limit.

From the impurity entropy (\ref{multichannelentropy}), we again can compute the specific heat. For brevity, we only include the $T\ll T_M$ expression for the specific heat:
\beq
C_\text{imp}(T \ll T_M)  \approx \frac{4(T/T_M)^{4/K}\sin^2(2\pi/\kappa)}{Ke^2}. \label{Cassmulti}
\eeq 
Additionally, the impurity susceptibility is 
\beq 
\chi_\text{imp} = \frac{s^2}{3T}\left[(1+\kappa \delta s_M)^2 + \mathcal{O}(1/s)\right] = \frac{s^2}{3T}\left[\left(1 - \frac{\kappa \rho}{2\pi}\right)^2 + \mathcal{O}(1/s)\right].\label{multichannelsusc}
\eeq 
Again, after RG impoving $\rho \to \rho(T)$, for $T\ll T_M$,
\beq 
\chi_\text{imp}(T\ll T_M) \approx \frac{K^2(T/T_M)^{4/K}\sin^2(2\pi/\kappa)}{12e^2 \pi^2 T}.  \label{chiassmulti}
\eeq 
 The scaling with respect to $T$ of both (\ref{Cassmulti})  and (\ref{chiassmulti}) agrees with previous results on the overscreened strong coupling fixed point \cite{AFFLECK1991641,affleck1993} in the limit $K\to \infty$ (the exact scaling exponent of $C_{\rm imp}(T)$ and $T \chi_{\rm imp}(T)$ is $4/(K+2)$). Additionally, the ratio of these results reproduces the  overscreened fixed point Wilson ratio \cite{AFFLECK1991641} in the large $K$ limit:
 \beq
 R_{{\rm imp},W}= \left(\frac{\chi_\text{imp}(T\ll T_M)}{C_\text{imp}(T\ll T_M)}\right)/ \left(\frac{\chi_{\rm bulk}(T)}{C_\text{bulk}(T)}\right) \approx \frac{K^3}{36}, \quad K\to \infty,
 \eeq 
 where $\chi_\text{bulk}(T)$ and $C_\text{bulk}(T)$ are the bulk susceptibility and specific heat of the free Fermi gas. This should be compared to  the exact expression (see App.\ \ref{app:exact_bcft})  
 \beq
 R_{{\rm imp},W} = \frac{(K/2+2)(K+2)^2}{18}.
 \eeq 
 
 Here we have only compared the $T \ll T_M$ limit of the entropy and susceptibility to the exact solution. In appendix \ref{app:FmultiBethe}, we analyze the Bethe ansatz equations that yield the free energy and show that in the large spin limit they reduce to a Liouville equation in the $(\log(T_M/T), s)$ plane. We then show that our expression for the entropy (\ref{multichannelentropy}) satisfies this Liouville equation.

\subsection*{The fully screened fixed point: breakdown of RG}
We now briefly comment on the fully screened case $\kappa = 2$. As we already noted, in this case we expect the $1/s$ expansion to break down at the largest length scales, so that the fully screened Fermi-liquid fixed point that controls the physics at $\omega \ll T_M$ is inaccessible. An analysis of Eq.~(\ref{Hrho}) shows that for $\kappa = 2$, $\rho(\omega)$ reaches the value $\rho = \pi$ at $\omega = T_M$, i.e. in a finite RG time. This unphysical behavior is a signature of the breakdown of $1/s$ expansion. Indeed, from the form of the $\pi$-propagator, Eq.~(\ref{eq:DPiK}), we see that the $1/s$ expansion breaks down at $|\pi - \rho| \sim O(1/\sqrt{s})$. For $\omega \gg T_M$, $|\pi - \rho| \gg O(1/\sqrt{s})$ and so our RG  analysis is valid. Eq.~(\ref{eq:rhoW}) holds for $\omega \gg T_M$, where the $W$ function with negative argument $z < 0$ should be understood as $W_0(z+i \epsilon)$. Likewise, Eqs.~(\ref{MagK}), (\ref{multichannelentropy}) and (\ref{multichannelsusc}) apply.

Even though the $1/s$ expansion breaks down for $T \lesssim T_M$, the ``run away" flow $\rho \to \pi$ as $T \to T_M$ and the associated screening of  the impurity Berry phase in Eq.~(\ref{sBactK}) are highly suggestive of a flow to a fully screened fixed point, as argued in Ref.~\cite{hu2022kondo}.

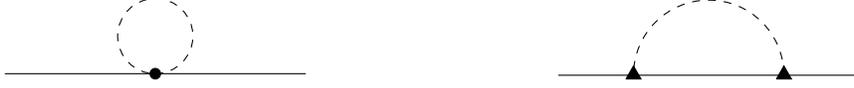
\begin{figure}
    \centering
        \begin{tikzpicture}

 \draw (0,0) -- (4,0);
 \draw[dashed] (2,0) arc (270:630:0.5);
\filldraw (2,0) circle (2pt);
 \end{tikzpicture}
  \hspace{0.2\textwidth}
    \begin{tikzpicture}
 \draw (0,0) -- (4,0);
 \draw[dashed] (3,0) arc (0:180:1);
\filldraw (0.9,-0.05)--(1.1,-0.05) -- (1,0.123)--(0.9,-0.05);
\filldraw (2.9,-0.05)--(3.1,-0.05) -- (3,0.123)--(2.9,-0.05);

 \end{tikzpicture}
 \caption{One-loop corrections to the electron two point function in the multichannel Kondo model. The solid lines are electron propagators, the dashed lines are $\pi$ propagators. The circle vertex corresponds to the 4-point interaction arising from $r \vec{\pi}^2 \psi^\dagger_{Rw}\sigma^3 \psi_{Rw}(0^+, \tau)$ in the action, and the triangle vertices correspond to the 3-point interaction arising from $\epsilon^{ij} \pi^i(\tau) j^j_x(0^+,\tau)$ in the action.} \label{fig:SelfEnergy}
\end{figure}

\section{Impurity spectral function and resistivity} \label{sec:spectral_function}
We now use our result for the RG flow of $\rho$ to compute the impurity spectral function and electrical conductivity of the $K$-channel Kondo model.
There are two spin-1/2 impurity operators with lowest scaling dimension: $\psi_{R\alpha w}(0)$ and
\beq f_{\alpha w} = \vec{S} \cdot \vec{\sigma}_{\alpha \beta} \psi_{R \beta w}(0). \eeq 
Here we define these operators in the UV at the decoupled impurity fixed point. Note that these operators transform oppositely under the combined particle-hole symmetry and time reversal:
\beq CT: \quad \psi_{R \alpha w}(x,t) \to \psi^{\dagger}_{L \alpha w}(x,-t), \quad  \psi_{L \alpha w}(x,t) \to \psi^{\dagger}_{R \alpha w}(x,-t),  \quad \vec{S} \to - \vec{S}, 
\quad\quad i \to -i.\eeq
Under $CT$, $\psi_{R \alpha w}(0) \to \psi^{\dagger}_{R \alpha w}(0)$ and $f_{\alpha w} \to - f^{\dagger}_{\alpha w}$, where we have used the boundary condition $\psi_R(0) = \psi_L(0)$ of the UV theory. The transformation properties have been written in real time. We focus on the spectral function of $f$, which, as we discuss below, is related to the resistivity of the full 3d Kondo model.\cite{Suhl, affleck1993}

The two-point function of $f(t)$ is related to the two-point function of $\psi(x,t)$ by equations of motion. Here, we work in an ``unfolded" picture, where $\psi_{Rw}(x)$ is extended from $x > 0$ to the entire real line via, $\psi_{R\alpha w}(-x) = \psi_{L\alpha w}(x)$. Then
\beq (\d_t + \d_x) \psi_{R\alpha w}(x,t) = - i \frac{J}{2}\delta(x) f_{\alpha w}(t),  \label{EOM}\eeq
where $J$ denotes the bare Kondo coupling.
Let us define the retarded two point functions:
\bea G^R_f(t_1, t_2) \delta_{\alpha \beta} \delta_{ww'} &=& - i \theta(t_1 - t_2) \langle \{ f_{\alpha w}(t_1), f^{\dagger}_{\beta w'}(t_2)\}\rangle. \nn\\
G^R(x_1, t_1; x_2, t_2) \delta_{\alpha \beta}\delta_{w w'} &=&  -i \theta(t_1 - t_2) \langle \{\psi_{R\alpha w}(x_1, t_1), \psi^{\dagger}_{R\beta w'}(x_2, t_2)\} \rangle. 
\label{eq:green_f}
\eea
Here the brackets on the right-hand-side denote thermal average.
Then from (\ref{EOM}), 
\beq G^R(x_1, x_2, \omega) = G^R_0(x_1-x_2, \omega) + \frac{J^2}{4} G^R_0(x_1, \omega) G^R_f(\omega) G^R_0(-x_2, \omega ),\eeq
where we work in mixed position-frequency representation, and
$G^R(\omega) = \int d t\,  e^{i \omega t} G^R(t)$, 
\beq G^R_0(x, t) = \int \frac{d \omega}{2 \pi} \frac{dk}{2\pi}\frac{1}{\omega - k + i \epsilon} e^{-i \omega t} e^{i k x} 
= -i \theta (t) \delta(x-t) \implies G^R_0(x, \omega) = - i \theta(x) e^{i \omega x}.\eeq
Therefore,
\beq G^R(x_1, x_2, \omega) = - i e^{i \omega(x_1 -x_2)} \left[\theta(x_1 -x_2) - \frac{i J^2}{4} \theta(x_1) \theta(-x_2) G^R_f(\omega)\right]. \label{GGf} \eeq 
We now compare this result to our effective theory calculation. To leading order in $1/s$, the fermion two point function in real time is \eqref{Psit}
\beq \langle \{ \psi_{R \alpha w }(x_1, t_1), \psi^{\dagger}_{L \beta w'}(x_2,t_2\}\rangle \approx e^{- i \alpha \rho} \delta_{\alpha \beta} \delta_{w w'} \delta(x_1 + x_2 - (t_1 - t_2)). \eeq 
Tracing over spin indices in order to consider an $SU(2)$ invariant quantity,
\beq G^R(x_1, x_2, \omega) \approx - i \cos \rho\big({\rm max}(|\omega|, T)\big)\, e^{i \omega (x_1 - x_2)}, \quad\quad x_1 > 0, x_2 < 0. \eeq
Here, we have used the fact that the bulk fermion operators do not receive an anomalous dimension to RG improve the expression. 

Thus, from \eqref{GGf},
\beq \frac{J^2}{4}G^R_f(\omega) \approx - 2 i \sin^2 \left(\rho\big({\rm max}(|\omega|, T)\big)/{2}\right). \eeq
The  $T= 0$ spectral density corresponding to $f$ is:
\beq A_f(\omega ) = -\frac{1}{\pi} {\rm Im} (G^R_f(\omega)) \approx \frac{8}{\pi J^2}  \sin^2 \frac{\rho(|\omega|)}{2}. \eeq
Explicitly, for $K=1$,
\beq
2\sin^2 \frac{\rho(|\omega|)}{2} = 1 - \frac{\log(|\omega|/T_0)/\pi s}{\sqrt{1 + \log^2(|\omega|/T_0)/\pi^2 s^2}},
\eeq 
from which we see the formation of a Kondo resonance. 
The underscreened multichannel case has qualitatively the same behavior. In the overscreened case $K > 2s$,
\beq
2\sin^2 \frac{\rho(|\omega|)}{2} = 1 - \frac{\Re W\left(\left(\frac{|\omega|}{T_M}\right)^{2/K} e^{\frac{2\pi i }{\kappa} - 1}\right)}{\abs {W\left(\left(\frac{|\omega|}{T_M}\right)^{2/K} e^{\frac{2\pi i }{\kappa} - 1}\right)}}.
\label{eq:sinrho}
\eeq 
This expression also displays a Kondo peak at $\omega = 0$.

We now come back to the original 3d Kondo model. From the $f$ spectral function, we can pass to the $T$-matrix of 3d electrons and electrical conductivity. Following Refs.~\cite{Suhl, affleck1993}  the $T$-matrix for elastic scattering in the full 3d Kondo model is 
\beq
T(\omega) = \frac{1}{2 \pi \nu} \frac{J^2_0}{4} G^R_f(\omega) \approx\frac{-i}{\pi \nu}  \sin^2 \left(\rho\big({\rm max}(|\omega|, T)\big)/{2}\right),
\eeq 
where $\nu = k_F^2(2\pi^2 v_F)^{-1}$ is the density of states per spin per flavor per volume and we've restored the Fermi-velocity $v_F$. Now consider a system with a dilute density of impurities $n_i$. Then, the scattering time $\tau_s$ is \cite{affleck1993}
\beq
\frac{1}{\tau_s(\omega)} = -2 n_i \Im T(\omega) \approx \frac{2 n_i}{\pi \nu} \sin^2 \frac{\rho\big({\rm max}(|\omega|, T)\big)}{2}.
\eeq 
Finally, the electrical conductivity as a function of temperature is \cite{affleck1993}
\beq
\sigma(T)  = -K \frac{2q_e^2}{3m^2} \int \frac{\dd[3]p}{(2\pi)^3} \dv{n_F}{\epsilon_p} p^2 \tau_s(\epsilon_p) \approx - \frac{2K q_e^2 v_F^2 \nu}{3} \int_{-\infty}^\infty \dd \epsilon_p \dv{n_F}{\epsilon_p} \tau_s (\epsilon_p), \label{sigmaint}
\eeq 
where $k_F$ is the Fermi momentum, $n_F(\epsilon_p)$ is the Fermi distribution at temperature $T$, $m$ is the electron mass (which satisfies the relationship $v_F = k_F/m$), $\epsilon_p \approx v_F (p - k_F)$, and $q_e$ is the electron charge.
Note that 
$\dv*{n_F}{\epsilon_p}$ varies on scales of order $\epsilon_p \sim T$ over which $\tau_s(\epsilon_p)$ is nearly constant. As a result, in the large $s$ limit we may replace $\tau_s(\epsilon_p) \to \tau_s(\epsilon_p = T)$ in the integral (\ref{sigmaint}), obtaining
the  resistivity 
\beq
r_\text{imp}(T) = \sigma(T)^{-1} = \frac{3 n_i}{\pi K (\nu q_e v_F)^2} \sin^2 \frac{\rho(T)}{2}. \label{eq:res}
\eeq 

We now focus on the overscreened regime and compare our result (\ref{eq:res}) to the exact CFT results for the impurity resistivity in the regime $T \ll T_M$ \cite{affleck1993}. At $T = 0$, our expression (\ref{eq:res}) reduces to
\beq
r_\text{imp}(T = 0) = \frac{3 n_i}{\pi K (\nu q_e v_F)^2 } \sin^2 \left(\frac{\pi}{\kappa}\right).
\eeq 
In the large $s$ limit, this agrees with the exact CFT result \cite{affleck1993} 
\beq
r^{\rm CFT}_\text{imp}(T = 0) =\frac{3 n_i}{2\pi K (\nu q_e v_F)^2 } \left[1 - \cos\left(\frac{(2s+1)\pi}{2+K}\right) \sec\left(\frac{\pi}{2+K}\right)\right].
\eeq 
Additionally, for $T \ll T_M$, we expand \eqref{eq:sinrho} and find
\beq
r_\text{imp}(T \ll T_M) -r_\text{imp}(0)= - \frac{3n_i}{2\pi K (\nu q_e v_F)^2} \left(\frac{T}{T_M}\right)^{2/K} e^{-1} \sin^2\left(\frac{2\pi}{\kappa}\right).
\eeq 
The $(T/T_M)^{2/K}$ scaling should be compared to the exact CFT result of $(T/T_M)^{2/(K+2)}$ \cite{affleck1993}. Moreover, we compute the ratio
\beq
\frac{(r_\text{imp}(T \ll T_M) - r_\text{imp}(0))^2}{c_\text{imp}(T\ll T_M)} = \frac{9 n_i\sin^2(2\pi/\kappa)}{16 \pi^2 K (\nu q_e v_F)^4 },
\eeq 
where the low-temperature behavior of the impurity specific heat per unit volume $c_\text{imp} = n_i C_\text{imp}$ is given by Eq.~\eqref{Cassmulti}. This result is in agreement, in the large $K$ limit, with the exact BCFT result for the low-temperature impurity resistivity from Ref.\ \cite{affleck1993} and the exact BCFT result for the low-temperature impurity specific heat from Ref.\ \cite{AFFLECK1991641} (see App.\ \ref{app:exact_bcft}).



\section{Kondo screening cloud} \label{sec:cloud}

A potential application of our RG treatment is the investigation of the profile of the Kondo screening cloud.  RG arguments imply the existence of a dynamically generated length scale $\xi_K = 1/T_K$, where $T_K$ is the Kondo temperature, that characterizes the Kondo screening.  
Spatial dependence of correlation functions is typically difficult to access with Bethe ansatz and so our approach has a unique advantage.

One observable characterizing the spatial profile of the Kondo cloud is the  equal time impurity-electron spin correlation function in a 3D metal:
\begin{equation}
    {\rm K}(\vb{x}) = \frac{1}{3}\ev{\Psi_w^\dagger \frac{\sigma^a}{2} \Psi_w (\vb{x},0) S^a(0)},
\end{equation}
where $\vb{x}$ is the distance from the impurity and $\Psi$ is the 3D electron field operator. The 
behavior of ${\rm K}$ in the short distance, $|\vb{x}| \ll \xi_K$, and long distance, $|\vb{x}| \gg \xi_K$, regimes was studied in Refs.~\cite{cloud_1996, cloud_1998, cloud_review}. Below, we describe the full crossover between these regimes in the large spin limit in both the single channel and multichannel cases.  

We follow the standard reduction from the 3D to the 1D problem.\cite{Affleck:1995ge} Assuming $s$-wave scattering, the 3D electron field $\Psi$ satisfies 
\begin{equation}
    \Psi_{w}(\vb{x}) \sim \frac{1}{2\sqrt{\pi}|\vb{x}|}[e^{-ik_F |\vb{x}|}\psi_{Lw}(|\vb{x}|) - e^{ik_F|\vb{x}|}\psi_{Rw}(|\vb{x}|)],
\end{equation}
where $k_F$ is the Fermi momentum. Then,\footnote{We choose the normalization of ${\rm K}_{un/2k_F}$ to agree with Refs. \cite{cloud_1996, cloud_1998,cloud_review}.}
\begin{align}
{\rm K}(\vb{x}) &= \frac{{\rm K}_{2k_F}(|\vb{x}|)}{4\pi^2 \vb{x}^2}\cos(2k_F |\vb{x}|) + \frac{{\rm K}_{un}(|\vb{x}|)}{8\pi^2 \vb{x}^2},  \\
{\rm K}_{2k_F}(x) &= -(2\pi/3)\ev{\psi_R^\dagger\frac{\sigma^a}{2} \psi_L(x,0) S^a(0) }, \label{eq:k2kfdef} \\
{\rm K}_{un}(x) &= (2\pi/3)\ev{j^a_0(x,0) S^a(0) }. \label{eq:kundef}
\end{align}

We evaluate both correlation functions ${\rm K}_{un}$ and ${\rm K}_{2k_F}$ to leading order in $1/s$ at zero temperature  (see App.\ \ref{app:imp_spin}). For the single channel case,
\begin{align}
    {\rm K}_{2k_F}(x) &= s \frac{\sin (\rho(\ell))}{6 x} + \mathcal{O}(s^0) \label{eq:k_2kf} \\
    {\rm K}_{un}(x) &= -\frac{2\sin^2 (\rho(\ell))}{3\pi x} + \mathcal{O}(s^{-1}),\label{eq:k_un}
\end{align}
where $\ell = \log(x/a)$, $a$ is a short distance cutoff, and $\rho(\ell)$ is given by Eq.~(\ref{rhoell}). Equivalently, we may express $\rho(\ell)$ in terms of the Kondo scale $T_0$ as in Eq.~(\ref{eq:dim_transmutation}),
\begin{equation}
    \rho(\ell) = \frac{\pi}{2} +\tan^{-1} \left(\frac{\log(xT_0)}{\pi s}\right).\footnote{Recall that $T_0$ and the Kondo temperature $T_K$ differ by at most a $\mathcal{O}(1)$ constant.}
\end{equation}
These results at small $\rho \approx sJ/2$ agree with the leading order perturbation theory results in the Kondo coupling $J$ \cite{cloud_1996, cloud_1998, cloud_review}:
\begin{align}
    {\rm K}_{2k_F}(x) &= \frac{s(s+1)J}{12 x} + \mathcal{O}(J^2) \\
    {\rm K}_{un}(x) &= -\frac{s(s+1)J^2}{6\pi x} + \mathcal{O}(J^3).
\end{align}


For the multi-channel case, 
\begin{align}
    {\rm K}_{2k_F}(x) &= \frac{Ks \sin(\rho(\ell))}{6x} \frac{1 -\kappa \rho(\ell)/2\pi}{1-\kappa \rho_0/2\pi} + \mathcal{O}(s^1) \label{eq:k_2kf_multi}, \\
    {\rm K}_{un}(x) &= -\frac{2K d(\rho(\ell)) \sin^2 (\rho(\ell))}{3\pi x} \frac{(1 - \kappa \rho(\ell)/2\pi)^2}{1- \kappa \rho_0/2\pi} + \mathcal{O}(s^{0}),\label{eq:k_un_multi}
\end{align}
where $\rho_0 = \rho(\ell =0)$ is the bare phase shift, see \eqref{rhoell}.
Using (\ref{eq:rhoW}), we obtain the following long distance asymptotic behavior in the overscreened case:
\begin{align}
    {\rm K}_{2k_F}(x \gg T_M) &\propto \frac{1}{T_M^{2/K}x^{1 + 2/K}} \\
    {\rm K}_{un}(x \gg T_M) &\propto \frac{1}{T_M^{4/K} x^{1 + 4/K}}.
\end{align}
These results agree with Ref.\ \cite{cloud_1998}, where it was shown that ${\rm K}_{2k_F}(x) \propto x^{-1 -\Delta}$ and  ${\rm K}_{un}(x) \propto x^{-1 - 2 \Delta}$ as $x \to \infty$ with the exact exponent $\Delta = 2/(K+2)$.

\subsection*{Sum rule}
It was shown in Refs.~\cite{cloud_1996, cloud_1998,cloud_review} that the uniform part of the impurity-electron spin correlation function, ${\rm K}_{un}(x)$, obeys a sum rule.  Slightly refining the arguments in Ref.~\cite{cloud_1998},  we show in App. \ref{app:imp_spin} that in the limit when the bare Kondo coupling $J \to 0$, the finite temperature ${\rm K}_{un}(x, T)$ satisfies, 
\beq \int_0^{\infty} dx  \, {\rm K}_{un}(x, T) = 2 \pi \left(T \chi_{{\rm imp}}(T) - \frac{1}{3} s(s+1)\right). \label{eq:sumrule}\eeq
In the fully screened or overscreened case, $T \chi_{{\rm imp}}(T) \to 0$ as $T \to 0$, so at zero temperature 
\beq \int_0^{\infty} dx \, {\rm K}_{un}(x, T =0) = -\frac{2 \pi}{3} s(s+1). \label{sumruleover}\eeq
In the underscreened case $2s > K$, $T \chi_{{\rm imp}}(T) \to \frac{1}{3}(s-K/2)(s-K/2+1)$ as $T \to 0$, so the right-hand-side of Eq.~(\ref{eq:sumrule}) reduces to 
\beq \int_0^{\infty} dx \, {\rm K}_{un}(x, T =0) = -\frac{2 \pi }{3} K (s-K/4 + 1/2). \label{sumruleunder}\eeq

We find that our result for ${\rm K}_{un}(x)$ in both the single channel and multichannel case obeys the sum rule above. Indeed, in the single channel case, from Eq.~(\ref{eq:k_un}),
\beq \int_a^{\infty} dx \, {\rm K}_{un}(x) = -\frac{2}{3\pi} \int_0^{\infty}d \ell\, \sin^2(\rho(\ell)) = \frac{2}{3\pi} \int_{\rho_0}^{\pi} \frac{d\rho}{\beta(\rho)} \sin^2 \rho = -\frac{2s}{3} \int_{\rho_0}^{\pi} d\rho = - \frac{2s}{3} (\pi - \rho_0) \to -\frac{2 \pi s}{3},\eeq
where we took the bare coupling $\rho_0 \to 0$ in the last step. This agrees with Eq.~(\ref{sumruleunder}) with $K  =1$ in the large $s$ limit.

In the multichannel case, similar manipulations give 
\beq \int_a^{\infty} dx \, {\rm K}_{un}(x) = - \frac{2 K s}{3} \int_{\rho_0}^{\rho_*} d\rho \,\frac{1 - \kappa \rho/2\pi}{1 - \kappa \rho_0/2\pi} \to -\frac{2 \pi s^2}{3} \left(1 - \left(1 - \frac{\kappa \rho_*}{2\pi}\right)^2\right). \label{eq:ossr}\eeq
Here $\rho_*$ stands for $\rho(\ell = \infty)$ and we took $\rho_0 \to 0$ in the last step. 
In the overscreened case,  $\rho_* = 2 \pi/\kappa$, and the right-hand-side of (\ref{eq:ossr}) becomes $-\frac{2 \pi s^2}{3}$, in agreement with Eq.~(\ref{sumruleover}). In the underscreened case, $\rho_* = \pi$, and the right-hand-side becomes $-\frac{2 \pi K(s- K/4)}{3}$ in agreement with Eq.~(\ref{sumruleunder}).

\section{Outlook}\label{sec:future}
In this paper, we demonstrate that the Kondo model becomes analytically tractable via renormalization group in the regime of large impurity spin. Our treatment adds to a number of other limits in which generalizations of the Kondo model can be solved without recourse to Bethe ansatz, such as the limit of large number of channels $K$ (with spin $s$ fixed)\cite{NozierBlandin, GanAndrei} and various generalizations of the $SU(2)$ spin to $SU(N)$ \cite{ParcolletSym, ParcolletAsym}.

Beyond the single impurity Kondo problem, our renormalization group treatment may provide a good starting point for analyzing chains of magnetic impurities patterned on a metallic surface (see Refs. \cite{HeinrichArray, GroverAsaadArray} and references therein).

\section*{Acknowledgements}
We are grateful to Paul Fendley, Matthew P. A. Fisher, Tarun Grover, Patrick Lee, Simon Martin,  Seth Musser, Adam Nahum, Chetan Nayak, Philip Phillips,  R. Shankar, Zhengyan Darius Shi, Ashvin Vishwanath and Paul Wiegmann for discussions. We also thank Natan Andrei,  Gabriel Cuomo, Leonid Glazman, Haoyu Hu, Zohar Komargodski and Qimiao Si for comments on the manuscript. M.M. is supported by the National Science Foundation under Grant No. DMR-1847861. The work of A.K. was supported
by the National Science Foundation Graduate Research Fellowship under Grant No. 1745302. A.K.
also acknowledges support from the Paul and Daisy Soros Fellowship and the Barry M. Goldwater
Scholarship Foundation.
\appendix
\section{Relation between the Kondo exchange coupling and Kondo phase shift}\label{app:unfolded}
To determine the relation between $\rho$ and $J$, we replace $\psi_{R\alpha}(x)$ with $\psi_{L\alpha}(-x)$ in \eqref{eq:folded1}:
\begin{equation}
    S = S_B + \int \dd \tau \int_{-\infty}^\infty \dd x \left[\psi_{L\alpha}^\dagger(\partial_\tau + i \partial_x) \psi_{L\alpha} \right]+  \int \dd \tau J \psi_{L\alpha}^\dagger(0,\tau)\frac{\sigma^a_{\alpha \beta}}{2} \psi_{L\beta}(0,\tau)  S^a.
\end{equation}
Here, the fields in the second integral live exactly at $x=0$, not an infinitesimal distance away from the impurity. We again take the $s = \infty$ limit and freeze the impurity spin while holding $sJ$ fixed.
The frozen action (i.e., the action for infinite $s$) is then 
\begin{align}
    S_\text{frozen}= \int \dd \tau \int_{-\infty}^\infty  \dd x \psi_{L\alpha}^\dagger(\partial_\tau + i \partial_x) \psi_{L\alpha} + \int \dd \tau  sJ {\psi}^\dagger_{L\alpha}(0,\tau) \frac{\sigma^3_{\alpha \beta}}{2} {\psi}_{L\beta}(0,\tau) .
\end{align}
We compute the relation between $\rho$ and $J$ using the equation of motion following \cite{KondoReview}. For the up spins, the equation of motion for a mode of energy $k$ is 
\begin{equation}
    i\partial_x \psi_{L\uparrow} + \frac{sJ}{2}\delta(x) \psi_{L\uparrow} = k \psi_{L\uparrow}.
\end{equation}
The general  solution to the equation of motion for $x\ne 0$ is
\begin{align}
    \psi_{L\uparrow}(x) = \begin{cases}
       A e^{-ikx}  &x >0  \\
        B e^{-ikx}  & x<0,
    \end{cases}
\end{align}
where $A$ and $B$ are to-be-determined complex constants. 
We regulate the wavefunction at $x=0$ by choosing 
\begin{equation}
    \psi_{L\uparrow}(0) = \frac{1}{2}(\psi_{L\uparrow}(0^+) + \psi_{L\uparrow}(0^-)) = \frac{1}{2}(A+B).
\end{equation}
The equation of motion is thus
\begin{equation}
    i(\psi_{L\uparrow}(0^+) - \psi_{L\uparrow}(0^-)) = -\frac{sJ}{4}(\psi_{L\uparrow}(0^+) + \psi_{L\uparrow}(0^-))
\end{equation}
We choose an overall phase for $\psi_{L\uparrow}$ such that $\psi_{L\uparrow}(0) \in \mathbb{R}$. Then,
\begin{align}
    \psi_{L\uparrow}(x) = \begin{cases}
       e^{i\rho/2} e^{-ikx}  &x >0  \\
        e^{-i\rho/2} e^{-ikx}  & x<0 \\
        \cos(\rho/2) & x=0,
    \end{cases}
    \label{eq:electron_phase_shift}
\end{align}
where $\rho = 2 \arctan(sJ/4)$.  The down spins have the opposite phase shift.

\section{Order \texorpdfstring{$\vec{\pi}^2$}{pi squared} contribution to \texorpdfstring{$S_\text{int}$}{Sint}}\label{app:kappa_term}
We compute $S_\text{int}$ order by order in $\pi^i$. When the impurity spin is aligned with the North pole, 
\begin{equation}
    \ev{{\psi}_{R}^\dagger\sigma^a \psi_L (x,0)} = -\frac{\sin \rho \delta_{a3}}{2\pi x}.
\end{equation}
Instead, suppose the impurity spin points in the fixed direction $\vec{n} = (\pi_1, \pi_2, \sqrt{1-\vec{\pi}^2})$. Then, 
\begin{equation}
\label{eq:su(2)_symmetrized_correlator}
    \ev{{\psi}_{R}^\dagger\sigma^a \psi_L (x,0)} = -\frac{\sin \rho n^a}{2\pi x}.
\end{equation}
For the total effective action to be $SU(2)$ invariant, $S_\text{frozen}(\rho) + S_\text{int}$ must reproduce the above correlation function. 
Recalling the allowed marginal boundary operators above \eqref{eq:particle_hole}, we express 
\begin{equation}
    S_\text{int} = \int \dd \tau f^b(\vec{\pi}) \psi^\dagger_{R\alpha} \sigma^b \psi_{R\beta}(0^+,\tau),
\end{equation}
where $f^b(\pi^i)$ are to be determined functions of the spin fluctuations $\vec{\pi}$.

We first consider $a \in \{1,2\}$. To leading order in $f^b$,
\begin{align}
     \ev{{\psi}_{R}^\dagger\sigma^i \psi_L (x,0)} &= -\int \dd \tau f^b(\vec{\pi})\ev{{\psi}_{R\alpha}^\dagger\sigma^i_{\alpha \beta} \psi_{L\beta} (x,0)  \psi^\dagger_{R\gamma} \sigma^b_{\gamma \delta} \psi_{R\delta}(0^+,\tau)}_F \nonumber \\
     &= -\frac{f^b(\vec{\pi})}{4\pi x }\tr[\sigma^i (\cos \rho \mathbb{1} + i \sin \rho \sigma^3)\sigma^b] \nonumber\\
     &= -\frac{f^b(\vec{\pi})}{2\pi x } (\cos \rho \delta_{ib} +\sin \rho \epsilon_{ib}),
\end{align}
where the subscript $F$ means the correlation function is evaluated with the frozen action. Note that $f^3$ does not enter this equation. The values of $f^j(\vec{\pi})$ that reproduce \eqref{eq:su(2)_symmetrized_correlator} are, to leading order in $\pi$,
\begin{equation}
    f^i(\vec{\pi}) = \sin \rho(\pi_i \cos \rho - \epsilon_{ij}\pi_j \sin \rho).
\end{equation}
Thus, we have reproduced \eqref{Sint1}. 

We now compute $f^3(\vec{\pi})$. We go to second order in $f^i(\vec{\pi})$ and first order in $f^3(\vec{\pi})$ (as shown below, $f^3(\vec{\pi}) \sim \mathcal{O}(\vec{\pi}^2)$). The second order contribution is 
\begin{align}
    \delta_{(2)}  \ev{{\psi}_{R}^\dagger\sigma^3 \psi_L (x,0)} &= \frac{1}{2}\mathcal{P}\int \dd \tau_1 \dd \tau_2 f^i(\vec{\pi})f^j(\vec{\pi})\ev{{\psi}_{R}^\dagger\sigma^3 \psi_{L} (x,0)  \psi^\dagger_{R} \sigma^i \psi_{R}(0^+,\tau_1)\psi^\dagger_{R} \sigma^j \psi_{R}(0^+,\tau_2)}_{F,{\rm c}} \nonumber \\
    &= -i f^i(\vec{\pi})f^j(\vec{\pi})\int \dd \tau_1  \frac{1}{16\pi^2(\tau_1^2 + x^2)}\tr[\sigma^3(\cos \rho \mathbb{1} + i \sin \rho \sigma^3)\{\sigma^j,\sigma^i\}] \nonumber\\
    &= \frac{\vec{\pi}^2\sin^3 \rho }{4\pi x}.
\end{align}
Likewise, the first order contribution is 
\begin{align}
    \delta_{(1)}  \ev{{\psi}_{R}^\dagger\sigma^3 \psi_L (x,0)} &= -\int \dd \tau f^b(\vec{\pi})\ev{{\psi}_{R\alpha}^\dagger\sigma^3_{\alpha \beta} \psi_{L\beta} (x,0)  \psi^\dagger_{R\gamma} \sigma^3_{\gamma \delta} \psi_{R\delta}(0^+,\tau)}_{F,{\rm c}} = -\frac{f^3(\vec{\pi})(\cos \rho)}{2\pi x }.
\end{align}
To find $f^3(\vec{\pi})$, we solve 
\begin{equation}
    \ev{\psi_R^\dagger \sigma^3 \psi_L(x,0)}_F +  \delta_{(1)}  \ev{{\psi}_{R}^\dagger\sigma^3 \psi_L (x,0)}  +  \delta_{(2)}  \ev{{\psi}_{R}^\dagger\sigma^3 \psi_L (x,0)} =   -\frac{\sin \rho }{2\pi x}(1 - \vec{\pi}^2/2 + \cdots).
\end{equation}
Solving yields $f^3(\vec{\pi}) = -[\sin (2\rho)/4] \vec{\pi}^2 + \mathcal{O}(|\vec{\pi}|^3) $, and so we have derived \eqref{eq:kappa_term}.

\section{Contact terms in the Kondo impurity effective action}
\subsection{Berry phase contact term} \label{app:berry_phase_counter}
As explained in Sec.\ \ref{sec:contact_terms}, the full effective action \eqref{Sefffull} has a Berry phase contact term (\ref{sBact}) that is fixed by a quantization condition. Specifically, consider a finite system of length $L$ with a boundary condition $\psi_{R\alpha} = -\psi_{L\alpha}$ at $x=L$.\footnote{As explained in Sec.\ \ref{sec:contact_terms}, this boundary condition ensures that the ground state is separated from the excited states by a finite gap of order $1/L$. Therefore, we can integrate out the fermions for energies $\omega \sim 1/L$.} Integrating out the fermion degrees of freedom must produce a quantized Berry phase term in the effective action for the boundary spin $\vec{n}(\tau)$. We carry out the integration over the fermions with the partial effective action $S = S_{\rm frozen}(\rho) + S_\text{int}$, as defined in (\ref{Sfrozen1}) and (\ref{Sint1}). The result of the integration yields a non-quantized contribution to the total Berry phase of the finite system and must be compensated by a Berry phase contact term proportion to $\delta s_B$ in (\ref{sBact}). We only fix the Berry phase contact term at $\mathcal{O}(\vec{\pi}^2)$, which suffices for our large-$s$ analysis.

Integrating out the fermions using the action $S_{\rm frozen}(\rho) + S_\text{int}$ yields the following correction to the Berry phase action of the spin:
\begin{align}
    \Delta S_\text{B} \approx  
    -\frac{1}{2}\int  \dd\tau_1 \dd \tau_2 \,\, \epsilon^{ij} \epsilon^{kl} \pi^i(\tau_1) \pi^k(\tau_2) \ev{j^j_x(0^+, \tau_1) j^l_x(0^+, \tau_2)} + r \int \dd\tau \, \vec{\pi}^2(\tau) \langle \psi^{\dagger}_R \sigma^3 \psi_R(0^+, \tau)\rangle , \nn\\ \label{DeltaS}
\end{align}
where $j^{a}_x$ is the spin current \eqref{eq:spin_current}. We use Eq.~(\ref{eq:spin_current}) and Wick's theorem to evaluate the correlator above.
For the finite size system, we write 
\begin{equation} \label{eq:finite_L_psi}
    \left(\begin{array}{c} \psi_{R\alpha}(x,\tau) \\ \psi_{L\alpha}(x,\tau)\end{array}\right) = \frac{1}{\sqrt{2L}} \sum_{k \in K_\alpha}  \left(\begin{array}{c} e^{i (k x -  \alpha \rho)} \\ e^{-i k x}\end{array}\right) e^{- k \tau} c_{k\alpha},
\end{equation}
where 
\beq
K_\alpha = \left\{\frac{n\pi}{L} + \frac{\pi + \alpha \rho}{2L}\mid n\in\mathbb{Z}\right\}
\eeq 
and $c_{k\alpha}$ is the annihilation operator for  electrons with momentum $k$ and spin index $\alpha \in \{1,-1\}$. We only need the propagator for $\psi_{R\alpha}$ at $x = 0$:
\beq \langle \psi_{R \alpha}(x = 0^+, \tau) \psi^{\dagger}_{R \beta}(x = 0^+, 0)\rangle = \frac{\delta_{\alpha \beta}}{4 L \sinh(\pi\tau/ 2L)} e^{-\alpha \rho \tau/2L}, \label{PsiPropx0} \eeq
which gives
\beq \langle j^i_{x}(\tau) j^j_{x}(0)\rangle = \frac{\sin^2 \rho}{8L^2 \sinh^2(\pi \tau/2L)} (\delta^{ij} \cosh (\rho \tau/L) + i \epsilon^{ij} \sinh (\rho \tau/L))\eeq
After fourier transforming,
\beq \Pi^{ij}(\omega) = \int \dd \tau \, \langle j^i_{x}(\tau) j^j_{x}(0)\rangle e^{i \omega \tau} = \delta^{ij} \Pi_1(\omega) +  \epsilon^{ij} \Pi_2(\omega).\eeq
\bea \Pi_1(\omega) &=& \frac{\sin^2 \rho}{8L^2} \int_{|\tau| > a} \dd \tau  \, \csch^2(\pi \tau/2L) \cosh (\rho \tau/L) \cos \omega \tau \nn\\
\Pi_2(\omega) &=& - \frac{\sin^2 \rho}{8L^2} \int \dd \tau  \, \csch^2(\pi \tau/2L) \sinh (\rho \tau/L) \sin \omega \tau 
\eea
Note that the integral for $\Pi_1(\omega)$ is UV divergent and we have regularized  it. We are only interested in the $\omega \to 0$ behavior of $\Pi_{1,2}$:
\bea \Pi_1(\omega) = \frac{\sin^2 \rho}{\pi^2 a} - \frac {\rho \sin 2 \rho}{4\pi L} + O(\omega^2), \label{Pi1o}\\
\Pi_2(\omega) = -\frac{\omega}{2\pi} \left(\rho - \frac{\sin 2 \rho}{2}\right) + O(\omega^3). \label{Pi2o}\eea
Finally, we may evaluate the last term in Eq.~(\ref{DeltaS}) using point-splitting regularization:
\beq \langle \psi^{\dagger}(0^+, \tau) \sigma^3 \psi(0^+,0)\rangle = \frac{\sinh (\rho \tau/2L)}{2L \sinh(\pi \tau/2L)} \to \frac{\rho}{2\pi L}, \quad\quad \tau \to 0. \label{eq:spin_density}\eeq
After putting all the contributions together, the low energy effective action for $\pi(\tau)$ in (\ref{DeltaS}) becomes:
\beq \Delta S_{\rm spin} = \frac{1}{2} \int \dd \tau \left[ - \frac{\sin^2 \rho}{2 \pi^2 a} \, \vec{\pi}^2 + \frac{1}{2\pi}\left(\rho - \frac{1}{2}\sin 2 \rho\right) \epsilon^{ij} \pi^i  \d_\tau \pi^j \right].\eeq
The contribution to $\Delta S_{\rm spin}$ from the last term in $\Pi_1(\omega)$ in Eq.~(\ref{Pi1o}) has canceled the contribution from the $r$ term in Eq.~(\ref{DeltaS}) after using Eq.~(\ref{eq:kappa_term}). The $\pi$ independence of the effective potential fixes $\delta m_\pi$ in the contact term, Eq.~(\ref{sBact}), to be the non-universal value 
\beq
\delta m_\pi = \frac{\sin^2 \rho}{4 \pi^2 a}.
\eeq 
The quantization of the Berry phase term for $\vec{n}$ yields the coefficient of the Berry phase contact term, Eq.~(\ref{deltasB}).

\subsection{Spin density contact term} \label{app:finite_density}
As noted in Sec.~\ref{sec:magnet}, in the frozen spin limit described by $S_{\rm frozen}(\rho)$ the fermion spin density receives a contribution localized at the impurity; we capture this contribution with 
a contact term $\delta s_M$ in our effective action (\ref{Seffh}).  To determine $\delta s_M$, we again consider a finite system of length $L$ and compute the spin density away from the impurity. Since the total spin must be quantized, the localized spin near the impurity must cancel the total spin away from the impurity. We again use the point-splitting regularization
\beq \ev{j_0^3(x,\tau)} = \frac{1}{2} \langle \psi^{\dagger}_R(x, \tau + \eta) \sigma^3 \psi_R(x,\tau)\rangle + \frac{1}{2} \langle \psi^{\dagger}_L(x, \tau + \eta) \sigma^3 \psi_L(x,\tau)\rangle, \eeq
where the limit $\eta \to 0$ is understood. The fermion propagator at coincident spatial points is identical to Eq.~(\ref{PsiPropx0}):
\beq \langle \psi_{R \alpha}(x , \tau) \psi^{\dagger}_{R \beta}(x , 0)\rangle = \langle \psi_{L \alpha}(x , \tau) \psi^{\dagger}_{L \beta}(x , 0)\rangle = \frac{\delta_{\alpha \beta}}{4 L \sinh(\pi\tau/ 2L)} e^{-\alpha \rho \tau/2L}.\eeq
Therefore,
\beq \ev{j_0^3(x,\tau)} = \langle \psi^{\dagger}(x, \tau + \eta) \sigma^3 \psi(x,\tau)\rangle = \frac{\sinh (\rho \eta/2L)}{2L \sinh(\pi \eta/2L)} \to \frac{\rho}{2\pi L}, \quad\quad \eta \to 0. \eeq
Thus, a total spin $S^z = \rho/2\pi$ is accumulated in the bulk of the system away from the impurity. To restore spin quantization, a spin $\delta s_M = -\rho/2\pi$ must be localized near the impurity.

\subsection{Spin current induced by adiabatic tuning of phase shift} \label{app:spin_current}
Here, we use another method, complementary to the calculation in appendix \ref{app:finite_density}, to compute the fermion spin accumulated near the impurity at a finite phase shift $\rho$. 
Working in real time, we consider the action $S_{\rm frozen}(\rho)$, adiabatically tune the phase shift $\rho$, and compute the resulting spin current that flows to the impurity. We assume that the phase shift induced by the impurity is initially $\rho_0$ and increases monotonically by a small value $\Delta \rho$ adiabatically.
The resulting Hamiltonian is
\begin{equation}
H(t) = H_\text{frozen}(\rho_0) +  \delta \rho(t)\, \psi_{R}^\dagger \sigma^3 \psi_{R} (0^+).
\end{equation}
where $\delta \rho(t = -\infty) = 0$, $\delta \rho (t = \infty) = \Delta \rho \ll 1$.

We are after the amount of spin that accumulates at the impurity during this process:
\begin{equation}
\ev{S^z} = -\int_{-\infty}^\infty\ev{j^3_x(x,t)} \dd t,
\end{equation}
where $j^3_x(x,t)$ is defined in \eqref{eq:spin_current}.
By the Kubo formula, 
\begin{equation} \label{eq:kubo}
    \ev{j^3_x(x,t))} = -i\int_{-\infty}^t \dd t'\ev{[j^3_x(x,t),\psi_{R}^\dagger \sigma^3 \psi_{R} (0^+,t')]}_0 \delta \rho(t'),
\end{equation}
The real time (non-time ordered) fermion correlators evaluated with $H_\text{frozen}(\rho_0)$ are:
\bea \langle \psi_{R\alpha}(x,t) \psi^{\dagger}_{R\beta}(x',t')\rangle  &=& \langle \psi^{\dagger}_{R\alpha}(x,t) \psi_{R\beta}(x',t')\rangle = \frac{i \delta_{\alpha \beta}}{2\pi(x-x'-(t-t') + i \epsilon)},\nn\\
 \langle \psi_{L\alpha}(x,t) \psi^{\dagger}_{L\beta}(x',t')\rangle &=&  \langle \psi^{\dagger}_{L\alpha}(x,t) \psi_{L\beta}(x',t')\rangle = \frac{-i \delta_{\alpha \beta}}{2\pi(x-x'+(t-t') - i \epsilon)},\nn\\
  \langle \psi_{R\alpha}(x,t) \psi^{\dagger}_{L\beta}(x',t')\rangle  &=& \frac{i e^{-i \alpha \rho_0} \delta_{\alpha \beta}}{2\pi(x+x'-(t-t') + i \epsilon)},\nn\\
  \langle \psi^{\dagger}_{L\alpha}(x,t) \psi_{R\beta}(x',t')\rangle &=& \frac{-i e^{-i \alpha \rho_0} \delta_{\alpha \beta}}{2\pi(x+x'+(t-t') - i \epsilon)},\nn\\
   \langle \psi_{L\alpha}(x,t) \psi^{\dagger}_{R\beta}(x',t')\rangle  &=& \frac{-i e^{i \alpha \rho_0} \delta_{\alpha \beta}}{2\pi(x+x'+(t-t') - i \epsilon)},\nn\\
   \langle \psi^{\dagger}_{R\alpha}(x,t) \psi_{L\beta}(x',t')\rangle  &=& \frac{i e^{i \alpha \rho_0} \delta_{\alpha \beta}}{2\pi(x+x'-(t-t') + i \epsilon)}. \label{Psit}
\eea
Using Wick's theorem we find
\begin{align}
   \ev{[j^3_x(x,t),\psi^\dagger_{R}\sigma^3 \psi_{R}(0^+,t')]}_0  = \frac{1}{4\pi^2(t-t'-x+i\epsilon)^2}- \frac{1}{4\pi^2(t-t'-x-i\epsilon)^2}\nn\\
   +\frac{1}{4\pi^2(t-t'+x-i\epsilon)^2} - \frac{1}{4\pi^2(t-t'+x+i\epsilon)^2}.
\end{align}
Because we are only interested in $t > t'$ in Eq.~(\ref{eq:kubo}) and because $x>0$, the last two terms in the equation above cancel. Then
\begin{align}
    \ev{[j^3_x(x,t),\psi^\dagger_{R}\sigma^3 \psi_{R}(0^+,t')]}_0 &= \frac{i}{2\pi^2} \Im\left(\frac{1}{(t-t'-x+i\epsilon)^2}\right) \nonumber \\
     &= \frac{-i}{2\pi^2} \pdv{t} \Im\left(\frac{1}{t-t'-x+i\epsilon}\right) \nonumber \\
     &= \frac{i}{2\pi}\pdv{t}  \delta(t-t'-x).
\end{align}
Evaluating the integral in Eq.~(\ref{eq:kubo}) yields
\begin{equation}
    \ev{j^3_x(x,t))} = \frac{1}{2\pi}  \frac{\d}{\d t} \delta \rho (t-x) \implies 
    \ev{\delta S^z }= -\int_{-\infty}^\infty  \ev{j^3_x(x,t))}\dd t = -\frac{\Delta \rho}{2\pi}.
\end{equation}
Thus, the total spin accumulated at the impurity is 
\begin{equation} \label{eq:deltas}
    \Delta S^z(\rho) = -\frac{\rho}{2\pi}.
\end{equation}

\section{Finite temperature calculations}
\subsection{Current-current correlator at the boundary at finite temperature} \label{app:current_current}
We compute the correlation function $\langle j^i_x(0^+, \tau) j^j_x(0^+, \tau')\rangle_0$ at finite temperature, which enters the calculation of thermodynamic properties in section \ref{sec:thermo}.

The finite temperature fermion correlation functions as $L \to \infty$ {follow by a conformal mapping from} 
\eqref{psipsi1}:
\begin{align}
    \ev{\psi_{R\uparrow}(x,\tau) \psi^\dagger_{L\uparrow}(x',0)} &=  \frac{Te^{-i\rho}}{2\sin (\pi T(\tau - i(x+x')))},\quad x,x' > 0, \nn\\
    \ev{\psi_{L\uparrow}(x,\tau) \psi^\dagger_{R\uparrow}(x',0)} &=  \frac{Te^{i\rho}}{2\sin (\pi T(\tau + i(x+x')))},\quad x,x' > 0, \nn\\
    \ev{\psi_{R\uparrow}(x,\tau) \psi^\dagger_{R\uparrow}(x',0)} &=  \frac{T}{2 \sin (\pi T(\tau - i(x-x')))},\quad x,x' > 0, \nn \\
    \ev{\psi_{L\uparrow}(x,\tau) \psi^\dagger_{L\uparrow}(x',0)} &=  \frac{T}{2 \sin (\pi T(\tau + i(x-x')))},\quad x,x' > 0, \label{psipsi1T}
\end{align}
When $L$ is finite and $LT \gg 1$, this expression still holds up to an error $\mathcal{O}(\exp{-2L\pi T})$.
We use \eqref{jpsiR} to simplify the correlator $\ev{j^i_x(0^+,\tau) j^i_x(0^+,0)}$:
\begin{align}
\ev{j^i_x(0^+,\tau) j^j_x(0^+,0)} = \frac{T^2}{4} \sin^2 \rho \csc^2(\pi T \tau)  \big[&\sin^2 \rho \tr(\sigma^i \sigma^j) -\sin \rho \cos \rho \epsilon^{ik} \tr(\sigma^k \sigma^j) \\&-\sin \rho \cos \rho \epsilon^{jl} \tr(\sigma^i \sigma^l) + \cos^2 \rho \epsilon^{ik} \epsilon^{jl} \tr(\sigma^k \sigma^l)\big].
\end{align}
Simplifying yields 
\begin{align}
\ev{j^i_x(0^+,\tau) j^j_x(0^+,0)} = \frac{T^2}{2}\sin^2 \rho \csc^2(\pi T \tau) \delta_{ij}.
\end{align}
\subsection{Gradient formula for boundary entropy}\label{app:g_theorem}
We now compute the impurity entropy by using a gradient formula relating the boundary entropy to the boundary $\beta$-function of a $1+1$D quantum system \cite{boundaryentropy}. 
We first recall the theorem of Ref.~\cite{boundaryentropy}. Consider a $1+1$D CFT with a boundary. Let the complete set of boundary coupling constants be $\lambda^a$, let $Z$ be the total partition function and let $\phi^a$ be local boundary fields satisfying
\begin{equation}
    \pdv{\lambda^a}\log Z = \int \dd \tau \ev{\phi^a}. 
\end{equation}
Then, the boundary entropy $S_\text{bdry}$ satisfies 
\begin{equation}
    \pdv{S_\text{bdry}}{\lambda^a} = g_{ab}(\lambda) \beta^b(\lambda), 
\end{equation}
where $\beta^b(\lambda) = - (\pdv*{\lambda^b}{\ell})$ is the $\beta$-function for coupling constant $\lambda^b$, and $g_{ab}$ is a metric on the space of boundary coupling constants defined as  
\begin{equation}
    g_{ab}(\lambda) =2 \int_0^{1/T} \dd \tau_1 \int_0^{1/T} \dd \tau_2 \ev{\phi^a(\tau_1) \phi^b(\tau_2)}_c \sin^2(\pi T (\tau_1 - \tau_2)).
\end{equation}

In the case of the Kondo impurity, the coupling constant of interest is $\rho$, and we evaluate the metric $g_{ab}$ at lowest order in $1/s$. 
As noted below \eqref{drho}, the boundary operator $\psi_R^\dagger \sigma^3 \psi_R(0^+)$ tunes the phase shift $\rho$ in $S_{\rm frozen}$. Therefore, to leading order in $1/s$ the operator conjugate to $\rho$ is:
\begin{equation}
    \phi(\tau) = -\psi_R^\dagger \sigma^3\psi_R.
\end{equation}
With this definition, we compute the metric at lowest order in $1/s$.
From Wick's theorem and Eq.~(\ref{psipsi1T}), 
\begin{equation}
     \ev{\phi(\tau_1) \phi(\tau_2)}_c = \frac{T^2}{2} \csc^2(\pi T(\tau_1 - \tau_2)).
\end{equation}
We thus find that $g(\rho) = 1$, and using the $\beta$-function (\ref{beta}),
\begin{equation}
    \pdv{S_\text{bdry}}{\rho} = -\frac{\sin^2\rho}{\pi s},
\end{equation}
in agreement with \eqref{eq:imp_entropy} at large $s$.

\section{Bethe ansatz in the large \texorpdfstring{$s$}{s} limit: the single channel case} \label{app:larges}
\subsection{Zero temperature magnetization}
We evaluate \eqref{MhBethe0} in the large $s$ limit to $\mathcal{O}(1)$ in $s$.
We first substitute $y \to y/s$:
\begin{equation}
M(h) = s-\frac14+ \frac{1}{2 \pi^{3/2}} \int_0^{\infty} \frac{d y}{y} {\rm Im}\left\{ \Gamma\left(\frac12-\frac{iy}{s}\right) e^{i y/s \left(\log y/s - 1 + 2 \log \frac{h}{2 T_1}\right)} \right\} e^{-2 \pi (1-1/(4s)) y }.
\end{equation}
We assume $s \gg 1$ 
and we choose a cutoff $\Delta$ such that $1 \ll \Delta \ll s$. For $0<y<\Delta$, the expression 
\begin{equation}
\label{eq:gammaexp}
    \Gamma\left(\frac12-\frac{iy}{s}\right) e^{i y/s (\log y/s - 1)} \approx \sqrt{\pi} + i\mathcal{O}((y/s)\log|y/s|) + \mathcal{O}((y/s)^2).
\end{equation}
is slowly varying and can be treated as a constant. For $y > \Delta$, the expression remains $\mathcal{O}(1)$ and decreases for $y \gtrsim s$. Then, 
\begin{align}
M(h) \approx s-\frac14&+ \frac{1}{2 \pi^{3/2}} \int_0^{\Delta} \frac{dy}{y} \sqrt{\pi} e^{-2 \pi (1-1/(4s)) y} \sin\left(2 y/s \log \frac{h}{2T_1} \right)\nonumber \\
&+ \frac{1}{2 \pi^{3/2}} \int_\Delta^{\infty} \frac{d y}{y} {\rm Im}\left\{ \Gamma\left(\frac12-\frac{iy}{s}\right) e^{i y/s \left(\log y/s - 1 + 2 \log \frac{h}{2 T_1}\right)} \right\} e^{-2 \pi y (1-1/(4s))}.
\end{align}
Because the first integrand is oscillating and suppressed by $\exp{-2\pi \Delta}/\Delta$ for $\omega > \Delta$, we expand the first integral from $[0,\Delta]$ to $[0,\infty]$ 
and we replace $\exp{-2 \pi  y(1-1/(4s))}$ with $\exp{-2 \pi y}$
. 
Because the second integrand is oscillating, decreasing, and suppressed by $\exp{-2\pi \Delta}/\Delta$ 
we can neglect the second integral. Therefore, 
\begin{align}
M(h) \approx s-\frac14+ \frac{1}{2 \pi^{3/2}} \int_0^{\infty} \frac{d y}{y} \sqrt{\pi} e^{-2 \pi y} \sin\left(2 y/s \log \frac{h}{2T_1} \right) = s - \frac14 + \frac{1}{2\pi} \tan^{-1}\left(\frac{1}{\pi s} \log \frac{h}{2T_1}\right).
\end{align}

\subsection{Free energy at finite temperature}
Now, we evaluate \eqref{F_B} in the large $s$ limit.
The solution $\eta_n$ has the following asymptotic behavior \cite{KondoReview}:
\beq \eta_n(\ell = \infty) = \frac{\sinh^2 (n x_0)}{\sinh^2 x_0} - 1, \quad \quad \eta_n(\ell = -\infty) = \frac{\sinh^2 (n+1) x_0)}{\sinh^2 x_0} - 1, \label{etabc} \eeq
where $x_0 = h/2T$. Furthermore, $\eta_n(\ell)$ is a monotonically decreasing function of $\ell$.

We use the recursive equations (\ref{eq:recurse}) to derive an expression for the free energy in the large $s$ limit. For this, we need the form of $\eta_n(\ell)$ for $n \to \infty$.  We follow Ref.~\cite{KondoReview} and write,
\beq \eta_n(\ell) = e^{2 \xi_n(\ell)} \frac{\sinh^2 (n x_0)}{\sinh^2 x_0} - 1, \label{xindef}\eeq
and $\xi_n$ satisfies
\beq \xi_n = G \star (\xi_{n+1} + \xi_{n-1}) + b_n, \quad\quad n > 1, \label{xiG}\eeq
where
\beq  b_n = -\frac{1}{2} \log\left(1+ \epsilon_n (1- e^{-2 \xi_n})\right), \quad\quad \epsilon_n = \left(\frac{\sinh^2(n x_0)}{\sinh^2 x_0} - 1\right)^{-1}.\eeq

\subsubsection*{Zero field}
We first consider the case when $h = 0$. Then 
\beq \xi_n(\ell=\infty) = 0, \quad\quad \xi_n(\ell = -\infty) = \log(1+1/n) \approx 1/n,\eeq 
where we took the large $n$ limit in the last equation.  Since $\xi_n(\ell)$ is a uniformly decreasing function of $\ell$, it remains $O(1/n)$ for all $\ell$. We may then expand $b_n \approx - \xi_n/n^2$ for large $n$. 

To simplify the convolution in (\ref{xiG}), we may Fourier transform,
\beq G(k) = \int \dd \ell G(\ell) e^{-i k \ell} = \frac 12 \sech \pi k/2 \stackrel{k \to 0}{\approx} \frac12  - \frac{\pi^2}{16} k^2 +O(k^4). \label{eq:Gk} \eeq
As shown later, $\xi_n(\ell)$ varies over a length-scale $\ell \sim n$ for $n \gg 1$, thus, the Fourier transform $\xi_n(k) = \int \dd \ell \, \xi_n(\ell) \exp{-i k \ell}$ is appreciable only for $k \ll 1$. So in Fourier space and for $k \ll 1$, \eqref{xiG} becomes:
 \beq - (\xi_{n+1}(k) + \xi_{n-1}(k) - 2 \xi_n(k)) + \frac{2}{n^2} \xi_n(k) = - \frac{\pi^2}{4} k^2 \xi_n. \eeq
 Since $n \gg 1$, we replace the discrete second derivative with respect to $n$ with the continuous second derivative in the above equation:
 \beq - \frac{\d^2 \xi}{\d n^2} + \frac{2}{n^2} \xi_n(k) = - \frac{\pi^2}{4} k^2 \xi_n \label{xinhomo}\eeq
Fixing $k$, the above equation has two linearly independent solutions: 
\beq
\xi_n(k) = \left(1 \pm \frac{2}{ \pi |k| n}\right) e^{\mp \pi |k| n/2}.
\eeq 
To have a sensible large-$n$ limit, we must choose the solution that decays with $n$:
\beq \xi_n(k) = \left(1 + \frac{2}{\pi |k| n}\right)\left(1 + \frac{2}{\pi |k|n_*}\right)^{-1} e^{-\pi |k| (n-n_*)/2} \xi_{n_*}(k), \eeq
where $n_* \gg 1$ is some reference $n$. We see that $\xi_n(k)$ is only appreciable for $k \lesssim 1/n$, justifying our initial assumption. Taking $n \gg n_*$ and focusing on $k \lesssim 1/n$ yields
\beq \xi_n(k) \approx \frac{n_*}{n} \left(1 + \frac{\pi |k| n}{2}\right) e^{-\pi |k| n/2} \xi_{n^*}(k).\eeq
Converting this to real space,
\beq \xi_n(\ell) = n_* \int \dd \ell' \frac{(n \pi/2)^2}{((\ell - \ell') +(n \pi/2)^2)^2} \xi_{n_*}(\ell')\eeq
Due to the behavior of the kernel, the integral above is saturated over $\ell' \sim n$. On the other hand, we expect $\xi_{n_*}(\ell')$ to vary over $\ell' \sim n_*$. Therefore, we may replace $\xi_{n_*}(\ell')$ by its asymptotic values: $\xi_{n_*} (\ell') \approx \theta(-\ell')/n_\star$. Then performing the integral:
\beq \xi_n(\ell) \approx \frac{1}{2n} - \frac{1}{\pi n} \left(\tan^{-1}(2 \ell/\pi n) + \frac{2 \ell/\pi n}{1 + (2 \ell/\pi n)^2}\right). \label{xinx0}\eeq
Inserting this into the expression for $F_B$ (\ref{F_B}) and noting that $G(\ell)$ varies over the scale $\ell \sim 1$,
\bea F_B &\approx& - \frac{T}{2} \log\left[1 + \eta_{2s}(\ell = \log T_0^B/T)\right] \approx - T(\log 2s + \xi_{2s}(\ell = \log T_0^B/T)) \nn\\
&=& -T \left[\log 2 s + \frac{1}{4s} - \frac{1}{2\pi s} \left(\tan^{-1}(\ell/\pi s) + \frac{\ell/\pi s}{1 + (\ell/\pi s)^2}\right)\right], \eea
where $\ell = \log T_0^B/T$ in the last equation.

\subsubsection*{Small finite field}
We now add a small external magnetic field (i.e., $x_0 \ll 1/n$) in order to obtain the finite temperature susceptibility. As a result, the functions $\xi$, $\eta$, and $F_B$ all change by some value $\delta \xi$, $\delta \eta$, and $\delta F_B$ respectively. We denote the $x_0 = 0$ solution $\xi_n$ in \eqref{xinx0} by $\xi^{(0)}_n$. 
We recall that $\eta_n$ satisfies the boundary conditions (\ref{etabc}), which translate to 
  \beq \delta \xi_n(\ell = \infty) = 0, \quad\quad \delta \xi_n(\ell = -\infty) \approx \frac{n}{3} x^2_0. \label{dxinbc}\eeq
We work to leading order in $x_0$ and in the limit $n \gg 1$. Going into Fourier space, \eqref{xiG} gives: 
  \beq - \frac{\d^2 \delta \xi_n(k)}{\d n^2} + \frac{2}{n^2} \delta \xi_n(k) +  \frac{\pi^2}{4} k^2 \delta \xi_n =  \frac{2x_0^2}{3} \xi^{(0)}_n(k), \label{xinx}\eeq
where using  \eqref{xinx0},
  \beq \xi^{(0)}_n(k) = \frac{i}{n(k+i \epsilon)} \left(1 + \frac{\pi n |k|}{2}\right) e^{-\pi n |k|/2}. \eeq
 Because $\xi^{(0)}_n(k)$ is a function of $kn$, we can look for a particular solution to  \eqref{xinx} of the form: $\delta \xi_n(k) = q(kn)/k^2$, where $q$ is a to be determined function. Substituting yields 
  \beq q''(\bar{k}) + \frac{2}{\bar{k}^2} q(\bar{k}) + \frac{\pi^2}{4} q(\bar{k}) = \frac{2 x^2_0}{3} \frac{i}{\bar{k} + i \epsilon}\left(1+ \frac{\pi \bar{k}}{2}\right) e^{-\pi \bar{k}/2}, \label{qde}\eeq
  where $\bar{k} = nk$. A particular solution to (\ref{qde}) is $q(\bar{k}) = (ix_0^2/3) \bar{k} \exp{- \pi |\bar{k}|/2}$, so that
  \beq \delta \xi_n^{\rm par}(k) = \frac{i x^2_0}{3} \frac{n}{k+ i \epsilon} e^{- \pi n |k|/2}. \label{dxink}\eeq
 Note that the $k \to 0$ behavior of the particular solution (\ref{dxink}) matches the boundary conditions (\ref{dxinbc}). The general solution $\delta \xi_n(k)$ is composed of both the particular solution $\delta \xi_n^{\rm par}(k)$ and a solution to the homogenous equation (\ref{xinhomo}).  Choosing the solution that decays with $n$, we find
\beq
 \delta \xi_n(k) = \delta \xi_n^{\rm par}(k) + \delta \xi_n^{\rm hom}(k), \quad \quad \delta \xi^{\rm hom}_n(k) = \left(1 + \frac{2}{\pi |k| n} \right) e^{-\pi n |k|/2} c(k),
\eeq
for some to-be-determined function $c(k)$.
Again, because $\delta\xi^{\rm hom}_n(k)$ is only appreciable for $|k| \lesssim n$, we can replace $c(k)$ by its leading behavior for $k \to 0$. Since $\delta\xi_n(\ell)$ is bounded for $|\ell| \to \infty$, we must have $c(k) \to c x^2_0 {\rm sgn}(k)$ for $k \to 0$. The resulting $\delta\xi^{\rm hom}_n(k)$ is then down by two powers of $n$ compared to $\delta\xi^{\rm par}_n(k)$ in (\ref{dxink}). Thus, we conclude that $\delta \xi_n(k)$ is given by $\delta\xi^{\rm par}_n(k)$ to leading order in $n$. After Fourier transforming, we obtain
\beq
 \delta \xi_n (\ell) = \frac{x_0^2 n}{6}\left(1 - \frac{2}{\pi}\tan^{-1}\left(\frac{2\ell}{n \pi}\right)\right).
 \eeq
 
  Substituting into the expression for the free energy (\ref{F_B}) and using \eqref{xindef} with $x_0 = h/(2T)$,
  \beq
  \delta F_B \approx  -\frac{h^2}{6T}\left(s^2 + \frac{s}{2}\left\{1 - \frac{2}{\pi}\tan^{-1}\left(\frac{\ell}{\pi s}\right)\right\} +  \mathcal{O}(s^0)\right),
  \eeq 
  where $\ell = \log T^B_0/T$.
 
\section{Numerical computation of Bethe ansatz free energy}\label{app:numerics}

Recall the recursion relations in \eqref{eq:recurse}. Per \cite{KondoReview}, for a suitable choice of initial $\eta_n$s, one can successively update each $\eta_n$ as follows:
\begin{equation}\label{eq:update}
    \log \eta_n^i = G\star \log[(1 + \eta_{n+1}^{i-1})(1 + \eta_{n-1}^i)] - 2\delta_{n,1}e^{\ell},
\end{equation}
where $i$ labels the update step. Assuming that the initial $\eta_n$s are chosen to respect the boundary condition $\lim_{n\to \infty} (\log \eta_n)/n = h/T$, after sufficiently many update steps, the $\eta_n$s converge to their true values. For the given recursion relation and boundary condition, we find that 
\begin{equation}
    \lim_{\ell \to -\infty} \log(1 + \eta_{n}(\ell)) = 2 \log \left(\frac{\sinh[(n+1) x_0]}{\sinh(x_0)}\right), \quad \lim_{\ell \to \infty} \log(1 + \eta_{n}(\ell)) = 2 \log \left(\frac{\sinh(n x_0)}{\sinh(x_0)}\right),
\end{equation}
where $x_0 = h/2T$ \cite{numerics}. We define the left limit and right limit as $L^l$ and $L^r$ respectively. We thus define our initial $\eta_n^0$ seed values to be 
\begin{equation}
    \log(1 + \eta_n^0(\ell)) = \frac{1}{2}(L^l + L^r) - \frac{1}{2}(L^l-L^r) \frac{\ell}{\ell_\text{max}},
\end{equation}
where $[-\ell_\text{max},\ell_\text{max}]$ is the bounds of the numerical recursion.
This choice leads to quick convergence.

Computing $\eta_n$ numerically requires some practical constraints. Because the convolutions are done numerically, we only define $\eta_n(\ell)$ in a range $[-\ell_\text{max}, \ell_\text{max}]$ for values of $\ell$ spaced by a granularity $\ell_g$. We additionally only update a finite number of $\eta_n$s between $n=1$ and $n = N_\text{max}$. In Figs.\ \ref{fig:sp_heat} and \ref{fig:susc}, we choose $\ell_\text{max} = 500$, $\ell_g = 0.2$, and $N_\text{max} = 400$.  We complete a total of $4 \times 10^4$ updates.

We additionally consider edge effects. Suppose we are computing $G \star f$ numerically for some function $f(\ell)$, and we want the value of 
$G\star f(\ell)$ where $|\ell|$ is close to $\ell_\text{max}$. Because the support of $f$ and $G$ is limited to $[-\ell_\text{max}, \ell_\text{max}]$, when we sum to compute the convolution, we must estimate the value of $f(\ell)$ for $|\ell| >\ell_\text{max}$ to compute $G\star f(\ell)$ without dropping any terms.

In our case, $f(\ell)=\log[(1 + \eta_{n+1}^{i-1}(\ell))(1 + \eta_{n-1}^i(\ell))]$. Per the asymptotic behavior of $\eta_n$ described in \cite{KondoReview}, $f(\ell)$ is roughly constant outside $[-\ell_\text{max}, \ell_\text{max}]$, so we can set $f(\ell) \approx f(\sign(\ell) \ell_\text{max})$ for $|\ell| > \ell_\text{max}$. This approximation allows us to compute the convolution without dropping terms in the numerical convolution, but it also suppresses the difference between the maximum and minimum values of $\eta_n$. Thus, we first numerically compute an intermediate $(\eta_n^i)_\text{int}$  by using \eqref{eq:update} and treating $f(\ell)$ as constant for $|\ell| > \ell_\text{max}$.  We then rescale $(\eta_n^i)_\text{int}$ as follows. 
The finite $\ell$ corrections to $L^l$ and $L^r$ are 
\begin{equation}
L^l(-\ell_\text{max}) \approx L^l  + 2((n+1) \coth((n+1) x_0) - \coth(x_0))x_0 \left[-\frac{1}{2\ell_\text{max}} - \frac{\log|\ell_\text{max}|}{4\ell_\text{max}^2}\right] - \frac{\pi^2 n^2}{6 \ell_\text{max}^3},\quad  \ell_\text{max} \to \infty,
\end{equation}
\begin{equation}
L^r(\ell_\text{max}) \approx L^r  + 2(n \coth(n x_0) - \coth(x_0))x_0 \left[\frac{1}{2\ell_\text{max}} - \frac{\log|\ell_\text{max}|}{4\ell_\text{max}^2}\right] + \frac{\pi^2 n^2}{6 \ell_\text{max}^3},\quad  \ell_\text{max} \to \infty.
\end{equation}
The first correction parameterized by $x_0$ is in \cite{KondoReview}. The $\mathcal{O}(\ell_\text{max}^{-3})$ correction
arises from the asymptotic form of the impurity specific heat, i.e.,  that $C(T) \sim s^2 \pi^2 \log(T/T_0^B)^{-4}$ for $|\log(T/T_0^B)|\gg s$ \cite{Kondothermo}. We use these limits to rescale the intermediate $(\eta_n^i)_\text{int}$ as 
\begin{equation}
    \eta_n^i(\ell)  = \left[(\eta_n^i)_\text{int}(\ell) - \overline{(\eta_n^i)}_\text{int}\right] \cdot \frac{L^l(-\ell_\text{max}) - L^r(\ell_\text{max})}{(\eta_n^i)_\text{int}(-\ell_\text{max}) - (\eta_n^i)_\text{int}(\ell_\text{max})}
    +\overline{(\eta_n^i)}_\text{int},
\end{equation}
where
    \beq \overline{(\eta_n^i)}_\text{int} = \frac{1}{2} \left[(\eta_n^i)_\text{int}(-\ell_\text{max}) + (\eta_n^i)_\text{int}(\ell_\text{max})\right].
\eeq
This procedure is an ad hoc approximation to compensate for edge effects, and its accuracy is determined by how constant $f(\ell)$ is close to $\pm \ell_\text{max}$. As long as $\ell_\text{max} \gg n^2$, we can make this assumption. 

\section{Phase shift and magnetization in the multichannel Kondo problem} \label{app:multichannelW}
In this section of the appendix, we solve the RG equation (\ref{Hrho}) for the phase shift in the multichannel Kondo problem using the Lambert W function.\cite{NIST:DLMF} The Lambert W functions are a family of solutions $\{W_k(z)\}$ labeled by integer $k$ for the following implicit equation:
\beq
W_k(z) e^{W_k(z)} = z.
\eeq 
The principal branch of this family, $W_0(z)$ satisfies the condition $z \in \mathbb{R}^{\ge 0} \implies W(z) \in \mathbb{R}^{\ge 0}$. 

\subsection*{The Lambert W function}\label{app:Product_Log}
\begin{figure}
\centering
\includegraphics[width = 0.48\textwidth]{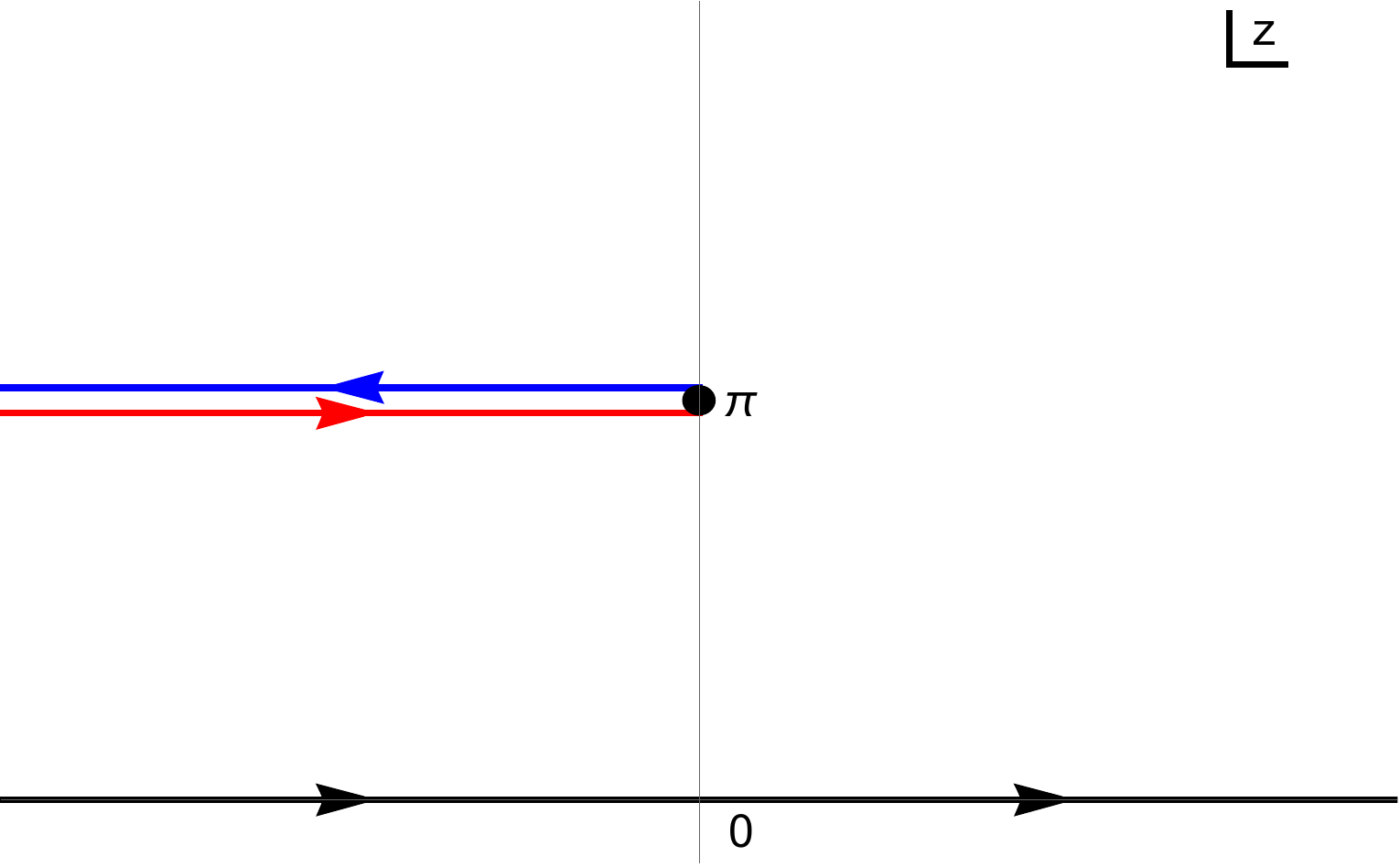}
\hfill
\includegraphics[width = 0.48\textwidth]{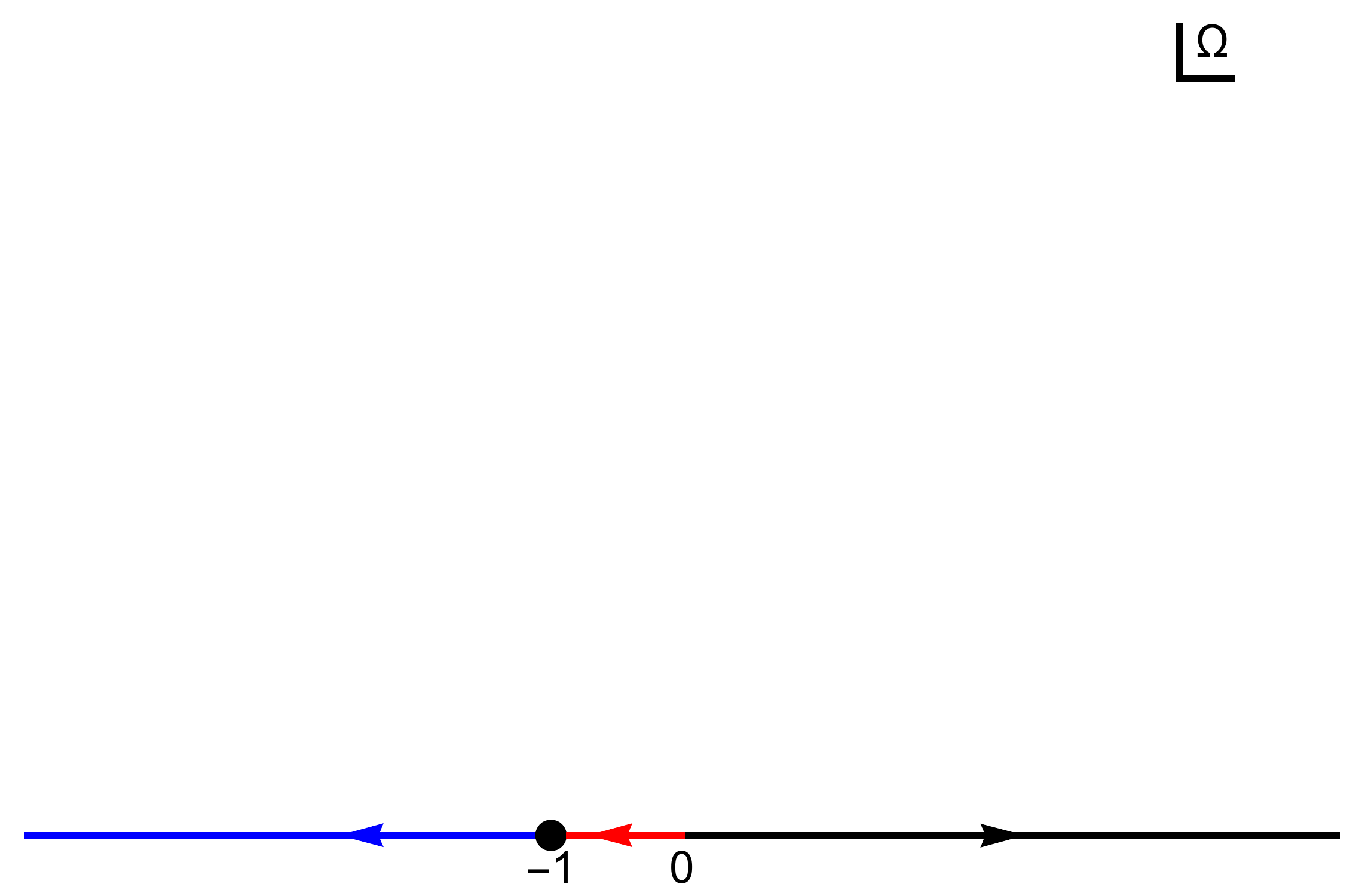}\\
\caption{
The function $\Omega(z)$ in Eq.~(\ref{eq:Omegadef}) defined for $z$ in the upper half plane by stitching together various branches of the $W$ function. Left: the domain of $\Omega$, right: the image of $\Omega$. Two sides of the branch cut are shown in red and blue. }\label{fig:W}
\end{figure}

We write $W_k(z) = u_k(z) + i v_k(z)$, where $u_k(z)$, $v_k(z) \in \mathbb{R}$. Moreover, we write $z$ in polar form as $re^{i\theta}$, where $r >0$ and $-\pi\le \theta < \pi$. Then,
\beq 
(u_k+iv_k) e^{u_k+iv_k} = re^{i\theta}.
\eeq 
Multiplying by $e^{-i \theta}$ and taking the real and imaginary parts of the resulting equation, 
 we find 
\bea
&u_k = v_k \cot (\theta - v_k), \\
&v_k \csc(\theta -v_k) e^{v_k \cot(\theta -v_k)} = r \label{ImplicitProductLog}
\eea
Consider the left hand side of \eqref{ImplicitProductLog} as a function of $v_k$. The left hand side must be positive, so
\beq
v_k \in
\begin{cases}
     [\theta + (2k-1)\pi,\theta + 2k\pi] & k >0 \\
     [\theta + 2k\pi,\theta + (2k+1)\pi] & k <0 \\
     [0,\theta] & k= 0.
\end{cases}
\eeq
For future convenience, we define a function in the upper half-plane 
\beq \Omega(z) = W_{k(z)}(e^{z-1}), \label{eq:Omegadef} \eeq
where the branch $k \ge 0$ is chosen so that ${\rm Im}(z) \in ((2k-1) \pi, (2k+1) \pi)$. This function has a single branch cut along the ray $z = t + i \pi$, $t \in {\mathbb R}_{-}$. The image of the upper half plane under $\Omega$ is the upper half plane, with the two sides of the branch cut mapping to the negative real axis (see Fig.~\ref{fig:W}).

\subsection*{RG flow of the multichannel phase shift}
Note that after exponentiating Eq.~(\ref{Hrho}), we obtain
\beq \left(\frac{2 \pi}{\kappa} -\rho\right) \csc \rho \,e^{(\frac{2\pi}{\kappa} -\rho)\cot \rho} = \left(\frac{\omega}{T_M}\right)^{2/K} e^{-1}.\eeq
This equation is of the same form as Eq.~(\ref{ImplicitProductLog}), with the identification $v = 2\pi/\kappa - \rho$, $\theta = 2\pi/\kappa$, $r = \left(\omega/T_M\right)^{2/K} e^{-1}$. Thus,
\beq \rho(\omega) = \frac{2\pi}{\kappa} - {\rm Im} \,W\left[\left(\frac{\omega}{T_M}\right)^{2/K} e^{2 \pi i/\kappa -1}\right]. \label{eq:rhoWapp}\eeq
The appropriate branch of the $W$ function in Eq.~(\ref{eq:rhoWapp}) follows from the condition $\rho \in [0, \pi].$ In the overscreened case $\kappa > 2$, $\rho \in [0,{2\pi}/\kappa]$, so we must choose the zeroth branch. More generally, the condition $\rho \in [0,\pi]$ implies
\beq \rho(\omega) = \frac{2\pi}{\kappa} - {\rm Im} \, \Omega\left(\frac{2}{K} \log\frac{\omega}{T_M} + \frac{2 \pi i}{\kappa}\right) \label{eq:rhomulti},\eeq
where $\Omega(z)$ is defined in and below Eq.~(\ref{eq:Omegadef}). We see that $\rho$ is a smooth function of $\omega$ and $\kappa$, except for a weak singularity along the branch cut  $\kappa = 2$, $\omega < T_M$. As already noted in section \ref{sec:multi}, the branch cut is a consequence of the breakdown of the large $s$ expansion in this region.

\subsection*{Magnetization in the overscreened multichannel case}
In this section we prove the equivalence of Eqs.\ \eqref{BetheMagK} and \eqref{MagK} at large $s$ and $K>2s$. We let $\ell = (2/K) \log T_M/h$ and observe that in the large spin limit Eq.~\eqref{BetheMagK} can be written as  
\beq M_{\rm Bethe}(\ell) = \int_{-\infty}^{\infty} dy \,P(y) e^{- i y \ell}, \label{MBethel} \eeq
with 
\beq P(y) =  \frac{-i K}{4 \pi (y-i \epsilon)}  \frac{1}{\Gamma(1+iy)} \left(\frac{i y +\epsilon}{e}\right)^{iy} \frac{\sinh(\frac{2 \pi y}{\kappa})}{\sinh \pi y}. \label{eq:Pdef}\eeq
Now our renormalization group treatment at large $s$, Eq.~(\ref{MagK}), gives in the overscreened regime $\kappa >2$
\beq M(\ell) = \frac{K}{2\pi}\Im \left\{W_0\left(e^{-\ell - 1} e^{2\pi i/\kappa}\right)\right\}. \label{eq:Ml}
\eeq 
To show that $M$ in Eq.~(\ref{eq:Ml}) is identical to the Bethe ansatz result (\ref{MBethel}), we show that its Fourier transform with respect to $\ell$ is equal to $P$. We thus compute the Fourier transform:
\begin{align}
\tilde{M}(y) &= \frac{K}{2\pi}\int_{-\infty}^\infty \frac{\dd \ell}{2\pi} \, e^{i\ell y} M(\ell) \nonumber\\
&= \frac{K}{8\pi^2 i} \int_{-\infty}^\infty \dd \ell\, e^{i\ell(y - i\epsilon)}\left[W_0\left(e^{-\ell - 1} e^{2\pi i/\kappa}\right) - W_0\left(e^{-\ell - 1} e^{-2\pi i/\kappa}\right)\right] \nonumber \\
&= \frac{K}{8 \pi^2 i} \left[\int_{-\infty -2\pi i /\kappa}^{\infty -2\pi i /\kappa} \dd \ell\,  W_0(e^{-\ell}) e^{i(\ell - 1 + 2\pi i/\kappa)(y-i\epsilon)} -  \int_{-\infty +2\pi i /\kappa}^{\infty +2\pi i /\kappa} \dd \ell \, W_0(e^{-\ell}) e^{i(\ell - 1 - 2\pi i/\kappa)(y-i\epsilon)} \right]
\end{align}
where $\epsilon >0$ regulates the integral. Because $\epsilon >0$ and because $W_0(e^{-\ell})$ is analytic for $|\Im \ell| \le 2\pi /\kappa$, we can deform both integrals to the real line: 
\begin{align}
\tilde{M}(y) &= -\frac{K}{4\pi^2 i}\int_{-\infty}^{\infty} \dd \ell\,  W_0(e^{-\ell}) e^{i(\ell - 1)(y-i\epsilon)}\sinh\left(\frac{2\pi}{\kappa}(y - i\epsilon)\right) \nonumber \\
&=-\frac{K e^{-i y-\epsilon}}{4\pi^2 i}\sinh\left(\frac{2\pi}{\kappa}(y - i\epsilon)\right) \int_{0}^{\infty} \dd z \, W_0(z) z^{-\epsilon - iy - 1} 
\end{align}
Now, we substitute $u = W(z)$, and rewrite the integral:
\begin{align}
\tilde{M}(y) &= -\frac{K e^{-iy-\epsilon}}{4\pi^2 i}\sinh\left(\frac{2\pi}{\kappa}(y - i\epsilon)\right) \int_{0}^{\infty} \dd u (1+u) (ue^{u})^{-\epsilon - iy} \nonumber\\
&= -\frac{K e^{-iy-\epsilon}}{4\pi^2 i}\sinh\left(\frac{2\pi}{\kappa}(y - i\epsilon)\right) (iy + \epsilon)^{-2 + \epsilon + iy}\Gamma(1 -\epsilon - iy).
\end{align}
After applying the Euler reflection formula, $\Gamma(z) \Gamma(1-z) = \pi \csc (\pi z)$, to   $\Gamma(1-\epsilon - iy)$ and taking $\epsilon\to 0$ wherever we can avoid poles, we find $\tilde{M}(y) = P(y)$ in Eq.~(\ref{eq:Pdef}).

A similar calculation shows that the result of RG calculation in the underscreened regime,
\beq M(h) = \frac{K}{2\pi}\Im \Omega\left[\frac{2}{K}\log \left(h/T_M\right)+ \frac{2\pi i}{\kappa}\right] \eeq
agrees with the Bethe ansatz result in the large spin limit\cite{WiegmannMultiShort, WiegmannMultiLong}:
\beq
M_\text{Bethe}(h) = s-\frac{K}{2} -\frac{iK}{4\pi^{3/2}}\int_{-\infty}^\infty \dd y \frac{e^{2i y \frac{\log(h/T_H)}{K}}}{y -i \epsilon}\frac{\Gamma(1 + i y/K)\Gamma(1/2 - i y/K)}{\Gamma(1 + iy)} \left(\frac{iy + \epsilon}{e}\right)^{i y} e^{- \pi |y| \left(\frac{2}{\kappa} - 1\right)}.\label{BetheMagKover}
\eeq 


\section{Impurity entropy in the multichannel Kondo model}\label{app:entropyK}
We compute the partition function for the action \eqref{SefffullK}. The partition function is 
\beq 
\mathcal{Z} \approx \mathcal{Z}_\text{frozen} \mathcal{Z}_\text{imp},
\eeq 
where $\mathcal{Z}_\text{imp}$ is the partition function for the action \eqref{eq:nonlocalK}, i.e., the action after integrating out the fermions. This partition function is, to leading order in $1/s$ (see \cite{Hasenfratz} and the discussion above \eqref{Jacobian}), 
\begin{equation}
\mathcal{Z}_\text{imp}  \approx \int \dd \vec{n}_0 \int \mathcal{D} \vec{\pi} \prod_i\delta\left(\frac{1}{\beta}\int_0^{\beta}\pi^i(\tau)\right) e^{-S_2[\vec{\pi}]}.
\end{equation}
where $S_2[\vec{\pi}]$ is defined in \eqref{eq:PiPropK}, $\vec{n}_0$ is defined in \eqref{spinfluctuations}. 
Fourier transforming the $\pi^i$ fields as 
\begin{equation}
    \pi^i(\tau) = \frac{1}{\sqrt{\beta}} \sum_n e^{-i\omega_n \tau} \tilde{\pi}^i(i\omega_n),
\end{equation}
we find that 
\begin{equation}
    \mathcal{Z}_\text{imp}  \approx 2\beta  \int' \, \mathcal{D} \vec{\pi} e^{-\tilde{S}_2[\tilde{\pi}^1, \tilde{\pi}^2]},\quad \tilde{S}_2[\tilde{\pi}^1, \tilde{\pi}^2] = \frac{1}{2} \sum_{n\ne 0} \tilde{\pi}^i (-i\omega_n) D^{-1}_{\pi, ij}(i\omega_n) \tilde{\pi}^j(i\omega_n).
\end{equation}
where 
\beq
\int' \, \mathcal{D} \vec{\pi} = \int \prod_{n = 1}^{\infty} \prod_i \frac{ d {\rm Re} \tilde{\pi}^i (i\omega_n)  d {\rm Im} \tilde{\pi}^i (i\omega_n)}{\pi}.
\eeq
Then,
\begin{equation}
\mathcal{Z}_\text{imp} \approx2\beta \prod_{n \ne 0} \sqrt{\det D_{\pi}(i\omega_n)} =2 \beta \prod_{n\ne 0} \frac{\sqrt{ d(\rho)}}{s|\omega_n|} = 2\beta \prod_{n =1}^\infty \frac{\beta^2d(\rho)}{s^2(2\pi n)^2}. 
\end{equation}
After computing the product with zeta-function regularization, we find 
\beq
\mathcal{Z}_\text{imp} = \frac{2s}{\sqrt{d(\rho)}} + \mathcal{O}(1/s) \implies \mathcal{S}_\text{imp} = \log(2s) - \frac{1}{2}\log d(\rho) + \mathcal{O}(1/s). \label{eq:Simpmultiapp}
\eeq
As usual, we RG improve the expression for the entropy by substituting $\rho \to \rho(T)$. For future reference, we  write an explicit expression for $\mathcal{S}_\text{imp}(T)$. Noting that
\beq d(\rho)^{-1} = \left(1 - \frac{\kappa \rho}{2\pi}\right)^2 + \left(\frac{\kappa}{2\pi}\right)^2\sin^2\rho + \frac{\kappa}{2\pi}\sin (2\rho) \left(1 - \frac{\kappa \rho}{2\pi}\right) \eeq
and using Eqs.~(\ref{eq:rhomulti}), (\ref{ImplicitProductLog}), we obtain
\beq \mathcal{S}_\text{imp}(T) \approx \frac{1}{2} \log \eta^{\rm rg}(z, \bar{z}), \quad\quad z = \frac{2}{K} \log\frac{T}{T_M} + \frac{2\pi i}{\kappa} \label{eq:Setarg},\eeq
and
\beq \eta^{\rm rg}(z, \bar{z}) =  \left(\frac{K}{\pi}\right)^2 [{\rm Im} \,\Omega(z)]^2\frac{|1 + \Omega(z)|^2}{|\Omega(z)|^2}. \label{eq:etarg}\eeq


\section{Bethe ansatz in the large \texorpdfstring{$s$}{s} limit: the multichannel case}
\label{app:FmultiBethe}
The impurity free energy expression from the multichannel Bethe ansatz is \cite{PhysRevLett.52.364}
\begin{equation}
    F = -T f_{2s}(T/T_0^B),\footnote{We add an overall factor of $-T$ missing from \cite{PhysRevLett.52.364}.} \quad f_j(T/T_0^B) = \int_{-\infty}^\infty \dd \ell \, G(\ell + \log(T/T_0^B)) \log(1 + \eta_j(\ell)),
\end{equation}
where $G(\ell) = (1/2\pi) \sech(\ell)$ as in the single channel case. $T_0^B$ is a microscopic energy scale  related to the energy scale $T_M$ defined in \eqref{tmdef} by some yet-to-be-determined constant $c$ 
\begin{equation}
    \log(T_M/T_0^B)=  c.
\end{equation}
The functions $\eta_j(\ell)$ satisfy the recursion relations
\begin{equation}
    \log \eta_j = G\star [\log(\eta_{j+1}  +1) + \log(\eta_{j-1} + 1)] - 2\delta_{j,K}e^{\ell} \label{multichannelrecursion}
\end{equation}
with boundary conditions
\begin{equation}
   \lim_{\ell \to -\infty} \eta_j(\ell)  = \frac{\sinh^2((j+1)x_0)}{\sinh^2(x_0)}, \quad  \lim_{\ell \to \infty} \eta_j(\ell)=
\begin{cases}
\sin^2\left(\frac{(j+1)\pi}{K+2}\right)\csc^{2}\left(\frac{\pi}{K+2}\right) - 1
 & j < K\\
\sinh^2((j+1-K)x_0)\csch^{2}(x_0) - 1 & j \ge K.\\
\end{cases}\footnote{We correct a typo in the argument of the $\csc^{2}$ function in \cite{PhysRevLett.52.364} using the entropy result from \cite{AffleckLudwigg}.}
\label{multichannelboundary}
\end{equation}
Here, $x_0 = h/2T$. Throughout the appendix, we work at large $j$. As a consequence, as we show below, $\eta_j(\ell)$ varies on scales of size $\ell \sim j$, and so its Fourier transform $\tilde{\eta}_j(k)$ is only appreciable for $k \sim 1/j$. As in the single channel case, Eq.~(\ref{eq:Gk}), we use that for small $k$
\beq
\tilde{G}(k) \approx \frac{1}{2} - \frac{\pi^2}{16} k^2.
\eeq 
Substituting this into (\ref{multichannelrecursion}) yields
\beq
\log \eta_j(\ell) = \frac{1}{2}\left[\log(1 + \eta_{j+1}(\ell)) + \log(1 + \eta_{j-1}(\ell))\right] + \frac{\pi^2}{8} \dv[2]{\ell} \log \eta_j(\ell) - 2\delta_{j,K} e^{\ell}. \label{eq:multichannelBethe}
\eeq 
For large $j$, we treat $j$ as a continuous variable. Then for $j \neq K$, 
\beq
\log \eta_j(\ell) - \log(1 + \eta_j(\ell)) = \frac{1}{2} \dv[2]{j} \log(1 + \eta_{j}(\ell)) + \frac{\pi^2}{8} \dv[2]{\ell} \log \eta_j(\ell) 
\label{continuousrecursionsmalleta}
\eeq 
Finally, using that $\eta_j \gg 1$,\footnote{This assumption holds (as we show later) as long as $\kappa$ is not near $2$.} we find that 
\beq 
\left(\dv[2]{j} + \frac{\pi^2}{4} \dv[2]{\ell}\right) \log \eta_j(\ell) = -\frac{2}{\eta_j(\ell)}, \quad j \neq K.
\eeq 
We now convert to complex coordinates by identifying 
\begin{equation}
    z = -\frac{2}{K}\left(\ell - \frac{ \pi i j}{2}\right), \quad \bar{z} = -\frac{2}{K}\left(\ell + \frac{ \pi i j}{2}\right),
    \label{eq:defz}
\end{equation} 
and also rescale $\eta  = \left(K/2\pi\right)^2 \hat{\eta}$. Then, the recursion relation reduces to
\beq 
\partial_{\bar{z}} \partial_z \log \hat{\eta}(z, \bar{z})= -\frac{2}{\hat{\eta}(z, \bar{z})}, \quad   {\rm Im}(z) \neq \pi.  
\label{continuousrecursion}
\eeq 
Note that (\ref{continuousrecursion}) is just the Liouville equation. 

\subsection*{Zero magnetic field}
We first present the functional form of $\eta_j$ as predicted by our large $s$ RG results in section \ref{sec:multi} 
, and then we show that 
it satisfies the Bethe ansatz Eq.\ \eqref{continuousrecursion}. 
As already noted, for large $j$, $\eta_j(\ell)$ varies over $\ell \sim O(j)$. 
Then, the free energy and entropy are 
\begin{equation}
    F \approx -\frac{T}{2} \log(1 + \eta_{2s}(\log(T_0^B/T))), \quad \mathcal{S} \approx \frac{1}{2} \log \eta_{2s}(\log(T_0^B/T)),
\end{equation}
where we have again used $\eta_j \gg 1$ for large $j$ away from $j = K$. Matching with Eq.\ \eqref{eq:Simpmultiapp} RG improved by $\rho \to \rho(T)$, we find that the large $j$ prediction for $\eta_j$ is
\begin{equation}
    \eta_j(\log(T_0^B/T)) \approx j^2 d(\rho(T))^{-1}
\end{equation}
Therefore,  we expect the solution to Bethe ansatz equations $\eta^{{\rm BA}}(z, \bar{z})$ to be related to $\eta^{\rm rg}$ in Eq.~(\ref{eq:etarg}) coming from our large $s$ RG solution via 
\beq \eta^{\rm BA}(z, \bar{z}) = \eta^{\rm rg}\left(z-\frac{2c}{K}, \bar{z} - \frac{2c}{K}\right),\eeq
where $z, \bar{z}$ are given by Eq.~(\ref{eq:defz}).
This function, indeed, satisfies the Liouville equation Eq.\ \eqref{continuousrecursion} and the boundary conditions in Eq.\ \eqref{multichannelboundary}.

From \eqref{multichannelrecursion}, we expect $\eta^{\rm BA}(z, \bar{z})$ to be non-smooth across $\Re z < 0$ and $\Im z = \pi$. Note that in the limit considered, we expect a smooth behavior  of $\eta^{\rm BA}(z, \bar{z})$ across $\Re z > 0$ and $\Im z = \pi$. Indeed, we are considering $|\ell| \sim O(s)$, so the $e^{\ell}$ term on the right-hand-side of Eq.~(\ref{eq:multichannelBethe}) is exponentially small for $\ell < 0$.  Since $\Omega(z)$ has a branch cut along the same ray $z \in {\mathbb R}_- + \pi i$, we conclude that $c/K < \mathcal{O}(1)$, i.e $\log T_M/T^B_0 \sim O(s^0)$.


\subsection*{Small finite field}
We now add a small external magnetic field $h$ such that $x_0 \ll 1/K$ and $x_0 \ll 1/s$. As a result, the function $\eta(z,\bar{z})$ change by some value $\delta \eta(\bar{z},z)$. Working again in the regime $\eta (z, \bar{z}) \gg 1$ and $\delta \eta (\bar{z}, z) \ll 1$, we find that the differential equation $\delta \eta$ satisfies is 
\beq
\partial_z \partial_{\bar{z}} (\delta \hat{\eta}/ \hat{\eta}) = 2\delta \hat{\eta}/\hat{\eta}^2. \label{continuousdeltaeta}
\eeq 

We once again find the functional form of $\delta \eta$ predicted by the large $s$ results and check that it satisfies Eq.\ \eqref{continuousdeltaeta}.
For small, finite $x_0$, the free energy
\beq
F(T, x_0) = -\frac{T}{2} \log(1 + \eta_{2s}(\ell) + \delta \eta_{2s}(\ell)) = F(T, x_0=0) - \frac{T}{2} (\delta \eta_{2s}/\eta_{2s}), \quad\quad \ell = \log T^B_0/T.
\eeq 
Per \eqref{susccalc}, the magnetic susceptibility satisfies
\beq
\chi = \frac{1}{8T \eta_{2s}} \pdv[2]{x_0}(\delta \eta_{2s}).
\eeq 
After matching to \eqref{multichannelsusc}, we find that 
\beq
\frac{\delta \eta(z, \bar{z})}{\eta(z, \bar{z})} =
\frac{K^2 x_0^2}{3\pi^2}\left( \Im \Omega(z)\right)^2.
\eeq 
This function indeed satisfies \eqref{continuousdeltaeta} and the boundary conditions in \eqref{multichannelboundary}.

\section{Exact BCFT results for observables near the overscreened fixed point}\label{app:exact_bcft}
In this appendix, we briefly provide the exact BCFT results from Refs.\ \cite{affleck1993, AFFLECK1991641} for the impurity specific heat, impurity susceptibility, and impurity resistivity near the overscreened fixed point.

The leading irrelevant operator at the overscreened fixed point is $\mathcal{J}^a_{-1} \phi^a$, where $\phi^a$ is the spin-$1$ primary field of dimension $\Delta = 2/(2+K)$ and $\mathcal{J}^a_{-1}$ is the Kac-Moody raising operator  \cite{AFFLECK1991641}. 

Thus, to obtain the low temperature behavior of observables at the over-screened fixed point, we consider the action 
\beq
S = S_\text{over-screened} + \lambda \int \dd \tau \mathcal{J}_{-1}^a \phi^a. 
\eeq 
The resulting low temperature impurity specific heat and susceptibility were computed in Ref.\ \cite{AFFLECK1991641}:
\begin{align}
    C_\text{imp}(\lambda T^\Delta \ll 1) &= \frac{6 \Delta^2 \Gamma (1/2 - \Delta)}{\Gamma(1-\Delta)}\lambda^2 T^{2\Delta} \pi^{2\Delta + 3/2} (2 + K/2)\\
    \chi_\text{imp}(\lambda T^\Delta \ll 1) &=  \frac{\Gamma(1/2 - \Delta)}{\Gamma(1-\Delta)} \lambda^2 T^{2\Delta - 1} \pi^{2\Delta - 1/2} (2 + K/2)^2 
\end{align}
The bulk specific heat and bulk susceptibilities are 
\beq
    C_\text{bulk} = \frac{2 \pi^2 K \nu}{3}  T, \quad \chi_\text{bulk} = \frac{K \nu}{2},
\eeq 
where $\nu = \frac{k^2_F}{2 \pi^2 v_F}$ is the bulk density of states per spin, per channel, and so we arrive at the exact value of the Wilson ratio:
\beq
R_W = \lim_{T \to 0} \frac{\chi_\text{imp}(T)/C_\text{imp}(T)}{\chi_\text{bulk}(T)/C_\text{bulk}(T)} = \frac{(K+2)^2 (K/2 + 2)^2}{18}.
\eeq 

The low temperature resistivity for a density $n_i$ of impurities was computed in Ref.\ \cite{affleck1993}:
\beq
r_\text{imp}(\lambda T^\Delta \ll 1) = \frac{3n_i(1 - S_{1})}{2K\pi (q_e \nu v_F)^2}\left[1 - \frac{2 \lambda N_0 \sin (\pi \Delta)}{1 - S_{1}} (2\pi T)^\Delta I(\Delta)\right],
\eeq 
where 
\beq S_1 = \frac{\cos(\pi(2s+1)/(K+2))}{\cos(\pi/(K+2))} \eeq
and
\beq
N_0 = \left[\frac{9}{8} \frac{\Gamma\left(\frac{K}{K+2}\right)^2}{\Gamma\left(\frac{K+1}{K+2}\right)\Gamma\left(\frac{K-1}{K+2}\right)\cos\left(\frac{\pi}{K+2}\right)}\frac{\cos\left(\frac{2\pi}{K+2}\right)-\cos\left(\frac{2\pi (2s + 1)}{K+2}\right)}{1 + 2 \cos\left(\frac{2\pi}{K+2}\right)}\right]^{1/2},
\eeq 
and 
\beq 
I(\Delta) = \int_0^1 \dd u \left[|\log u| (1-u)^{\Delta - 1} {}_2F_{1}(1+\Delta, 1+\Delta; 1;u) - \frac{\Gamma(1 + 2\Delta)}{\Gamma(1+\Delta)^2}u^{\Delta - 1}(1-u)^{-1-\Delta}\right].
\eeq 
The quantity 
\beq
\frac{[r_\text{imp}(\lambda T^\Delta \ll 1) - r_\text{imp}(0)]^2}{n_i C_\text{imp}(\lambda T^\Delta \ll 1)}
\eeq 
eliminates the non-universal coefficient $\lambda$ and thus allows us to compare our result for the resistivity with the exact BCFT results. In the large $K$ limit, 
$N_0 \approx \sqrt{3}/2 \sin(2\pi/\kappa)$ and $I(\Delta) \approx -1/\Delta$, and so we find that 
\beq
\frac{[r_\text{imp}(\lambda T^\Delta \ll 1) - r_\text{imp}(0)]^2}{n_i C_\text{imp}(\lambda T^\Delta \ll 1)}  \approx \frac{9n_i \sin^2(2\pi/\kappa)}{16 \pi^2 K (\nu q_e v_F)^4}
\eeq 

\section{Impurity-electron spin correlation function} \label{app:imp_spin}
In this appendix, we calculate the equal time correlation functions $\ev{\psi_R^\dagger \sigma^a \psi_L (x,0) S^a(0)}$ and $\ev{j^a_0(x,0) S^a(0)}$ to leading order in $s$.
\subsection*{Single channel case}
We first do the calculation in the single channel case to leading order in $1/s$.
First, we calculate 
\begin{equation}
    \ev{\psi_R^\dagger \sigma^3 \psi_L(x,0) S^3(0)} = s\ev{\psi_R^\dagger \sigma^3 \psi_L(x,0)} + \mathcal{O}(s^0) = -\frac{s \sin \rho}{2\pi x} + \mathcal{O}(s^0).
\end{equation}
Additionally, $ \ev{\psi_R^\dagger \sigma^i \psi_L(x,0) S^i(0)}$ does not contribute at $\mathcal{O}(s^1)$, so 
\begin{equation}
    \ev{\psi_R^\dagger \sigma^a \psi_L(x,0) S^a(0)} = -\frac{s \sin \rho}{2\pi x} + \mathcal{O}(s^0).
\end{equation}
Because the anomalous dimension of the impurity spin operator $\vec{S}$ is $\sim \mathcal{O}(s^{-2})$, we can RG improve this expression via $\rho \to \rho(\ell)$, where $\ell = \log(\Lambda x)$.

We now calculate $\ev{j^a_0(x,0) S^a(0)}$. First, to $\mathcal{O}(s^0)$, 
$\ev{j^3_0(x,0) S^3(0)} = 0$ because the contributions from $S_\text{int}$ to $\ev{j^3_0(x,0)}$, corresponding to the diagrams in Fig.\ \ref{fig:SelfEnergy} with the fermion line closed off at the point $(x,0)$, vanish.\footnote{In particular,  the contribution of the diagram in Fig.~\ref{fig:SelfEnergy}b) vanishes for the same reason as the diagram in Fig. 12a) in Ref.~\cite{cloud_1998}.}
On the other hand, 
\begin{align}
    \ev{j^i_0(x,0) S^i(0)} &= -s\epsilon^{jk} \int \dd \tau \ev{\pi^i(0)\pi^j(\tau)}_0  \ev{j^i_0(x,0) j^k_x(0^+,\tau)}_0 + \mathcal{O}(1/s) \nonumber\\ 
    &= -\frac{i \delta_{ik}}{2}\int \dd \tau \ev{j^i_0(x,0) j^k_x(0^+,\tau)}_0 \text{sgn}(\tau) + \mathcal{O}(1/s) \nonumber \\
    &= -\frac{\sin^2 \rho}{\pi^2 x} + \mathcal{O}(1/s).
\end{align}
Here, $\ev{}_0$ denotes an expectation value in the decoupled theory. Thus, 
\begin{equation}
    \ev{j^a_0(x,0) S^a} = -\frac{\sin^2 \rho}{\pi^2 x} + \mathcal{O}(1/s).
\end{equation}
Again, because the anomalous dimension of the impurity spin is $\sim \mathcal{O}(s^{-2})$, we can RG improve this expression via $\rho \to \rho(\ell)$.

\subsection*{Multichannel case}
We repeat the calculation for the multichannel case. 
To leading order in $1/s$, 
\begin{equation}
    \ev{\psi^\dagger_{Rw} \sigma^a(x,0) \psi_{Lw} S^a} = -\frac{Ks \sin \rho}{2\pi x} + \mathcal{O}(s^1). \label{eq:multichannelspincorr}
\end{equation}
To RG improve this expression we need to take into account the impurity spin anomalous dimension, which in the multichannel case scales as $\eta_{s} \sim O(s^{-1})$.  We calculate $\eta_s$ from Eq.\ \eqref{eq:PiPropK}:
\begin{equation}
    \ev{n^a(\tau)n^a(0)} \approx 1  + D_{\pi, ii}(\tau) - D_{\pi, ii}(0) = 1 -  \frac{\kappa d(\rho) \sin^2 \rho  }{\pi^2 s} \log(\Lambda |\tau|),
\end{equation}
so 
\begin{equation}
    \eta_s = \frac{\kappa d(\rho) \sin^2 \rho}{2\pi^2 s} + {\cal O}(s^{-2}).
\end{equation}
To RG improve Eq.~(\ref{eq:multichannelspincorr}), we use the Callan-Symanzik equation
\beq \left(\Lambda \frac{\d}{\d \Lambda} + \beta(\rho) \frac{\d}{\d \rho} + \eta_s(\rho)\right) {\rm K}_{A}(x, \rho, \Lambda) = 0. \eeq
Here and below ${\rm K}_A$  stands for any of the two correlation functions in Eqs.~(\ref{eq:k2kfdef}), (\ref{eq:kundef}). Integrating the Callan-Symanzik equation,
\beq {\rm K}_{A}(x, \rho_0, \Lambda) = Z_s(\ell) {\rm K}_{A}(x , \rho(\ell), e^{-\ell} \Lambda), \label{CSsol}\eeq
where  
\beq Z_s(\ell) = \exp\left[- \int_0^{\ell} d \ell' \eta_s(\rho(\ell'))\right] = \exp\left[\int_{\rho_0}^{\rho(\ell)} \frac{d \rho}{\beta(\rho)} \eta_s(\rho) \right] \approx \frac{1- \kappa \rho(\ell)/2\pi}{1- \kappa \rho_0/2\pi}, \eeq
and $\rho_0 = \rho(\ell =0)$ as defined in \eqref{rhoell}.
Thus, using $\ell = \log \Lambda x$ and approximating ${\rm K}_{2k_f}$ on the right-hand-side of (\ref{CSsol}) by Eq.~(\ref{eq:multichannelspincorr}),
\begin{equation}
    \ev{\psi^\dagger_{Rw} \sigma^a \psi_{Lw}(x,0) S^a} = -\frac{Ks \sin(\rho(\ell))}{2\pi x} \frac{1 - \kappa \rho(\ell)/2\pi}{1 - \kappa \rho_0/2\pi}.
\end{equation}

We now calculate $\ev{j^a_0(x,0) S^a(0)}$. First, to $\mathcal{O}(s^0)$, 
$\ev{j^3_0(x,0) S^3(0)} = 0$ because again the contributions to  $\ev{j^3_0(x,0)}$ from diagrams in Fig.\ \ref{fig:SelfEnergy}  vanishes.
On the other hand, 
\begin{align}
    \ev{j^i_0(x,0) S^i(0)} &= -s\epsilon^{jk} \int \dd \tau \ev{\pi^i(0)\pi^j(\tau)}_0  \ev{j^i_0(x,0) j^k_x(0^+,\tau)}_0 + \mathcal{O}(s^0).
\end{align}
Performing the integral,
\begin{align}
    \ev{j^i_0(x,0) S^i(0)} &= -\frac{K d(\rho) \sin^2 \rho}{\pi^2 x} \left(1 - \frac{\kappa \rho}{2\pi}\right) + O(s^{-1}).
\end{align}
Upon RG improving as done for the $2k_f$ correlator, we arrive at 
\begin{align}
    \ev{j^a_0(x,0) S^a(0)} &= -\frac{K d(\rho(\ell)) \sin^2 \rho(\ell)}{\pi^2 x} 
    \frac{\left(1 - \kappa \rho(\ell)/2\pi\right)^2}{1 - \kappa \rho_0/2\pi}.
\end{align}

\subsection*{Sum rule}
We now follow Refs.~\cite{cloud_1996, cloud_1998,cloud_review} to derive the sum rule (\ref{eq:sumrule}), which holds in the limit of bare Kondo coupling $J \to 0$. We note that in the original paper Ref.~\cite{cloud_1998}, the first (susceptibility) term on the right hand side of (\ref{eq:sumrule}) was missing, even though the sum rule was written at finite temperature. 

Let's begin with the expression for the susceptibility of the full system at finite temperature:
\beq \frac{1}{3} \langle \vec{S}^2_{\rm tot} \rangle = T \chi_{\rm imp} + T L \chi_{\rm b}. \label{Ssq}\eeq
where $\vec{S}_{\rm tot} = \vec{S} + \vec{S}_e$, $\vec{S}_e$ is the electron spin, $\chi_{\rm b}$ is the bulk susceptibility density of a the free 1d fermion gas and $L$ is the size of the system. We can write $\vec{S}^2_{\rm tot} = s(s+1) + \vec{S}_e \cdot \vec{S}  + \vec{S}_e \cdot \vec{S}_{\rm tot}$. Now,
\beq S^a_e = \int_0^{L} dx \, j^a_0(x) = \int_0^{x_0} dx \, j^a_0(x) + \int_{x_0}^{L} dx \, j^a_0(x). \eeq
Here we will be careful to take any contact terms in the integral as $x \to 0$ into account. We  choose $x_0$ to be much smaller than any IR scale ($\xi_K$, $T^{-1}$), but bigger than the UV cut-off. We can then use an OPE,
\beq \int_0^{x_0} dx \, j^a_0(x) = c_1(J, x_0 \Lambda) S^a + \ldots. \eeq
Here $c_1$ can be computed in perturbation theory in $J$ and scales as $c_1 \sim {\cal O}(J)$. This OPE can be used in the equal time correlation function $\langle \int_0^{x_0} dx \, j^a_0(x) S^a_{\rm tot}\rangle$, since $S^a_{\rm tot}$ is a conserved quantity and can be moved to a large temporal separation from $j^a_0(x)$. In addition, it was shown in Ref.~\cite{cloud_1996, cloud_1998} that for $x$ much bigger than the UV cut-off scale
$ \langle j^a_0(x) {S}^a_{\rm tot} \rangle$ does not receive  corrections at any order in the Kondo coupling $J$. Thus,
\beq \langle j^a_0(x)  S^a_{\rm tot}\rangle = 3 T \chi_{\rm b}, \quad x \gg \Lambda^{-1} , \eeq
and
\beq \langle  \vec{S}_e \cdot \vec{S}_{\rm tot} \rangle \approx 3 T L  \chi_{\rm b} + c_1(J, x_0 \Lambda) \langle \vec{S} \cdot \vec{S}_{\rm tot}\rangle.\eeq
Since the left hand side is independent of $x_0$, $c_1$ must also be independent of $x_0$. Thus, from (\ref{Ssq}),
\beq \frac{1}{3} \langle \vec{S} \cdot \vec{S}_e \rangle = T (1+c_1)^{-1} \chi_{\rm imp}(T) -\frac{s(s+1)}{3}. \eeq
Finally,
\beq \langle \vec{S} \cdot \vec{S}_e \rangle = \int_0^{x_0} dx \, \langle S^a j^a_0(x)\rangle +  \int_{x_0}^{\infty} dx \, \langle S^a j^a_0(x) \rangle. \eeq
Using the OPE we may replace the first term in the equation above by a $T$-independent constant $c_0(J, x_0 \Lambda)$ that can be computed perturbatively in $J$. Thus,
\beq \frac{1}{3} \int_{x_0}^{\infty} dx \, \langle S^a j^a_0(x) \rangle =  T (1+c_1)^{-1} \chi_{\rm imp}(T) -\frac{s(s+1)}{3} - c_0(J, \Lambda x_0).\eeq
Taking $J \to 0$ while keeping $x_0$ fixed, $c_0$ and $c_1$ go to zero and we obtain our desired sum rule.

\bibliography{bibliography-short}
\bibliographystyle{JHEP}

\end{document}

%% file: defs.tex
\newcommand{\nn}[2]{\langle #1 #2 \rangle}

\renewcommand{\d}{\partial}
\renewcommand{\nn}{\nonumber}

\newcommand{\beq}{\begin{equation}}
\newcommand{\eeq}{\end{equation}}
\newcommand{\ba}{\begin{array}{ccc}}
\newcommand{\ea}{\end{array}}
\newcommand{\bea}{\begin{eqnarray}}
\newcommand{\eea}{\end{eqnarray}}

\usepackage{simpler-wick}